\documentclass[11pt,a4paper]{article}
\pdfoutput=1
\usepackage[pdftex]{graphics}
\usepackage{jheppub}
\usepackage{amsmath,amssymb,amsfonts}
\usepackage{enumitem}
\usepackage{multirow}
\usepackage{array,booktabs}
\usepackage{slashed}

\usepackage{accents}
\usepackage{mathrsfs}
\usepackage{mathtools}

\usepackage[usenames,dvipsnames]{xcolor}
\definecolor{labelcolor}{RGB}{194, 175, 116}
\definecolor{rmkcolor}{RGB}{215,30,5}
\definecolor{feyntext}{RGB}{20,125,233}
\definecolor{lred}{RGB}{255,130,130}
\definecolor{llred}{RGB}{255,160,160}
\definecolor{skyblue}{RGB}{34,139,230}
\definecolor{navy}{rgb}{0,0,0.7}
\definecolor{purple}{RGB}{171,1,207}
\definecolor{lgreen}{RGB}{231, 242, 43}
\definecolor{lgray}{RGB}{135, 135, 145}
\definecolor{recur}{RGB}{231, 242, 43}

\definecolor{rowactioncolor}{RGB}{135, 135, 145}
\definecolor{columnactioncolor}{RGB}{0,143,186}
\definecolor{domaincolor}{RGB}{63,72,204}

\usepackage[export]{adjustbox}
\usepackage[final]{showlabels} %

\newcommand{\fref}[1]{Fig.\,\ref{#1}}
\renewcommand{\eqref}[1]{Eq.\,(\ref{#1})}
\newcommand{\eqrefs}[2]{Eqs.\,(\ref{#1}) and (\ref{#2})}
\newcommand{\eqrefsor}[2]{Eqs.\,(\ref{#1}) or (\ref{#2})}
\newcommand{\eqrefss}[3]{Eqs.\,(\ref{#1}), (\ref{#2}), and (\ref{#3})}
\newcommand{\Sec}[1]{Sec.\,\ref{#1}}
\newcommand{\Secs}[2]{Secs.\,\ref{#1} and \ref{#2}}
\newcommand{\App}[1]{App.\,\ref{#1}}
\newcommand{\rcite}[1]{Ref.\,\cite{#1}}
\newcommand{\rrcite}[1]{Refs.\,\cite{#1}}
\newcommand{\transition}[1]{\qquad\adjustbox{scale=0.95}{\text{#1}}\qquad}

\DeclareMathOperator{\tr}{tr}

\DeclareMathOperator{\sgn}{sgn}

\DeclareMathOperator{\ad}{ad}
\DeclareMathOperator{\Ad}{Ad}

\def\mem{\hspace{0.1em}}
\def\hem{\hspace{0.05em}}
\def\nem{\hspace{-0.1em}}
\def\hnem{\hspace{-0.05em}}
\def\hhem{\hspace{0.025em}}
\def\hhnem{\hspace{-0.025em}}

\def\blank{{\,\,\,\,\,}}

\def\qiq{{\quad\implies\quad}}
\def\qfq{{\quad\iff\quad}}

\def\a{\alpha}
\def\b{\beta}

\def\d{{\delta}}

\def\ve{\varepsilon}
\def\m{\mu}

\def\s{\sigma}

\def\l{\lambda}

\def\bpsi{{\smash{\bar{\psi}}\kern0.02em\vphantom{\psi}}}
 \def\mathe{{\scalebox{1.01}[1]{$\mathrm{e}$}}}

\def\mwedge{{\mem\wedge\mem\hhem}}

\def\tensor\otimes

\def\MPl{{M_\text{Pl}}}

\newcommand{\wrap}[1]{{\smash{#1}\vphantom{\b}}}

\def\lsq{{
		\kern-0.037em
		\adjustbox{scale=0.90,valign=c}{$
			{
				\adjustbox{raise=-0.09em}{$\lfloor$}
				\llap{\reflectbox{\rotatebox[origin=c]{180}{$\lfloor$}}}
			}
			$}
		\kern-0.04em
}}
\def\rsq{{
		\kern-0.04em
		\adjustbox{scale=0.90,valign=c}{$
			{
				\rlap{\reflectbox{\rotatebox[origin=c]{180}{$\rfloor$}}} 
				\adjustbox{raise=-0.09em}{$\rfloor$}
			}
			$}
		\kern-0.037em
}}

\def\O{\mathcal{O}}
\def\P{\mathcal{P}}
\def\A{\mathcal{A}}

\def\G{\mathcal{G}}

\def\X{\mathcal{X}}

\newcommand{\BB}[1]{\Big(\,{#1}\,\Big)}
\newcommand{\bb}[1]{\bigg(\,{#1}\,\bigg)}
\newcommand{\bbsq}[1]{\bigg[\,{#1}\,\bigg]}
\newcommand{\bigbig}[1]{\big(\mem{#1}\mem\big)}
\newcommand{\lrp}[1]{\left(\,{#1}\,\right)}

\newcommand{\act}[1]{[\,{#1}\,]}

\newcommand{\Ket}[1]{{\hem\big|\hem{#1}\big\rangle}}
\newcommand{\Bra}[1]{{\big\langle{#1}\hem\big|\hem}}
\newcommand{\BraKet}[2]{{\big\langle{#1}\hem\big|\hem{#2}\big\rangle}}

\newcommand{\dbar}{
	d\kern-.20em\makebox[0pt][l]{\kern0.01em\adjustbox{raise=-0.01em}{\scalebox{1.4}[1.0]{$\bar{}$}}\kern-0.01em}\kern.20em
}
\newcommand{\deltabar}{
	\delta\kern-.20em\makebox[0pt][l]{\adjustbox{raise=-0.01em}{\scalebox{1.0}[1.0]{$\bar{}$}}}\kern.20em
}

\usepackage{bm}

\usepackage{dsfont}

\newcommand{\D}[1]{\mathcal{D}\hnem{#1}\mem}

\newcommand{\expval}[1]{
	\big\langle\hem{
		#1
	}\hem\big\rangle
}

\newcommand{\Bexpval}[1]{
	\Big\langle\,{
		#1
	}\,\Big\rangle
}

\let\oldcap\cap
\renewcommand{\cap}{{\,\oldcap\,}}

\setlist[itemize]{
	label=\adjustbox{scale=0.7}{$\bullet$}, itemsep=0pt,topsep=0px
}
\setlist[enumerate]{
	itemsep=0pt,topsep=3px
}

\usepackage{tikz}
\usetikzlibrary{calc} 
\usetikzlibrary{shapes.geometric} 
\usetikzlibrary{positioning} 
\usetikzlibrary{fit} 
\usepackage[a]{esvect} 
\tikzset{empty/.style = {inner sep = 0pt, outer sep = 0, minimum size = 0}}
\tikzset{b/.style = {inner sep = 2pt, outer sep = 4pt, minimum size = 12pt}}
\tikzset{c/.style = {inner sep = 2pt, outer sep = 4pt, minimum size = 12pt}}
\tikzset{w/.style = {inner sep = 1pt, outer sep = 2pt, minimum size = 12pt, anchor = west}}
\tikzset{s/.style = {inner sep = 2.5pt, outer sep =2.5pt, minimum size = 1pt, font = \small}}
\tikzset{lin/.style = {draw, line width = 0.5pt}}
\usepackage[export]{adjustbox}

\definecolor{skyblue}{RGB}{34,139,230}
\definecolor{indigo}{RGB}{63,72,204}
\definecolor{olg}{RGB}{100,154,0}
\definecolor{or}{RGB}{253,111,0}
\definecolor{pink}{RGB}{255, 69, 149}

\definecolor{emphcolor}{RGB}{0,143,186}

\usepackage[outline]{contour}
\contourlength{0.1pt}

\renewcommand{\emph}[1]{\textcolor{emphcolor}{\contour{emphcolor}{{\textit{#1}}}{\kern0.07em}}}

\newcommand{\defn}[1]{\textcolor{defcolor}{\contour{defcolor}{{\textit{#1}}}{\kern0.07em}}}

\definecolor{peierlscolor}{RGB}{255, 78, 139}

\newcommand{\pemph}[1]{\textcolor{peierlscolor}{\contour{peierlscolor}{{\textit{#1}}}{\kern0.07em}}}

\def\R{\mathbb{R}}

\def\tphi{\widetilde{\phi}}

\def\Diff{\mathrm{Diff}}

\newcommand{\Pexp}[1]{
    \mathrm{T}\kern-0.1em\exp\nem
    \bigg(\hem{
        #1
    }\bigg)
}

\def\lb{\{\kern-0.15em\{}
\def\rb{\}\kern-0.15em\}}

\newcommand{\pb}[2]{{\{\hem{#1},{#2}\hem\}}}
\newcommand{\spb}[2]{{\{\hem{#1},{#2}\hem\}}{}^\star }

\newcommand{\comm}[2]{[\hem{#1},{#2}\hem]}

\newcounter{alphnum}

\counterwithin*{alphnum}{subsubsection}

\def\Del{\mathit{\Delta}}

\DeclareMathOperator{\Aut}{Aut}
\DeclareMathOperator{\Der}{Der}

\def\trho{\tilde{\rho}}
\def\tH{\tilde{H}}
\def\tV{\tilde{V}}

\def\ba{{\bar{a}}}
\def\C{\mathbb{C}}

\def\tx{\tilde{x}}

\def\vk{{\vec{k}}}
\def\vx{{\vec{x}}}
\def\vex{{\vec{x}}}

\def\Cinfty{C^\infty\hnem}
\def\Hilb{\mathcal{H}}
\DeclareMathOperator{\Ops}{Ops}
\def\Q{\mathcal{Q}}
\def\bl{\bar{\lambda}}
\def\ba{\bar{a}}
\def\Delt{\hat{\Delta}}

\newcommand{\normal}[1]{{:\hnem{#1}\hnem:}}

\def\Uo{{U^\circ}\hnem}

\def\sX{X^\star}
\def\sG{{G^\star}\nem}
\def\sU{{U^\star}\hnem}
\def\sUo{{U^\star{}^\circ}\hnem}

\def\sS{{S^\star}\hnem}
\def\schi{{\chi^\star}\nem}

\def\hchi{\hat{\chi}}

\newcommand{\formal}[1]{[\nem[#1]\nem]}

\def\chia{\chi_{(1)}^{\vphantom{\star}}}
\def\chib{\chi_{(2)}^{\vphantom{\star}}}
\def\chic{\chi_{(3)}^{\vphantom{\star}}}
\def\chin{\chi_{(n)}^{\vphantom{\star}}}

\def\schia{\chi_{(1)}^\star}
\def\schib{\chi_{(2)}^\star}
\def\schic{\chi_{(3)}^\star}
\def\schin{\chi_{(n)}^\star}

\DeclareMathOperator{\Mag}{Mag}

\DeclareMathOperator{\Fuz}{Fuz}

\newcommand{\qwrap}[1]{{\quad #1 \quad}}
\newcommand{\bwrap}[1]{{\blank #1 \blank}}
\newcommand{\awrap}[1]{{\,\, #1 \,\,}}

\def\pen{\includegraphics[valign=c]{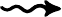}}
\def\qpq{\quad\pen\quad}
\def\qqpqq{\quad\quad\pen\quad\quad}

\newcommand{\inc}[2]{%
    {\includegraphics[#1]{figs/#2}}%
}
\newcommand{\qinc}[2]{%
    {\quad\includegraphics[#1]{figs/#2}\quad}%
}
\newcommand{\binc}[2]{%
    {\blank\includegraphics[#1]{figs/#2}\blank}%
}
\newcommand{\ainc}[2]{%
    {\,\,\includegraphics[#1]{figs/#2}\,\,}%
}

\DeclareMathOperator{\Desc}{Desc}
\DeclareMathOperator{\Prim}{Prim}

\def\and{{\,\,\texttt{\&}\,\,}}

\def\toT{{\to_\mathrm{T}}}
\def\toO{{\to_\mathrm{O}}}

\def\sp{\Omega}

\definecolor{customlink}{RGB}{0,31,133}

\hypersetup{
    citecolor=indigo,
    linkcolor=indigo,
    urlcolor=indigo
}

\title{
	Phase Space Formulation of S-matrix
}

\author[a]{Joon-Hwi Kim}

\affiliation[a]{Walter Burke Institute for Theoretical Physics,\\ California Institute of Technology, Pasadena, CA 91125}

\emailAdd{joonhwi@caltech.edu}

\abstract{
    We establish an exact relation between the S-symplectomorphism and the S-matrix 
    by means of the phase space formulation of quantum mechanics.
    The adjoint action of the S-matrix defines a fuzzy diffeomorphism on phase space whose classical limit is the S-symplectomorphism.
    The relation between classical and quantum eikonals is immediate via $\hbar$-deformation of each Poisson bracket in the Magnus formula.
    Diagrammatic computation of quantum eikonal is illustrated for quantizations in both symmetric and normal orderings.
}

\renewcommand*{\bibfont}{\footnotesize}

\begin{document}

\begin{flushright}
    \footnotesize
    CALT-TH 2025-033
\end{flushright}
\maketitle

\bibliographystyle{utphys-modified}
\renewcommand*{\bibfont}{\footnotesize}

\newpage

\section{Introduction}

\begin{figure}[t]
    \centering
    \includegraphics[scale=1.0]{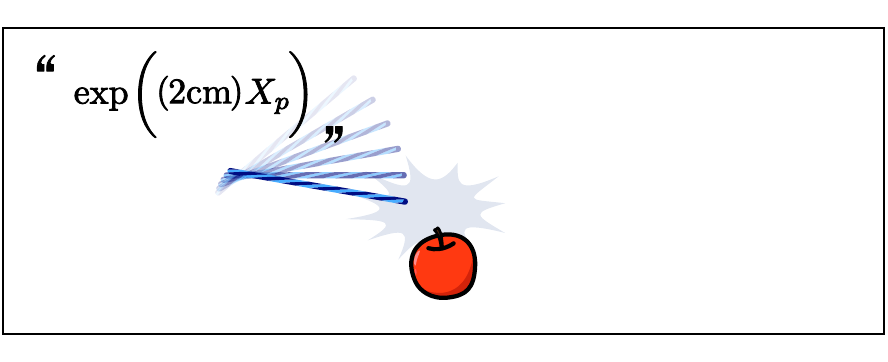}
    \includegraphics[scale=1.0]{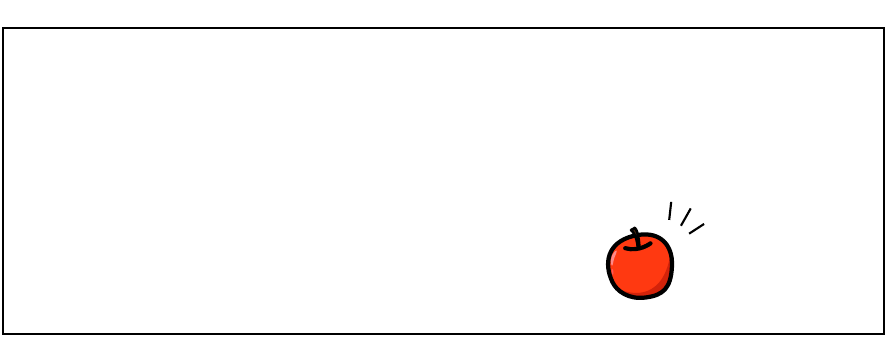}
    \caption{
        Translation is a symplectomorphism,
        whose generator is the momentum $p$.
    }
    \label{cartoonA}
\end{figure}

A variety of formulations
exist for classical mechanics.
Among them, Hamiltonian mechanics could stand out
in terms of its geometrical approach
based on phase space.
In a view,
the appeal of Hamiltonian mechanics
is that it naturally encourages an ``operator-like''---
or ``object-oriented''---
way of approaching classical mechanics.
Classical time evolution is an \textit{operator}
that acts on the phase space as an \textit{object}.
In \fref{cartoonA},
spatial translation 
is implemented by
exponentiating a Hamiltonian vector field
$X_p = \pb{p}{\blank}$
as an \textit{operator}.

Indeed,
this mode of thinking
provides glimpses into
the more fundamental framework,
i.e.,
quantum mechanics.
Quantum time evolution is a unitary \textit{operator} that acts on the Hilbert space as an \textit{object}.
Spatial translation
is generated by an exponential \textit{operator} that arises from the momentum.

In the fancy terms of category theory
\cite{eilenberg1945general,mac1998categories,Penrose:1956tensormethods,penrose1971negdim,joyal1991geometry,freyd1989braided,selinger2010survey,coecke2006kindergarten,coecke2015categorical},
we say that Hamiltonian mechanics
is based on the category $\mathsf{Symp}$.
\textit{Objects} are phase spaces as symplectic manifolds.
\textit{Morphisms} are maps between phase spaces that preserve the symplectic structure,
which are referred to as symplectomorphisms.
Classical time evolution is a symplectomorphism on phase spaces.
Transformations such as spatial translations are also symplectomorphisms on phase spaces.

Similarly, quantum mechanics is based on the category $\mathsf{Hilb}$.
\textit{Objects} are Hilbert spaces.
\textit{Morphisms} are operators.
Quantum time evolution is a unitary operator.
Spatial translation is also a unitary operator.


When applied to scattering theory,
the above analogy
between classical and quantum mechanics
elicits the idea of
``S-symplectomorphism.''
The S-matrix
is the unitary operator
mapping the initial Hilbert space of scattering states
to the final Hilbert space of scattering states.
Thus, its classical counterpart
shall be the symplectomorphism
mapping the initial phase space of scattering states
to the final phase space of scattering states,
which we dub S-symplectomorphism.

\begin{figure}[t]
    \centering
    \includegraphics[scale=1.0]{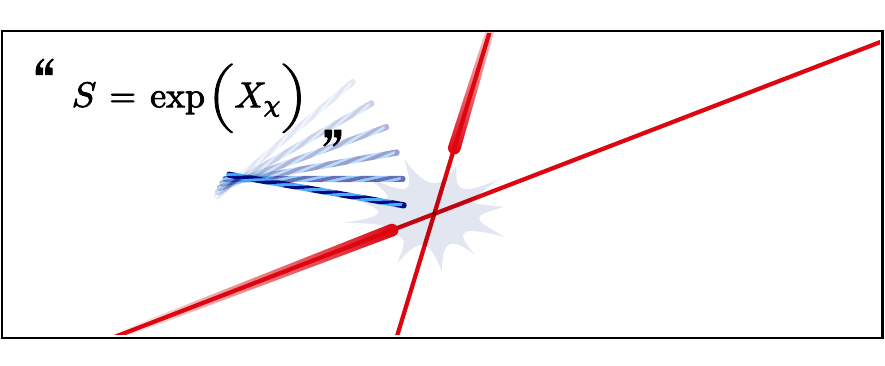}
    \\
    \includegraphics[scale=1.0]{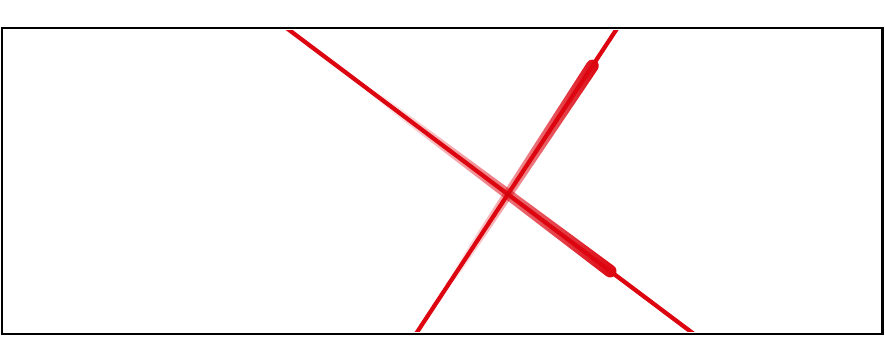}
    \caption{
        Scattering is a symplectomorphism,
        whose generator is the eikonal $\chi$.
    }
    \label{cartoonS}
\end{figure}

A cartoon is provided in \fref{cartoonS},
depicting the action of an S-symplectomorphism
for a two-particle system.
Just like the translation operator in \fref{cartoonA},
the S-symplectomorphism $S$ can be represented 
as the exponentiation of a Hamiltonian vector field
$X_\chi = \pb{\chi}{\blank}$.
This generator $\chi$ is referred to as the classical eikonal,
as it serves as the analog of the eikonal matrix $\hat{\chi}$ in quantum mechanics:
$\hat{S} = \exp(\hat{\chi}/i\hbar)$.

Historically,
the concept of S-symplectomorphism was proposed by
Hunziker \cite{hunziker1968s}
in 1968
and had been investigated in works
\cite{simon1971wave,herbst1974classical,sokolov1979classical,osborn1980levinson,narnhofer1981canonical,thirring1981classical} in the '70s and '80s
(called ``S-map,'' ``S-transformation,'' or ``canonical S-transformation'').
An impressive application of the idea
is given by \rcite{osborn1980levinson},
as is reviewed in
\rcite{thirring1981classical}.
It is shown that
the classical analog of Levinson's theorem \cite{levinson1949uniqueness}
follows from the volume-preserving property
of S-symplectomorphism,
relating the phase-space volume occupied by
bound orbits in an attractive potential
to an integral over time delay for scattering orbits.

The idea of S-symplectomorphism has been also
independently accessed by the present author
during the work \cite{ambikerr1},
which could be concretized through later works \cite{eikonaltwistor,eikonalsangmin1}.
The purely classical formulation of S-symplectomorphism
has been established in \rcite{Kim:2025sey}.

The identification of the classical eikonal $\chi$
as the generator of S-symplectomorphism
traces back to the work \cite{narnhofer1981canonical}
by Narnhofer and Thirring,
where it is found---yet for simple examples---that
``The quasiclassical phase shift is identified as the generator of the classical canonical S transformation.''
This observation has been 
established to full extent of generality
by \rrcite{eikonaltwistor,eikonalsangmin1},
where a nested bracket structure is essential.

The analogies between
the S-symplectomorphism $S$ and the S-matrix $\hat{S}$,
and also 
the classical eikonal $\chi$ and the eikonal matrix $\hat{\chi}$,
have been argued
from the correspondence principle
or related ideas
\cite{simon1971wave,herbst1974classical,sokolov1979classical,osborn1980levinson,narnhofer1981canonical,thirring1981classical,ambikerr1,eikonaltwistor,eikonalsangmin1}.
However, to our best understanding,
it has been not shown that
the S-symplectomorphism
arises as the faithful $\hbar\to0$ limit
of a quantum-mechanical construct.
The claims
have remained
at the levels of analogies or arguments,
which do not describe exact mathematical equalities.

The physical relevance of this problem
would be evaluated from the angle of
modern applications of scattering theory.
As the work \cite{KMOC} by
Kosower, Maybee, and O'Connell
has nicely reviewed and systematized,
the punchline has been 
``Extract classical observables
from the quantum S-matrix.''
A formalism based on the eikonal matrix
has also emerged,
proposed in \rcite{eikonaltwistor}
while tracing back to 
\rrcite{Gonzo:2024zxo,Damgaard:2023ttc}
as well.
The latter formalism
has shown that
the S-symplectomorphism provides
a purely classical framework for the classical eikonal,
the point of which is explicitly emphasized in \rcite{Kim:2025sey}.
Hence the punchline shifts to
``Extract classical observables from the S-symplectomorphism.''
To understand the precise relation 
between these two paradigms,
however,
an exact relationship needs to be clarified
between the S-symplectomorphism $S = \exp( X_\chi )$
and the S-matrix $\hat{S} = \exp(\hat{\chi}/i\hbar)$.

An immediate issue is 
a mismatch in ``data types.''
The S-matrix $\hat{S}$ is a linear transformation on a Hilbert space.
The literal $\hbar \to 0$ limit
cannot equate it to
the diffeomorphism $S$ on a phase space,
suddenly turning the Hilbert space into a symplectic manifold.
In the same way,
the eikonal matrix $\hat{\chi}$ 
cannot be directly equated with
a scalar-valued function $\chi$.
So to speak,
they are ``apples and oranges'' living in two separate worlds,
so any comparison is grammatically nonsense
unless a ``translator'' is provided.
Furthermore,
the classical limit of $\hat{S} = \exp(\hat{\chi}/i\hbar)$
is still divergent
even if 
the classical limit of $\hat{\chi}$ 
could 
describe
$\chi$.

In this paper,
we derive
the precise relation between
the S-matrix and the S-symplecto\-morphism
in terms of exact mathematical equalities.
Crucially,
this is facilitated by the phase space formulation of quantum mechanics,
which provides the ``translator''
that recasts
operators on Hilbert spaces
as scalar functions on phase spaces.

The phase space formulation of quantum mechanics
was
put forward by
Moyal and Groenewold
\cite{Moyal:1949sk,Groenewold:1946kp}
in the '40s,
based on earlier constructions by Weyl and Wigner \cite{weyl1927quantenmechanik,Wigner:1932eb}.
It is equivalent to the density matrix formulation of quantum mechanics,
where one focuses on operators instead of wavefunctions.
See \rrcite{Curtright:2011vw,zachos2005quantum,
leonhardt1997measuring}
for comprehensive expositions on the subject.
It
is also roughly synonymous to 
deformation quantization \cite{bayen1977quantum,bayen1978deformationI,bayen1978deformationII,berezin1975general,fedosov1994simple,kontsevich,weinstein1995deformation},
in which case
the geometrical picture
describes
a fuzzy (noncommutative) phase space.

In \Sec{PSFOR},
we review the phase space formulation of quantum mechanics in a friendly and concrete fashion,
since the formulation may not be standard or well-known 
depending on the community.
In \Sec{SCA},
we develop scattering theory in the phase space formulation of quantum mechanics,
which is new to our best knowledge.
In \Sec{LIMIT},
we establish the main statements about the classical limit.
The adjoint action of the S-matrix
is equivalent to the deformation of the S-symplectomorphism.
The eikonal matrix
is equivalent to the deformation of the classical eikonal.

Our formalism facilitates
a systematic diagrammatic evaluation of quantum eikonal
to all loop orders,
as shown in \Sec{MAGNUS}.
Explicitly, we compute
the all-loop quantum eikonal
at three vertices
in both symmetric and normal orderings.
Our approach and results are new,
despite the recent work \cite{Pichini:2025igz}
appeared during the preparation of this paper.
Mathematically, the implication is that
Magnus expansion \cite{magnus1954exponential}
in deformation quantization
defines two new graph functions
on directed acyclic kinds of graphs,
generalizing the one due to Murua \cite{Murua_2006}.

\newpage

\newpage

\section{Phase Space Formulation}
\label{PSFOR}

\subsection{Classical Mechanics}
\label{CLAS}

To initiate our journey in a friendly fashion,
let us suppose a simple concrete example:
a particle on a one-dimensional line $\R$.
The classical state of this particle
is represented as the pair $(x,p)$
of position $x$ and momentum $p$.
The \emph{phase space} is
the space $\P = \R^2$ of this pair $(x,p)$,
where each point represents a classical state.
Classical observables are smooth functions on the phase space,
the space of which is denoted as $\Cinfty(\P)$.

The mathematical model of phase space is \emph{symplectic geometry},
which we presume within this paper.
The symplectic form
is a nondegenerate closed two-form:
\begin{align}
    \label{xp.omega}
    \sp
    \,=\,
        dp \wedge dx
    \,.
\end{align}
The symplectic form $\sp$
induces the following geometric structures on the phase space.

First,
the phase space is equipped with the \emph{Poisson bracket}.
The Poisson bracket is a linear bi-differential operator
that takes two classical observables $f,g \in \Cinfty(\P)$ as input
and returns another classical observable $\pb{f}{g} \in \Cinfty(\P)$ as output.
For our one-dimensional particle, the Poisson bracket is defined by the canonical relations
\begin{align}
    \label{xp.pb}
    \pb{x}{x}
    \,=\,0
    \,,\quad
    \pb{x}{p}
    \,=\, 1
    \,,\quad
    \pb{p}{p}
    \,=\, 0
    \,.
\end{align}
The Poisson bracket in \eqref{xp.pb} is in an ``inverse'' relationship with the symplectic form in \eqref{xp.omega},
so knowing the former implies knowing the latter and vice versa.

Second,
the phase space is equipped with the \emph{Liouville measure},
a top-degree differential form
that provides an invariant integration measure
over the phase space.
It is constructed from the symplectic form as
\begin{align}
    \mu
    \,=\,
        \frac{dp \mwedge dx}{\varepsilon}
    \,.
\end{align}
Here, $\varepsilon$
is an ad-hoc constant
whose sole purpose is to make
$\mu$
dimensionless.

The time evolution of the particle's classical state
is described by the \emph{Hamiltonian} \emph{equations of motion}.
For a generic time-dependent Hamiltonian $H(t) \in \Cinfty(\P)$,
the Hamiltonian equations of motion read
\begin{align}
    \label{heom.xp}
    \dot{x}
    \,=\,
        \pb{x}{H(t)}
    \,,\quad
    \dot{p}
    \,=\,
        \pb{p}{H(t)}
    \,.
\end{align}
For example, $H(t) = p^2\nem/2m - F(t)\mem x$
will implement a particle of mass $m$
driven by
a time-dependent external force $F(t)$.
Evidently, \eqref{heom.xp}
is a first-order differential equation
that describes
how a point on $\P$
moves within time.

More generally, one can describe an ensemble of the same particle
in terms of the \emph{Liouville equation}:
\begin{align}
	\label{liouv}
	\dot{\rho}(t)
	\,=\,
		\pb{H(t)}{\rho(t)}
	\,.
\end{align}
Here, $\rho(t) \in \Cinfty(\P)$ is a time-dependent probability distribution
on $\P$.
It is unit-normalized 
with respect to the Liouville measure,
encoding the unity of total probability:
\begin{align}
	\label{1d.norm}
	\int \frac{dp \mwedge dx}{
		\varepsilon
	}\,\,
		\rho(t)
	\,=\,
		1
	\,.
\end{align}
\newpage

The classical time evolution can be studied by
either of \eqrefsor{heom.xp}{liouv}.
Physically, the former is a special case of the latter:
a single particle
is an extreme case of an ensemble of particles.
Geometrically, the former describes a map that sends a point in $\P$ to another,
whereas the latter describes
the evolution of test functions on $\P$.
These are two equivalent ways for
representing a diffeomorphism on $\P$
\cite{vershikgelfand1975diffreps}.

Intuitively speaking,
one can picture classical time evolution
as a fluid that fills in and streams through the phase space $\P$.
To grasp the flow of this fluid,
one could put a neutrally buoyant bubble at each point in $\P$
and track its trajectory.
Alternatively,
one could also study how splashes of inks
on the fluid would
change their shapes under time.

With this remark made,
let us stick to the latter description
in light of its
practical and conceptual advantages
as well as its relevance 
that will become evident shortly.
In this approach,
the classical time evolution
can be explicitly \textit{computed} as the following.

First of all,
for each function $f\in\Cinfty(\P)$
on phase space,
define $X_f$ as the first-order differential operator such that
\begin{align}
	\label{X}
	X_f\act{g}
	\,=\,
		\pb{f}{g}
	\,,\quad
	\forall\mem
		g \in \Cinfty(\P)
	\,.
\end{align}
$X_f$ is known as the \emph{Hamiltonian vector field} of $f$.
This terminology appeals to
the well-known mathematical equivalence between
vector fields
and first-order differential operators.
Importantly,
Hamiltonian vector fields
are differential operators
that preserve the algebraic structure of classical observables
due to pointwise addition, pointwise multiplication,
and Poisson bracket:
\begin{subequations}
\label{ag.diff}
\begin{align}
	\label{ag.diff.a}
	X_f\act{
		g + h
	}
	\,&=\,
		X_f\act{g} + X_f\act{h}
	\,,\\
	\label{ag.diff.b}
	X_f\act{
		g\hem h
	}
	\,&=\,
		(X_f\act{g})\hem h
		+
		g\hem (X_f\act{h})
	\,,\\
	\label{ag.diff.c}
	X_f\act{
		\pb{g}{h}
	}
	\,&=\,
		\pb{X_f\act{g}}{h}
		+
		\pb{g}{X_f\act{h}}
	\,.
\end{align}
\end{subequations}
This implies that Hamiltonian vector fields
preserve the symplectic structure.

Second of all,
provided the definition in \eqref{X},
the Liouville equation in \eqref{liouv} can be rewritten and solved as
\begin{align}
    \label{liouv.X}
    \dot{\rho}(t)
    \,=\,
        X_{H(t)}\act{
            \rho(t)
        }
    \qiq
    \rho(T)
    \,=\,
        U^*\act{
            \rho(0)
        }
    \,,
\end{align}
where we have arbitrarily set the initial and final times as $t = 0$ and $t = T$.
In \eqref{liouv.X},
$U^*$ is a differential operator that transforms the initial classical (ensemble) state $\rho(0)$
to the final classical state $\rho(T)$.
Well-established results
due to Dyson \cite{Dyson:1949ha}
and
Magnus \cite{magnus1954exponential}
facilitate the explicit formulae
\begin{align}
	\label{U.sol}
	U^*
	\,=\,
		\Pexp{\mem
			\int_0^T dt\,\,
				X_{H(t)}
		}
	\,=\,
		\exp\nem\BB{
			X_G
		}
	\,,
\end{align}
respectively.
In \eqref{U.sol},
$G \in \Cinfty(\P)$
is given by
\begin{align}
	\label{G+3}
	G
	\,=\,
		&
		\int dt_1\,\,
			H(t_1)
		+
		\frac{1}{2}\mem
		\int_{t_1>t_2}\nem d^2t\,\,
			\pb{H(t_1)}{H(t_2)}
\\
		&
		+
		\frac{1}{6}\mem
		\int_{t_1>t_2>t_3}\nem d^3t\,\,
			\BB{
				\pb{ H(t_1) }{ \pb{ H(t_2) }{ H(t_3) } }
				+
				\pb{ H(t_3) }{ \pb{ H(t_2) }{ H(t_1) } }
			}
		+ \cdots
	\,,
	\nonumber
\end{align}
where the default integral bounds are $[0,T]$.

\newpage

An important feature of
\eqref{G+3}
is that
each integrand is composed by nesting Poisson brackets.
This property
crucially counts on
the Jacobi identity
\begin{align}
	\label{Xcomm}
	\comm{X_{f}}{X_{g}}
	\,=\,
		X_{\pb{f}{g}}
	\,,
\end{align}
which encodes that Hamiltonian vector fields 
as first-order differential operators
are closed under the commutator.
Note that
\eqref{Xcomm} demands the closure of the symplectic form.

The formula
$U^* = \exp(X_G)$
in \eqref{U.sol}
establishes that
$U^*$ is the exponentiation of a Hamiltonian vector field.
This fact has the following implications.
Firstly, 
there exists a diffeomorphism $U : \P \to \P$
such that
$U^*$ implements its pullback:
$U^* = f \mapsto f \circ U^{-1}$.
This is because $U^*$ is the exponentiation of a first-order differential operator,
which can be checked by verifying the following 
conditions for a pullback:
\begin{subequations}
\label{ag.Diff}
\begin{align}
    \label{ag.Diff.a}
	U^*\act{
		f + g
	}
	\,&=\,
		U^*\act{f} + U^*\act{g}
	\,,\\
    \label{ag.Diff.b}
	U^*\act{
		f\hem g
	}
	\,&=\,
		U^*\act{f}\, U^*\act{g}
	\,.
\end{align}
Secondly,
the diffeomorphism $U$
preserves the symplectic structure.
This is because $X_G$ is not just an any first-order differential operator
but a Hamiltonian vector field,
which preserves the symplectic structure
as per
\eqref{ag.diff.c}.
Explicitly, for any classical observables $f,g \in \Cinfty(\P)$ it holds that
\begin{align}
    \label{ag.Diff.c}
	U^*\act{
		\pb{f}{g}
	}
	\,=\,
        \pb{ U^*\act{f} }{ U^*\act{g} }
	\,,
\end{align}
\end{subequations}
which is the exponentiated version of \eqref{ag.diff.c}.

Diffeomorphisms that preserve the symplectic structure
is called \emph{symplectomorphisms}.
The groups of diffeomorphisms and symplectomorphisms
on the phase space
are denoted as $\Diff(\P)$ and $\Diff(\P,\sp)$,
respectively.
The latter is a Lie subgroup of the former.

From this analysis,
we conclude that
the solution to the Liouville equation,
\eqref{liouv.X},
has computed a symplectomorphism on the phase space,
\begin{align}
    \label{U-in-Diff}
    U \,\,\,\in\,\,\, \Diff(\P,\sp)
    \,,
\end{align}
by representing it as the differential operator $U^*$
implementing the pullback.
To reiterate, $U^* : \rho(0) \mapsto \rho(T)$
transforms the initial classical state to the final classical state.
Therefore, the symplectomorphism $U$ in \eqref{U-in-Diff}
is the map from the initial phase space to the final phase space
due to the classical time evolution.
Hence, $U$ in \eqref{U-in-Diff} is referred to as the \emph{time-evolution symplectomorphism}.

Of course, the fact that classical time evolution is a symplectomorphism on phase space
is a well-known postulate of Hamiltonian mechanics.
The purpose of the above calculation and discussion
is to remind ourselves
how this postulate
is concretely approached
by solving the Liouville equation in \eqref{liouv}
(or the Hamiltonian equations of motion in \eqref{heom.xp}).

\subsection{Quantum Mechanics}
\label{QU}

Now let us concern the quantum mechanics of the particle.
In the standard formulation,
the starting point
is the declaration of the \emph{Hilbert space} as the space of square-integrable functions on the real line: $\Hilb = L^2(\R)$.
As is well-known, this is the space where well-behaved wavefunctions dwell in.

\newpage

An operator is a linear map
$\Hilb \to \Hilb$
that sends a wavefunction to another.
Especially, the position and momentum operators are defined as
\begin{align}
\begin{split}
	\label{xpdef.Hilb}
	\hat{x}
	&\,\,\,:\,\,\,
	\BB{
		x \,\mapsto\, \psi(x)
	}
	\,\,\mapsto\,\,
	\BB{
		x \,\mapsto\, x\mem \psi(x)
	}
	\,,\\
	\hat{p}
	&\,\,\,:\,\,\,
	\BB{
		x \,\mapsto\, \psi(x)
	}
	\,\,\mapsto\,\,
	\bb{
		x \,\mapsto\,
			-i\hbar\mem \frac{\partial}{\partial x} \psi(x)
	}
	\,.
\end{split}
\end{align}
The canonical commutation relations hold as
\begin{align}
	\label{xp.ccr}
    \comm{ \hat{x} }{ \hat{x} }
    \,=\,
        0
    \,,\quad
	\comm{ \hat{x} }{ \hat{p} }
	\,=\,
		i\hbar\, \hat{1}
	\,,\quad
    \comm{ \hat{p} }{ \hat{p} }
    \,=\,
        0
    \,.
\end{align}

\subsubsection{The Quantization Map}
\label{QMAP}


Let $\Ops(\Hilb)$ be the space of operators,
subject to one's favorite mathematical assumptions.
To construct the quantum theory fully,
one identifies the quantum-mechanical counterpart $\hat{f} \in \Ops(\Hilb)$ of
each function $f \in \Cinfty(\P)$ on the phase space $\P = \R^2$.
Mathematically, 
this means to define
a linear map
\begin{align}
	\Q
	\,\,\,:\,\,\,
	\Cinfty(\P) 
	\,\,\to\,\,
	\Ops(\Hilb)
	\,,
\end{align}
such that 
$\Q(x) = \hat{x}$ and $\Q(p) = \hat{p}$
\cite{shewell1959formation}.
This is
known as the \emph{quantization map}.


To this end, however,
it is well-known that
one must pick an \emph{ordering prescription}.
For example, consider the function $f(x,p) = x p^2$.
There are at least three different possibilities for its quantization:
$\hat{x} \hat{p}^2$,
$\hat{p} \hat{x} \hat{p}$,
and
$\hat{p}^2 \hat{x}$,
each differing from each other
by terms of $\O(\hbar^1)$.

A popular convention is the symmetric ordering,
which performs an equal-weight average:
$
\hat{f}
= (
	\hat{x} \hat{p}^2
	+ \hat{p} \hat{x} \hat{p}
	+ \hat{p}^2 \hat{x}
)/3
$.
A unique feature of
this prescription
is that all coordinates on the phase space,
namely position and momentum,
are put on an equal footing.

For simplicity,
suppose we stick to the symmetric ordering for the moment.
To recapitulate its definition,
the quantization map $\Q$ in the symmetric ordering prescription
sends $x^n p^m$
to the average of all possible orderings of $n$ factors of $\hat{x}$ and $m$ factors of $\hat{p}$.
Equivalently, a useful formula due to McCoy \cite{mccoy1932function} reads
\begin{align}
	\label{mccoy}
	\Q( x^n p^m )
	\,=\,
		\frac{1}{2^n}\,
		\sum_{k=0}^n\mem
			\binom{n}{k}\,
			\hat{x}^k\mem \hat{p}^m\mem \hat{x}^{n-k}
	\,=\,
		\frac{1}{2^m}\,
		\sum_{k=0}^m\mem
			\binom{m}{k}\,
			\hat{p}^k\mem \hat{x}^n\mem \hat{p}^{m-k}
	\,.
\end{align}
Using \eqref{mccoy},
it is not difficult to prove that \cite{weyl1927quantenmechanik}
\begin{align}
	\label{quantizer.sym}
	\Q(f)
	\,=\,
		\int\frac{dp \hhem\mwedge dx}{2\pi\hbar}\,\,
			f(x,p)\, \hat{\Delta}(x,p)
	\,,
\end{align}
where we have denoted
\begin{align}
	\label{str.sym}
	\hat{\Delta}(x,p)
	\,=\,
		\int dy\,\,
			\Ket{ x + y/2 }
			\,
				\mathe^{ipy/\hbar}
			\,
			\Bra{ x - y/2 }
	\,.
\end{align}

The above operator $\Delt(x,p)$,
labeled with phase space coordinates,
is
known as the \emph{Stratonovich kernel} \cite{stratonovich1957distributions,Varilly:1989sv}
for symmetric ordering.
The Stratonovich kernel
completely defines the quantization map $\Q$,
from which a unique operator $\hat{f}$ is paired to each phase-space function $f$ and vice versa.
The Stratonovich kernel
satisfies several axioms
such as Hermiticity,
normalization,
and orthocompleteness,
which one can direcly verify or derive
from its explicit definition in \eqref{str.sym}.


\subsubsection{The Inverse of Quantization Map}
\label{IQMAP}

Notably, the axioms of the Stratonovich kernel together ensures
a reinterpretation of quantum mechanics
reminiscent of statistical mechanics in phase space
\cite{stratonovich1957distributions,Varilly:1989sv},
as is explicitly advertized in the title of \rcite{Moyal:1949sk}:
``Quantum Mechanics as a Statistical Theory.''

The key idea is to pay attention to the inverse 
of the quantization map \cite{Wigner:1932eb}:
\begin{align}
	\label{dequantizer.sym}
	\Q^{-1}(\hat{f})(x,p)
	\,=\,
		\tr\nem\bigbig{ \hat{\Delta}(x,p)\mem \hat{f} }
	\,.
\end{align}
It is left as an exercise to check
that \eqref{dequantizer.sym}
is the inverse of \eqref{quantizer.sym}
by using \eqref{str.sym}.
Notably,
the inverse of the quantization map, $\Q^{-1}$,
designates a unique phase-space function $f \in \Cinfty(\P)$ to each operator $\hat{f} \in \Ops(\P)$:
\begin{align}
	\Q^{-1}
	\,\,\,:\,\,\,
	\Ops(\Hilb)
	\,\,\to\,\,
	\Cinfty(\P) 
	\,.
\end{align}

First of all,
consider two operators $\hat{f}$, $\hat{g}$
and their images $f$, $g$
under $\Q^{-1}$.
\eqrefs{quantizer.sym}{dequantizer.sym}
imply that
the operator product $\hat{f} \hat{g}$ corresponds to
the phase space function
\begin{align}
	\label{starQ}
	\Q^{-1}(\hat{f}\hat{g})
	\,=\,
		f \star g
	\,,
\end{align}
where $(f \star g)(x,p)$ is defined by the integral
\begin{align}
	\label{star-kernel}
	\int
	\frac{dp_1\hem dx_1}{2\pi\hbar}
	\frac{dp_2\hem dx_2}{2\pi\hbar}\,\,
		\tr\nem\bigbig{
			\Delt(x,p)\mem \Delt(x_1,p_1)\mem \Delt(x_2,p_2)
		}
		\,
		f(x_1,p_1)\mem g(x_2,p_2)
	\,.
\end{align}
By explicit evaluation of this integral,
it can be found that
\begin{align}
	\label{moyal-def}
	f \star g
	\,=\,
			f
			\,
			\exp\nem
			\bb{
				\frac{i\hbar}{2}\mem\bb{
					\overleftarrow{
						\frac{\partial}{\partial x}
					}
					\overrightarrow{
						\frac{\partial}{\partial p}
					}
				-
					\overleftarrow{
						\frac{\partial}{\partial p}
					}
					\overrightarrow{
						\frac{\partial}{\partial x}
					}
				}
			\nem
			}
			\,
			g
	\,,
\end{align}
where the arrows indicate
the directions on which derivatives act.
Mathematically,
\eqref{moyal-def} defines a noncommutative product between 
functions
$f$ and $g$,
known as the \emph{star product} due to Moyal \cite{Moyal:1949sk}.
Therefore,
the conclusion reads that
the quantum-mechanical operator product
is equivalent to
a noncommutative product on phase-space functions.

The noncommutativity of a star product is measured by
\begin{align}
	\label{star-pb}
	\spb{f}{g}
	\,=\,
		\frac{1}{i\hbar}\mem
		\bigbig{
			f \star g - g \star f
		}
	\,,
\end{align}
which will be referred to as the \emph{deformed Poisson bracket}.
It should be clear that \eqref{star-pb} is the phase space counterpart of the operator commutator,
divided by $i\hbar$:
$\spb{f}{g} = \Q^{-1}(\comm{\hat{f}}{\hat{g}}) / i\hbar$.
The correspondence principle
arises from the fact that
\eqref{star-pb} approaches to the Poisson bracket $\pb{f}{g}$
in the $\hbar \to 0$ limit.

Next, consider the image $\rho \in \Cinfty(\P)$
of a density matrix $\hat{\rho} \in \Ops(\Hilb)$
under $\Q^{-1}$
\cite{Wigner:1932eb}.
The axioms of the Stratonovich kernel imply that
\begin{align}
	\label{density-norm}
	\tr\nem\bigbig{ \hat{\rho} } \,=\, 1
	\qfq
	\int\frac{dp \hhem\mwedge dx}{2\pi\hbar}\,\,
		\rho
	\,=\,
		1
	\,.
\end{align}
Notably,
\eqref{density-norm} 
resembles \eqref{1d.norm},
the equation satisfied by the probability distribution
in classical statistical mechanics.
Specifically,
the ad-hoc constant $\ve$
will be replaced and identified with $2\pi\hbar$,
encoding the fundamental quantum of phase space volume.
Hence, it is suggested that the phase-space function $\rho$
might describe a probability distribution.

However,
explicit examples reveal that
the local value of
the phase-space function $\rho$
can be negative
in general;
see
\rrcite{raymer1997measuring,Curtright:2011vw,zachos2005quantum,Souza:2008nav,wheeler2000box}
for demonstrations.
In fact,
it is known that 
only Gaussian wavepackets
can achieve strict positiveness
\cite{hudson1974wigner};
the minimal-uncertainty wavepacket, for instance.
Therefore,
the precise statement reads that
$\rho$ defines a \emph{quasiprobability distribution} over the phase space.
Here, the prefix ``quasi-'' means that some of
the Kolmogorov probability axioms
are relaxed.



To establish the probabilistic interpretation of $\rho$,
one can consider 
the identities
\begin{align}
    \Bra{x} \hat{\rho} \Ket{x}
    \,=\,
    \int \frac{dp}{2\pi\hbar}\,\,
        \rho(x,p)
    \,,\quad
    \Bra{p} \hat{\rho} \Ket{p}
    \,=\,
    \int dx\,\,
        \rho(x,p)
    \,,
\end{align}
and their application to a pure state 
\smash{$\hat{\rho} = \Ket{\psi}\Bra{\psi}$}.
One can also examine the expectation value $\tr\nem\bigbig{ \hat{\rho}\hem \hat{f} }$
or the absolute-value-squared of a wavefunction overlap
$\BraKet{\psi_1}{\psi_2}$.

Finally,
suppose a time-dependent density matrix $\hat{\rho}(t)$,
whose evolution is governed by a time-dependent Hamiltonian $\hat{H}(t)$.
As is well-known,
the equation $\hat{\rho}(t)$ satisfies reads
\begin{align}
	\label{qliouv}
	\dot{\hat{\rho}}(t)
	\,=\,
		\frac{1}{i\hbar}\,
			\comm{ \hat{H}(t) }{ \hat{\rho}(t) }
	\,,
\end{align}
which is known as the \emph{quantum Liouville equation}.
From the results established above,
we find that
\eqref{qliouv} is equivalent to
a partial differential equation
on the domain $\P \times \R$:
\begin{align}
	\label{sliouv}
	\dot{\rho}(t)
	\,=\,
		\spb{H(t)}{\rho(t)}
	\,.
\end{align}
Here,
$\rho(t)$ and $H(t)$
are the images of $\hat{\rho}(t)$ and $\hat{H}(t)$
under $\Q^{-1}$, respectively.
The product manifold $\P \times \R$,
the three-dimensional space of $x$, $p$, and $t$,
is sometimes called the extended phase space \cite{arnold1989mathematical}.
Notably,
the partial differential equation in \eqref{sliouv}
is first-order in the time derivative
but is infinite-order in the $x,p$ derivatives.

To sum up,
the above analysis shows that
every operator equation
can be equivalently restated
as an equation about phase-space functions
via the map $\Q^{-1}$.
This establishes the so-called 
\emph{phase space formulation of quantum mechanics}
\cite{Moyal:1949sk,Groenewold:1946kp,weyl1927quantenmechanik,Wigner:1932eb},
summarized below.
\begin{enumerate}
\item 
	The point of departure is the density matrix formulation of quantum mechanics,
	built upon operator equations.
	(That is, do not talk about wavefunctions or kets/bras.)
\item 
	The quantization map is uniquely determined
	given a choice of ordering prescription.
\item 
	Provided the inverse of the quantization map,
	\begin{enumerate}[topsep=0pt]
		\item 
			\textit{operators} translate to \textit{phase-space functions},
		\item
			\textit{density matrices} translate to \textit{quasiprobability distributions} over phase space,
		\item
			the \textit{operator product} translates to a noncommutative product on phase space functions dubbed \textit{star product}.
	\end{enumerate}
\item 
	The quantum-mechanical time evolution
	is defined by the quantum Liouville equation
	as a partial differential equation on the extended phase space.
\end{enumerate}
\newpage

It shall be highlighted that the phase space formulation is
equivalent to the density matrix formulation:
every calculation viable in the former
should be possible in the latter
and vice versa.
This correspondence is exact and two-way,
provided a well-behaved, invertible
quantization map $\Q$
for a fixed ordering prescription.
For instance,
no approximation or truncation
has been made so far
for any parameter.
Note that the density matrix formulation 
is capable of describing
not only pure states but also mixed states (ensembles).

\subsubsection{Time Evolution as Fuzzy Diffeomorphism}
\label{TFUZ}

It remains to solve the quantum Liouville equation explicitly,
just like the treatment of the classical Liouville equation
in \Sec{CLAS}.

First of all,
for each function $f\in\Cinfty(\P)$
on phase space,
define $\sX_f$ as the first-order differential operator such that
\begin{align}
	\label{sX}
	\sX_f\act{g}
	\,=\,
		\spb{f}{g}
	\,,\quad
	\forall\mem
		g \in \Cinfty(\P)
	\,.
\end{align}
$\sX_f$ will be referred to as the \emph{deformed Hamiltonian vector field} of $f$,
as its classical limit is exactly the Hamiltonian vector field in \eqref{X}.
Importantly,
deformed Hamiltonian vector fields
are differential operators
that preserve the algebraic structure of phase-space functions
due to pointwise addition and star product:
\begin{subequations}
\label{ag.sdiff}
\begin{align}
	\label{ag.sdiff.a}
	\sX_f\act{
		g + h
	}
	\,&=\,
		\sX_f\act{g} + \sX_f\act{h}
	\,,\\
	\label{ag.sdiff.b}
	\sX_f\act{
		g \star h
	}
	\,&=\,
		(\sX_f\act{g}) \star h
		+
		g \star\hnem (\sX_f\act{h})
	\,.
\end{align}
\end{subequations}
In other words, deformed Hamiltonian vector fields
preserve the quantum operator algebra
boiled down into the phase space formulation.

Second of all,
provided the definition in \eqref{sX},
the quantum Liouville equation in \eqref{sliouv} can be rewritten and solved as
\begin{align}
    \label{sliouv.X}
    \dot{\rho}(t)
    \,=\,
        \sX_{H(t)}\act{
            \rho(t)
        }
    \qiq
    \rho(T)
    \,=\,
        U^\star\act{
            \rho(0)
        }
    \,,
\end{align}
where we have arbitrarily set the initial and final times as $t = 0$ and $t = T$.
In \eqref{sliouv.X},
$U^\star$ is a differential operator that transforms the initial quantum (ensemble) state $\rho(0)$
to the final quantum state $\rho(T)$
as quasiprobability distributions.
Again, well-established results
due to Dyson \cite{Dyson:1949ha}
and
Magnus \cite{magnus1954exponential}
facilitate the explicit formulae
\begin{align}
	\label{sU.sol}
	U^\star
	\,=\,
		\Pexp{\mem
			\int_0^T dt\,\,
				\sX_{H(t)}
		}
	\,=\,
		\exp\nem\BB{
			\sX_\sG
		}
	\,,
\end{align}
respectively.
In \eqref{sU.sol},
$\sG \in \Cinfty(\P)$
is given by
\begin{align}
	\label{sG+3}
	\sG
	\,=\,
		&
		\int dt_1\,\,
			H(t_1)
		+
		\frac{1}{2}\mem
		\int_{t_1>t_2}\nem d^2t\,\,
			\spb{H(t_1)}{H(t_2)}
\\
		&
		+
		\frac{1}{6}\mem
		\int_{t_1>t_2>t_3}\nem d^3t\,\,
			\BB{
				\spb{ H(t_1) }{ \spb{ H(t_2) }{ H(t_3) } }
				+
				\spb{ H(t_3) }{ \spb{ H(t_2) }{ H(t_1) } }
			}
		+ \cdots
	\,,
	\nonumber
\end{align}
where the default integral bounds are $[0,T]$.

\newpage

An important feature of
\eqref{sG+3}
is that
each integrand is composed by nesting deformed Poisson brackets.
This property
crucially counts on
the Jacobi identity
\begin{align}
	\label{sXcomm}
	\comm{\sX_{f}}{\sX_{g}}
	\,=\,
		\sX_{\spb{f}{g}}
	\,,
\end{align}
which encodes that deformed Hamiltonian vector fields 
as differential operators
are closed under the commutator.
Note that
\eqrefs{ag.sdiff.b}{sXcomm} demand associativity of the star product.

The formula
$\sU {\:=\,} \exp(\sX_\sG)$
in \eqref{sU.sol}
establishes that
$\sU$ is the exponentiation of a deformed Hamiltonian vector field.
Unlike 
as in the classical case,
this fact does not imply that
a diffeomorphism exists
such that $\sU$ implements a pullback,
as $\sX_G$ is not a first-order differential operator;
hence we have not put a superscript $^*$.
Still, however, it holds that
\begin{subequations}
\label{ag.sDiff}
\begin{align}
    \label{ag.sDiff.a}
	\sU\act{
		f + g
	}
	\,&=\,
		\sU\act{f} + \sU\act{g}
	\,,\\
    \label{ag.sDiff.b}
	\sU\act{
		f \star g
	}
	\,&=\,
		(\sU\act{f}) \star (\sU\act{g})
	\,,
\end{align}
\end{subequations}
for any phase-space functions $f,g \in \Cinfty(\P)$.
This is the exponentiated version of \eqref{ag.sdiff}.

To elicit the geometrical interpretation of $\sU$,
we might want to temporarily switch gears:
a brief mathematical digression
based on ideas in
noncommutative geometry.

Roughly speaking,
there is a sense in which 
examining the algebra of functions on a space
studies the geometry of that space itself
\cite{Connes:1994yd,gracia2013elements,Szabo:2001kg,gelfand1943imbedding,segal1947irreducible,rmont-RingStructB}.
This idea provides an algebraic perspective on geometry.
Especially, it facilitates the very definition of noncommutative geometry
as a space endowed with a noncommutative algebra 
of functions.

In an ordinary (classical) geometry,
smooth functions form 
a commutative ring $\A = (\Cinfty(\P),+,{\,\cdot\,})$
under the pointwise addition and multiplication.
In the algebraic approach to geometry,
diffeomorphisms are viewed as automorphisms of this ring $\A$.
Namely, they are transformations that preserve the pointwise addition and multiplication.
In fact, we have already explored this idea
in \eqrefs{ag.Diff.a}{ag.Diff.b}.

In a noncommutative geometry,
smooth functions form
a noncommutative ring $\A_\star = (\Cinfty,+,\star)$
due to the pointwise addition and a star product $\star$.
As the direct generalization of the identification made in the previous paragraph,
diffeomorphisms
of the noncommutative geometry
can be defined as automorphisms of this ring $\A_\star$.
Namely, they are transformations that preserve the pointwise addition and the star product.
Evidently, this statement is the exact content of
\eqrefs{ag.sDiff.a}{ag.sDiff.b}.

In this precise mathematical sense,
our map $\sU = \exp(\sX_\sG)$
describes and defines a diffeomorphism in a noncommutative geometry,
or a \emph{fuzzy diffeomorphism} in short.
Here, ``fuzzy'' is a technical term
referring to ``noncommutative'' in noncommutative geometry.
The space of fuzzy diffeomorphisms may be denoted as $\Diff(\P,\star)$,
so $\sU \in \Diff(\P,\star)$.

It is also instructive to approach this fact in terms of vector fields.
In an ordinary geometry,
vector fields are derivations on the commutative algebra $\A = (\Cinfty(\P),+,{\,\cdot\,})$.
Namely, they are differential operators that preserve the pointwise addition and multiplication.
In fact, we have already explored this idea in \eqrefs{ag.diff.a}{ag.diff.b}.
Similarly,
Hamiltonian vector fields
are derivations on the Poisson algebra
$\A_{\pb{\,\,}{\,\,}} = (\Cinfty(\P),+,{\,\cdot\,},\{\:\:\,\mem,\,\:\:\})$,
as they additionally preserve the Poisson bracket.
We have already adopted this view in
\eqref{ag.diff.c}.

In a noncommutative geometry,
vector fields are derivations on the noncommutative algebra $\A_\star = (\Cinfty(\P),+,\star)$.
Namely, they are differential operators that preserve the pointwise addition and the star product.
Evidently, this statement is the exact content of
\eqrefs{ag.sdiff.a}{ag.sdiff.b}.
Therefore, vector fields in a noncommutative geometry,
or \emph{fuzzy vector fields} in short,
are exactly 
the differential operators such as
the deformed Hamiltonian vector fields
defined in \eqref{sX}.

Exponentiating,
it follows that the map $\sU = \exp(\sX_\sG)$
preserves the algebra $\A_\star$:
$\sU$ is
a $\star$-preserving map
since $\sX_\sG$ is $\star$-preserving.
Specifically, one shows that
$\sU = \exp(\sX_\sG)$ satisfies \eqref{ag.sDiff}
if $\sX_\sG$ satisfies \eqref{ag.sdiff}.

Note how
\eqref{ag.diff}
has described the (semi)classical vestige of \eqref{ag.sdiff}.
Similarly, it should be clear that
$\sU = \exp(\sX_\sG)$ in \eqref{sU.sol}
reproduces $U = \exp(X_G)$ in \eqref{U.sol}
in the classical limit.
A $\star$-preserving diffeomorphism
$\sU \in \Diff(\P,\star)$
becomes an $\sp$-preserving diffeomorphism 
$U \in \Diff(\P,\sp)$
in the small fuzziness limit.
A $\star$-preserving vector field
$\sX_G$
becomes an $\sp$-preserving vector field 
$X_G$
in the small fuzziness limit.
$\sp$ is the vestige of $\star$.
Classical geometry of phase space emerges from
the quantum geometry of phase space.

Eventually,
let us return to the physics side and conclude.
In the phase space formulation of quantum mechanics,
the phase space $\P$ is endowed with a noncommutative product $\star$ on its functions,
encoding the quantum-mechanical operator algebra.
The pair $(\P,\star)$ will be called 
the \emph{fuzzy phase space},
which defines an instance of noncommutative geometry.

Intuitively, the fuzzy phase space is a geometry where
each point becomes dissolved or spread out a bit
due to the uncertainty principle,
the extent of which is characterized by the quantum of phase space volume $\varepsilon = 2\pi\hbar$.
In fact, the Poisson structure
of the classical phase space
shall be viewed as vestiges of this fuzziness,
emerging in the limit $\hbar \to 0$.

Provided this geometrical interpretation,
it follows that
the solution to the quantum Liouville equation,
\eqref{sliouv.X},
has computed a fuzzy diffeomorphism,
\begin{align}
    \label{sU-in-Diff}
    \sU \,\,\,\in\,\,\, \Diff(\P,\star)
    \,,
\end{align}
incarnated
as a $\star$-preserving differential operator.
As $\sU : \rho(0) \mapsto \rho(T)$
transforms the initial quantum state to the final quantum state
as quasiprobability distributions,
its interpretation
is the quantum-mechanical time evolution
reformulated
as a fuzzy diffeomorphism.
Hence, $\sU$ in \eqref{sU-in-Diff} would be referred to as the \emph{fuzzy time-evolution diffeomorphism}.

Of course,
from the perspective of the standard operator formalism,
this might be a merely pedantic way of assessing the fact that
the quantum-mechanical time evolution
is a map that
preserves the operator product.
Still,
its geometrical reinterpretation
in terms of noncommutative geometry
could be interesting
in light of its 
exotic semantics.


A few remarks are in order.
Firstly,
the consistency between
the normalization $\tr\nem\bigbig{ \hat{\rho}(t) } = 1$
and the quantum Liouville equation in \eqref{qliouv}
is ensured by the property
$
\tr\nem\bigbig{ \hat{f} \hem \hat{g} }
= \tr\nem\bigbig{ \hat{g} \hem \hat{f} }
$.
In the phase space formulation,
this consistency is ensured through the identity
\begin{align}
	\label{trace-swap-def}
	\int \frac{dp \hhem\mwedge dx}{2\pi\hbar}\,\,
		f \star g
	\,=\,
	\int \frac{dp \hhem\mwedge dx}{2\pi\hbar}\,\,
		g \star f
    \qfq
	\int \frac{dp \hhem\mwedge dx}{2\pi\hbar}\,\,
		\sX_f\act{g}
	\,=\,
        0
	\,,
\end{align}
which arises by the axioms of the Stratonovich kernel.
Formally,
\eqref{trace-swap-def}
is analogous to the vanishing total integral of a total derivative.

The above clarification establishes that
the action of a deformed Hamiltonian vector field
on quasiprobability distributions
preserves the unity of quantum-mechanical probability,
which reinterprets unitarity of quantum time evolution.
Compare this with the fact that
the action of a Hamiltonian vector field
on probability distributions
preserves the unity of classical probability,
which is
due to the invariance of the Liouville measure.
This encodes symplecticity of classical time evolution.

Next,
it is sometimes stated 
that
the Liouville theorem
breaks down at the quantum level.
At face value, this is correct since
$\sX_{H(t)}$ or $\sX_\sG$ are
not Hamiltonian vector fields
but deformed ones.
In fact,
they are not even first-order differential operators
and thus cannot be interpreted as a vector field or an infinitesimal diffeomorphism.
(What is the Lie derivative ``\mem$\pounds_{\sX_{H(t)}} \sp$\mem''?)
Namely, while their action on test functions
is well-defined,
their action on points in the phase space
is not defined;
see \rcite{Oliva:2018nno}, for instance.
Intuitively speaking, this means that they dissolve a point into a scattered cloud,
instead of mapping it to another solid point.
This precisely demonstrates their fuzzy nature.

In this regard,
a better comparison may be
evaluating the proper noncommutative geometry generalization of the Liouville theorem:
the preservation of the \textit{star product} by $\sX_{H(t)}$ or $\sX_\sG$.
This generalized Liouville theorem
is exactly satisfied
and encodes the very unitarity
as discussed above.
So probability is still conserved,
albeit quantum-mechanically.

\subsubsection{Adjoint Actions and Intertwining Identities}

Before moving on, it is helpful to establish a few mathematical identities regarding adjoint actions
to clarify the origin of fuzzy diffeomorphisms.

As is well-known,
the \emph{adjoint actions} are defined as
\begin{align}
    \label{ad-def}
    \ad_{\hat{f}}\act{
        \hat{g}
    }
    \,=\,
        \comm{\hat{f}}{\hat{g}}
    \,,\quad
    \Ad_{\hat{U}}\act{
        \hat{f}
    }
    \,=\,
        \hat{U}\hem \hat{f}\, \hat{U}^{-1}
    \,,
\end{align}
for any operators $\hat{f}, \hat{g}, \hat{U} \in \Ops(\Hilb)$
with the assumption that $\hat{U}$ is invertible.
In particular,
it is well-known that
exponentiation intertwines between
$\ad$ and $\Ad$ \cite{hall2013lie}:
\begin{align}
    \label{expadj}
    \Ad_{\exp(\hat{f})}
    \,=\,
        \exp\nem\BB{
            \ad_{\hat{f}}
        }
    \,.
\end{align}

To put it intuitively,
adjoint actions are ``operators on operators.''
Importantly,
adjoint actions
are maps that preserve the operator algebra.
This fact follows from the identities
\begin{subequations}
\label{ag.adj}
\begin{align}
	\label{ag.adj.a}
	\ad_{\hat{f}}\act{
		\hat{g} + \hat{h}
	}
	\,&=\,
		\ad_{\hat{f}}\act{\hat{g}} + \ad_{\hat{f}}\act{\hat{h}}
	\,,\\
	\label{ag.adj.b}
	\ad_{\hat{f}}\act{
		\hat{g} \hat{h}
	}
	\,&=\,
		(\ad_{\hat{f}}\act{\hat{g}}) \mem \hat{h}
		+
		\hat{g} \mem (\ad_{\hat{f}}\act{\hat{h}})
	\,,
\end{align}
\end{subequations}
as well as
\begin{subequations}
\label{ag.Adj}
\begin{align}
    \label{ag.Adj.a}
	\Ad_{\hat{U}}\act{
		\hat{f} + \hat{g}
	}
	\,&=\,
		\Ad_{\hat{U}}\act{\hat{f}} + \Ad_{\hat{U}}\act{\hat{g}}
	\,,\\
    \label{ag.Adj.b}
	\Ad_{\hat{U}}\act{
		\hat{f} \hat{g}
	}
	\,&=\,
		(\Ad_{\hat{U}}\act{\hat{f}})\mem
        (\Ad_{\hat{U}}\act{\hat{g}})
	\,.
\end{align}
\end{subequations}
In the mathematical language,
$\ad_{\hat{f}}$ describes a derivation on the operator algebra,
while
$\Ad_{\hat{U}}$ describes an automorphism of the operator algebra.
The astute reader will notice the parallel with
$\sX_f$ and $U^\star$ in \eqrefs{ag.sdiff}{ag.sDiff}.

To render
this parallel structure into precise mathematical statements,
we establish
a couple of identities
that would be called \emph{intertwining identities},
which assume a well-behaved quantization map $\Q$.
The first intertwining identity reads the following:
\begin{align}
    \label{tw1}
    \ad_{\Q(f)/i\hbar}
    \mem\circ\,
    \Q
    \,=\,
        \Q
        \,\circ
        \sX_f
    \,.
\end{align}
The proof is straightforward
from the definitions in \eqrefs{starQ}{star-pb}.
The second intertwining identity,
on the other hand,
is the exponentiated version of \eqref{tw1}:
\begin{align}
    \label{tw2}
    \Ad_{\exp(\Q(f)/i\hbar)}
    \mem\circ\,
    \Q
    \,=\,
        \Q
        \,\circ
        \exp\nem\BB{
            \sX_f
        }
    \,.
\end{align}
It can be seen that
\eqref{tw2}
is implied by \eqrefs{expadj}{tw1}.

To evaluate physical implications,
one solves the quantum Liouville equation 
in \eqref{qliouv}
in the operator language:
\begin{align}
    \label{qliouv.X}
    \dot{\hat{\rho}}(t)
    \,=\,
        \ad_{\hat{H}(t)/i\hbar}\act{
            \hat{\rho}(t)
        }
    \qiq
    \hat{\rho}(T)
    \,=\,
        \Ad_{\hat{U}}\act{
            \hat{\rho}(0)
        }
    \,.
\end{align}
In \eqref{qliouv.X}, the \emph{time-evolution unitary operator} $\hat{U}$ is given by
\begin{align}
	\label{qU.sol}
	\hat{U}
	\,=\,
		\Pexp{\mem
            \frac{1}{i\hbar}\mem
			\int_0^T dt\,\,
				\hat{H}(t)
		}
	\,=\,
		\exp\nem\BB{
			\hat{G}/i\hbar
		}
	\,,
\end{align}
where the operator $\hat{G}$ is given by
\begin{align}
	\label{G+3.op}
	\hat{G}
	\,=\,
		&
		\int dt_1\,\,
			\hat{H}(t_1)
		+
		\frac{1}{2(i\hbar)^1}\mem
		\int_{t_1>t_2}\nem d^2t\,\,
			\comm{\hat{H}(t_1)}{\hat{H}(t_2)}
\\
		&
		+
		\frac{1}{6(i\hbar)^2}\mem
		\int_{t_1>t_2>t_3}\nem d^3t\,\,
			\BB{
				\comm{ \hat{H}(t_1) }{ \comm{ \hat{H}(t_2) }{ \hat{H}(t_3) } }
				+
				\comm{ \hat{H}(t_3) }{ \comm{ \hat{H}(t_2) }{ \hat{H}(t_1) } }
			}
		+ \cdots
	\,.
	\nonumber
\end{align}
An important feature of
\eqref{G+3.op}
is that
each integrand is composed by nesting commutators.
This property
crucially counts on
the Jacobi identity,
\begin{align}
	\label{adjcomm}
	\comm{\ad_{\hat{f}}}{\ad_{\hat{g}}}
	\,=\,
		\ad_{\comm{\hat{f}}{\hat{g}}}
	\,.
\end{align}
Clearly, \eqrefs{ag.adj.b}{adjcomm} demand associativity of the operator algebra.

Evidently, \eqref{G+3.op}
is the image of \eqref{sG+3}
under $\Q$.
This establishes that
\begin{align}
    \label{tw.G}
    \hat{G} 
    \,=\,
        \Q(G^\star)
    \,.
\end{align}
Provided \eqref{tw.G},
the second intertwining identity in \eqref{tw2}
establishes a precise relationship
between 
the time-evolution unitary operator $\hat{U}$ in \eqref{qU.sol}
and the fuzzy time-evolution diffeomorphism $U^\star$ in \eqref{sU.sol}:
\begin{align}
    \label{tw.U}
    \Ad_{\hat{U}}
    \mem\circ\,
    \Q
    \,=\,
        \Q
        \,\circ\hem U^\star
    \,.
\end{align}
Therefore,
we conclude that
the quantization map $\Q$
intertwines between
the adjoint action of the time-evolution unitary operator,
$\Ad_{\hat{U}}$,
and
the fuzzy time-evolution diffeomorphism,
$U^\star$.
In other words,
$U^\star$
is equivalent to
the adjoint action $\Ad_{\hat{U}}$
via the change in formulations
(standard to phase space).

The above discussion 
should reveal the origin of the fuzzy time-evolution diffeomorphism
in a transparent fashion.
It might be also helpful to
carry out explicit calculations 
at the level of the Stratonovich kernel
such as
$
    \hat{U}\hem \Delt(x,p)\mem \hat{U}^{-1}
    =
        U^\star\act{
            \Delt(x,p)
        }
$,
which we leave as an exercise.

\newpage

\subsubsection{Normal Ordering}
\label{NORMAL}

In \Secs{QMAP}{IQMAP},
we have presumed the symmetric ordering prescription
for concreteness of our exposition.
However, another relevant prescription
is the normal ordering
for oscillator systems,
exclusively adopted in field theories.
For the reader's sake,
below, we carry out
the phase space formulation
in the normal ordering prescription
as well.

First,
the starting point is the identification of the particle's phase space
as a complex plane: 
$\P = \R^2 \cong \C^1$.
Two coordinate systems,
$(x,p)$ and $(a,\ba)$,
are introduced on $\P$
so that the symplectic structure 
in \eqrefs{xp.pb}{xp.omega}
described also as
\begin{align}
	\label{omega.1d}
	\sp
	\,=\,
		i\mem d\ba \wedge da
	\qfq
        \pb{a}{a} \mem=\mem 0
		\,,\quad
		\pb{a}{\ba} \mem=\mem -i
        \,,\quad
        \pb{\ba}{\ba} \mem=\mem 0
	\,.
\end{align}
Explicitly,
the coordinate transformations are given by
\begin{align}
\begin{split}
	\label{ab-xp}
	a
	\,=\,
		\frac{m\omega x + ip}{\sqrt{2m\omega}}
	&\,,\quad
	x\,=\,
		\frac{1}{\sqrt{2m\omega}}\,
		\bigbig{
			a + \ba
		}
	\,,\\
	\ba
	\,=\,
		\frac{m\omega x - ip}{\sqrt{2m\omega}}
	&\,,\quad
	p\,=\,
		-i\mem \sqrt{\frac{m\omega}{2}}\,
		\bigbig{
			a - \ba
		}
	\,.
\end{split}
\end{align}
The dimensionful constants
$m,\omega$
are supplied by the free harmonic oscillator Hamiltonian,
\begin{align}
	\label{1d.H0}
	H^\circ(a,\ba)
	\,=\,
		\frac{p^2}{2m} + \frac{1}{2}\, m\omega^2 x^2
	\,=\,	
		\omega\mem \ba a
	\,.
\end{align}

Second,
the Hilbert space can be kept as
the space 
$\Hilb = L^2(\R)$
of square-integrable functions on the position space,
so the definition of the position and momentum operators
remain the same as in \eqref{xpdef.Hilb}.
Then the lowering and raising operators are defined as
\begin{align}
\begin{split}
	\label{abdef.Hilb}
	\hat{a}
	&\,\,\,:\,\,\,
	\BB{
		x \,\mapsto\, \psi(x)
	}
	\,\,\mapsto\,\,
	\bb{
		x \,\mapsto\,
			\frac{1}{\sqrt{2m\omega}}\,
			\bb{
				m\omega x
				+ \hbar\mem \frac{\partial}{\partial x}
			}\,
			\psi(x)
	}
	\,,\\
	\hat{\ba}
	&\,\,\,:\,\,\,
	\BB{
		x \,\mapsto\, \psi(x)
	}
	\,\,\mapsto\,\,
	\bb{
		x \,\mapsto\,
			\frac{1}{\sqrt{2m\omega}}\,
			\bb{
				m\omega x
				- \hbar\mem \frac{\partial}{\partial x}
			}\,
			\psi(x)
	}
	\,,
\end{split}
\end{align}
due to \eqref{ab-xp}.
As desired,
\eqref{abdef.Hilb} realizes the canonical commutation relations
\begin{align}
	\label{ab.ccr}
    \comm{ \hat{a} }{ \hat{a} }
    \,=\,
        0
    \,,\quad
	\comm{ \hat{a} }{ \hat{\ba} }
	\,=\,
		\hbar\, \hat{1}
	\,,\quad
    \comm{ \hat{\ba} }{ \hat{\ba} }
    \,=\,
        0
    \,.
\end{align}

The coherent states are given as
\begin{align}
	\label{coherent.wf}
	\BraKet{x}{\l}
	\,=\,
		\bb{\nem\frac{m\omega}{\pi\hbar}\nem}^{\nem\nem1/4}
		\,
		\exp\nem\bb{\nem
			-\frac{1}{\hbar}\,
				\bb{
					\sqrt{\frac{m\omega}{2}}\, x
					- \l
				}^{\nem\nem2}
				+ \frac{\l^2}{2\hbar}
		}
	\qiq
	\hat{a} \Ket{\l}
	\,=\,
	\l\mem \Ket{\l}
	\,,
\end{align}
where $\Ket{\l} : x \mapsto \BraKet{x}{\l}$
is an element of $\Hilb$.
Our convention for their normalization is such that 
\begin{align}
	\label{coherent}
	\BraKet{\bl_2}{\l_1}
	\,=\,
		\mathe^{\bl_2 \l_1/\hbar}
	\,,\quad
	\int \frac{i\mem d\bl \mwedge d\l}{2\pi\hbar}\,\,
		\mathe^{-\bl\l/\hbar}\,
		\Ket{\l}\Bra{\bl}
	\,=\,
		\hat{1}
	\,,
\end{align}
where the integration employs the Liouville measure over the phase space.


Third,
the quantization map in the normal ordering prescription is defined as
\begin{align}
	\label{qzn.normal}
	\Q( a^n \ba^m )
	\,=\,
		\hat{\ba}^m \hat{a}^n
	\,.
\end{align}
The linearity of $\Q$ then implies that
each analytic function $f$ on $\P$ is mapped to
\begin{align}
	\Q(f)
	\,&=\,
		\int \frac{i\mem d\ba \mwedge da}{2\pi\hbar}
		\int \frac{i\mem d\bl \mwedge d\l}{2\pi\hbar}\,\,
			\exp\nem\bb{\hnem
				\frac{\l}{\hbar}\,
					(\hat{\ba}-\ba)
			\nem}
			\mem
			\exp\nem\bb{\nem
				-
				\frac{\bl}{\hbar}\,
					(\hat{a}-a)
			\nem}
		\,
		f(a,\ba)
	\,.
\end{align}
Direct computation shows that
\begin{align}
	\label{quantizer.normal}
	\Q(f)
	\,=\,
		\int \frac{i\mem d\ba \mwedge da}{2\pi\hbar}\,\,
			\Delt(a,\ba)\,
			f(a,\ba)
	\,,
\end{align}
where
\begin{align}
	\label{kernel.normal}
	\Delt(a,\ba)
	\,=\,
		\int \frac{i\mem d\bl \mwedge d\l}{2\pi\hbar}\,\,
			\Ket{ \l }
			\,
				\exp\nem\bb{\nem
					\frac{\ba a - \bl a - \ba \l}{\hbar}
				\nem}
			\,
			\Bra{ \bl }
	\,.
\end{align}
The kernel in \eqref{kernel.normal}
is attributed to
Glauber and Sudarshan
\cite{glauber1963quantum,sudarshan1963equivalence}.
It
satisfies several axioms
encoding Hermiticity and normalization:
\begin{align}
	\label{Delt-axioms.normal}
	\hat{\Delta}^\dagger(a,\ba)
	\,=\,
		\hat{\Delta}(a,\ba)
	\,,\quad
	\tr\nem\bigbig{
		\hat{\Delta}(a,\ba)
	}
	\,=\,
		1
	\,,\quad
	\int \frac{i\mem d\ba \mwedge da}{2\pi\hbar}\,\,
		\hat{\Delta}(a,\ba)
	\,=\,
		\hat{1}
	\,.
\end{align}


Fourth,
by using the properties of coherent states in \eqref{coherent},
the inverse of the quantization map
in \eqref{quantizer.normal}
is found as
\begin{align}
	\label{dequantizer.normal}
	\Q^{-1}(\hat{f})(a,\ba)
	\,=\,
		\mathe^{-\ba a/\hbar}\,
			\Bra{\ba} \hat{f} \Ket{a}
	\,.
\end{align}
This designates a unique phase space function $f \in \Cinfty(\P)$ to each operator $\hat{f} \in \Ops(\Hilb)$.
The phase space formulation of quantum mechanics
is facilitated by this map $\Q^{-1}$.

The star product is defined by \eqref{starQ}.
By using 
\eqrefs{quantizer.normal}{dequantizer.normal},
it follows that
$(f \star g)(a,\ba)$ is given by the integral
\begin{align}
	\label{star-kernel}
	\int
	\frac{i\mem d\ba_1 \mwedge da_1}{2\pi\hbar}
	\frac{i\mem d\ba_2 \mwedge da_2}{2\pi\hbar}\,\,
		\mathe^{-\ba a/\hbar}\,
		\Bra{\ba}
			\Delt(a_1,\ba_1)\mem \Delt(a_2,\ba_2)
		\Ket{a}
		\,
		f(a_1,\ba_1)\mem g(a_2,\ba_2)
	\,.
\end{align}
By explicit evaluation of this integral,
it follows that
\begin{align}
	\label{wick-def}
	f \star g
	\,=\,
			f
			\,
			\exp\nem
			\bb{\nem
				\hbar\,
				\overleftarrow{
					\frac{\partial}{\partial a}
				}
				\overrightarrow{
					\frac{\partial}{\partial \ba}
				}
			}
			\,
			g
	\,.
\end{align}
This star product would be referred to as the star product 
of the Wick kind.


The exponential in \eqref{wick-def}
exactly implements the
operator product between normal-ordered operators
via Wick's theorem.
To see this, observe the correspondences
\begin{align}
\begin{split}
    \label{wick-made-easy}
	\hat{\ba} \hat{a}
		\,=\, \normal{\hat{\ba}\hat{a}}
	&\quad\xleftrightarrow{\,\,}\quad
	\ba \star a
	\,=\,
		\ba a
	\,,\\
	\hat{a} \hat{\ba}
		\,=\, \normal{\hat{a}\hat{\ba}} + \hbar\hem \hat{1}
	&\quad\xleftrightarrow{\,\,}\quad
	a \star \ba
	\,=\,
		a\ba + \hbar
	\,.
\end{split}
\end{align}
Here, $\ba a = a \ba$ describes the commutative product
for $\Cinfty(\P)$.
Another helpful exercise is
\begin{align}
\begin{split}
	\normal{\hat{\ba}\hat{a}^2} \, \normal{\hat{\ba}^3\hat{a}}
	&\,=\, \normal{\hat{\ba}^4\hat{a}^3} 
		+ 6\hbar\mem \normal{\hat{\ba}^3\hat{a}^2}
		+ 6\hbar^2\mem \normal{\hat{\ba}^2\hat{a}}
	\,,\\
	\quad\xleftrightarrow{\,\,}\quad
	(\ba a^2)\hem \star (\ba^3 a)
	&\,=\,
		\ba^4 a^3 + 6\hbar\mem \ba^3 a^2 + 6\hbar^2 \ba^2 a
	\,,
\end{split}
\end{align}
where the power of $\hbar$
equals the number of Wick contractions.
It should be clear that
\begin{align}
    \Q^{-1}(\hat{f}) \,=\, f
    \qfq
    \hat{f}
    \,=\,   
        \Q(f)
    \,=\,
        \normal{f(\hat{a},\hat{\ba})}
    \,.
\end{align}

The deformed Poisson bracket is defined by
\eqref{star-pb}.
The Wick star product in \eqref{wick-def} gives
\begin{align}
	\label{spb.wick}
	\pb{f}{g}^\star
	\hem\,=\,\,
        \frac{1}{i}\mem
		\sum_{\ell=0}^\infty\,
			\frac{\hbar^{\ell}}{(\ell{\,+\mem}1)!}\,\,
		f
		\,
		\bbsq{
				\bb{\nem
					\overleftarrow{
						\frac{\partial}{\partial a}
					}
					\overrightarrow{
						\frac{\partial}{\partial \ba}
					}
				}^{\nem\nem\ell+1}
				-
				\bb{\nem
					\overleftarrow{
						\frac{\partial}{\partial \ba}
					}
					\overrightarrow{
						\frac{\partial}{\partial a}
					}
				}^{\nem\nem\ell+1}
		\mem}
		\,
		g
	\,,
\end{align}
where the integer $\ell$ can be interpreted as loop order
in a diagrammatic computation:
see \App{BFT}.
Also,
compare \eqref{spb.wick}
with the deformed Poisson bracket
due to the Moyal star product in \eqref{moyal-def},
\begin{align}
	\label{spb.moyal}
	\pb{f}{g}^\star
	\hem\,=\,\,
    \sum_{k=0}^\infty\,
        \frac{(i\hbar/2)^{2k}}{(2k{\,+\mem}1)!}\,\,
			f
			\,
            \bbsq{
			\bb{
					\overleftarrow{
						\frac{\partial}{\partial x}
					}
					\overrightarrow{
						\frac{\partial}{\partial p}
					}
				-
					\overleftarrow{
						\frac{\partial}{\partial p}
					}
					\overrightarrow{
						\frac{\partial}{\partial x}
					}
				}^{\nem\nem 2k+1}
            }
			\,
			g
	\,.
\end{align}
In \eqref{spb.moyal},
only the even powers of $\hbar$ occurs
due to symmetry properties.
Again, $2k$ in \eqref{spb.moyal} 
describes
the loop order
in the diagrammatic derivation
of star product
\cite{deformation-quantification}.

Finally,
a density matrix $\hat{\rho}$ is mapped to the phase-space function
\begin{align}
    \label{husimi}
    \rho(a,\ba)
    \,=\,
		\mathe^{-\ba a/\hbar}\,
			\Bra{\ba} \hat{\rho} \Ket{a}
    \,,
\end{align}
which is known as Husimi $Q$ representation \cite{husimi1940some}.
This is still a quasiprobability distribution,
since some of
the Kolmogorov probability axioms have to be relaxed
(due to the overcompleteness of coherent states, for instance).
Yet, it holds that \eqref{husimi} is non-negative and bounded
\cite{cartwright1976non}.
Physically, it is literally the probability for
obtaining a coherent state
in a measurement
due to the Born rule.
For a pure state 
\smash{$\hat{\rho} = \Ket{\psi}\Bra{\psi}$},
for instance,
one finds 
\smash{$\rho(a,\ba) = \big|\mem{
    \mathe^{-\ba a/2}\mem \BraKet{\ba}{\psi}
}\hem\big|^2$},
where 
the normalization factor
$\mathe^{-\ba a/2}$
is due to our convention.

The rest of the phase space formulation
unfolds in the exact same fashion as before.



\section{Scattering Theory}
\label{SCA}

In \Sec{PSFOR},
we have established the geometrical interpretations of
classical and quantum time evolutions
as
symplectomorphisms
and
fuzzy diffeomorphisms,
by means of the phase space formulations of classical and quantum mechanics.
In this section, we apply such ideas to scattering theory.

\Secs{CIP}{S-SYMP}
review the definition and computation of S-symplectomorphism
by following \rcite{Kim:2025sey}.
\Secs{QIP}{S-DIFF}
construct
quantum
scattering theory
in the phase space formulation of quantum mechanics
as the major achievement of this paper.

\subsection{Classical Interaction Picture}
\label{CIP}

To begin with,
suppose a classical system defined on a generic symplectic manifold $(\P,\sp)$ as the phase space.
Suppose the Hamiltonian is given in the form
\begin{align}
    \label{Hsplit}
    H(t)
    \,=\,
        H^\circ\hnem(t)
        + 
        V(t)
    \,.
\end{align}
The split in \eqref{Hsplit} defines time-dependent perturbation theory
in the phase space formulation of classical mechanics.
$H^\circ(t) \in \Cinfty(\P)$ is the free Hamiltonian,
while $V(t) \in \Cinfty(\P)$ is the interaction Hamiltonian.
The latter is treated perturbatively.

Let $U^\circ\hnem(t_1,t_2)$ and $U(t_1,t_2)$
be the time-evolution symplectomorphisms from time $t_2$ to time $t_1$
due to
the Hamiltonians
$H^\circ\hnem(t)$ and $H(t)$, respectively.
They satisfy the usual axioms:
$U(t_1,t_1) = \mathrm{id}$,
$U(t_1,t_2) = U(t_2,t_1)^{-1}$,
$U(t_1,t_2) \circ U(t_2,t_3) = U(t_1,t_3)$.

For any time-dependent function $f(t) \in \Cinfty(\P)$
on the phase space,
its \emph{classical} \emph{interaction picture} image is defined as
the pullback
\cite{Kim:2025sey,Campbell:1975nn}
\begin{align}
    \label{cip}
    \Uo(t_0,t)^*\act{
        f(t)
    }
    \,=\,
        f(t)
        \circ
        \Uo(t,t_0)
    \,,
\end{align}
where $t_0$ is a fixed time of one's choice,
often set to zero.

The first expression in \eqref{cip} 
shows that
the classical interaction picture
implements the pullback
due to the time-evolution symplectomorphism from $t$ to $t_0$.
At the same time,
the second expression in \eqref{cip}
shows that
the classical interaction picture 
means to simply insert the free-theory trajectory in the arguments of the function $f(t)$,
evolved from $t_0$ to $t$.
The reversal
in time direction here
is nothing more than
the usual wisdom that
shifting the argument of the function 
transforms the function in the reverse way.

\subsection{S-symplectomorphism}
\label{S-SYMP}

Given the above definition of classical interaction picture,
the idea of S-symplectomorphism arises.
The S-symplectomorphism
$S \in \Diff(\P,\sp)$
is a geometrical object
defined strictly within 
classical physics \cite{Kim:2025sey,hunziker1968s,simon1971wave,herbst1974classical,sokolov1979classical,narnhofer1981canonical,thirring1981classical}.
In short, it is the time-evolution symplectomorphism
in the classical interaction picture.

Firstly,
the point of departure is
the classical time-dependent perturbation theory
set up in \Sec{CIP}.
Let $\rho(t)$ be the probability distribution
governed by
the Liouville equation in \eqref{liouv}.
Let $\trho(t)$
be its classical interaction picture image.
Consider time evolution between times $t_\pm$, so
$
    \rho(t_+)
    =
    \rho(t_-) \circ U(t_-,t_+)   
$.
It follows that
\begin{align}
\begin{split}
    \label{rho-gym}
    \trho(t_+)
    \,&=\,
        \trho(t_-)
        \circ
        U^\circ\hnem(t_0,t_-)
        \circ
        U(t_-,t_+)
        \circ
        U^\circ\hnem(t_+,t_0)
    \,,\\
    \,&=\,
        \trho(t_-)
        \circ
        \BB{
            U^\circ\hnem(t_0,t_+)
            \circ
            U(t_+,t_-)
            \circ
            U^\circ\hnem(t_-,t_0)
        }^{\nem\nem-1}
    \,.
\end{split}
\end{align}
Provided a well-defined limit
\begin{align}
    \label{S}
    S
    \,=\mem
        \lim_{t_\pm \to \pm\infty}
            U^\circ\hnem(t_0,t_+)
            \circ
            U(t_+,t_-)
            \circ
            U^\circ\hnem(t_-,t_0)
    \,,
\end{align}
\eqref{rho-gym} implies
\begin{align}
    \label{S-act}
    \trho(+\infty)
    \,=\,
        S^*\act{
            \trho(-\infty)
        }
    \,,
\end{align}
meaning that the pullback $S^*$ transforms
the classical state at far past
to the classical state at far future
in the interaction picture.

Since $U^\circ(t_1,t_2)$ and $U(t_1,t_2)$
are symplectomorphisms for any $t_1,t_2$,
\eqref{S} defines a symplectomorphism as well,
dubbed \emph{S-symplectomorphism} \cite{Kim:2025sey}:
\begin{align}
    \label{S-in-Diff}
    S
    \,\,\,\in\,\,\,
    \Diff(\P,\sp)
    \,.
\end{align}

Secondly,
we desire to the evaluate the S-symplectomorphism
explicitly 
as a differential operator.
Let $\tH(t)$ and $\tV(t)$ be the classical interaction picture images of $H(t)$ and $V(t)$,
respectively.
By using \eqref{ag.Diff.c},
it follows that
\begin{align}
\begin{split}
    \label{liouv.gym}
    \text{\eqref{liouv}}
    &\qfq
    U^\circ\hnem(t_0,t)^*\act{
        \dot{\rho}(t)
    }
    \,=\,
    \pb{
        U^\circ\hnem(t_0,t)^*\act{
            H(t)
        }
    }{
        U^\circ\hnem(t_0,t)^*\act{
            \rho(t)
        }
    }
    \,,\\
    &\qfq
    \tilde{\dot{\rho}}(t)
    \,=\,
    \pb{
        \tH(t)
    }{
        \trho(t)
    }
    \,,\\
    &\qfq
    \dot{\trho}(t)
    \,=\,
    \pb{
        \tV(t)
    }{
        \trho(t)
    }
    \,,
\end{split}
\end{align}
where \smash{$\tilde{\dot{\rho}}(t) = 
U^\circ\hnem(t_0,t)^*\act{\dot{\rho}(t)}
$}.
The last line in \eqref{liouv.gym} uses
\begin{align}
    \label{intdot}
    \dot{\trho}(t)
    \,=\,
        \tilde{\dot{\rho}}(t)
        + \pb{ \trho(t) }{ \tilde{H}^\circ\hnem(t) }
    \,,
\end{align}
which is not difficult to deduce from $\trho(t) = \rho(t) \circ \Uo(t,t_0)$.

\eqref{liouv.gym} derives
the partial differential equation satisfied by $\trho(t)$,
whose solution
must reproduce the S-symplectomorphism in \eqref{S-act} as
\begin{align}
    \label{liouv.X.int}
    \dot{\trho}(t)
    \,=\,
        X_{\tV(t)}\act{
            \trho(t)
        }
    \qiq
    \trho(+\infty)
    \,=\,
        S^*\act{
            \trho(-\infty)
        }
    \,.
\end{align}
This derives
explicit formulae for $S^*$,
via
the Dyson \cite{Dyson:1949ha}
and
Magnus \cite{magnus1954exponential} series:
\begin{align}
	\label{S.sol}
	S^*
	\,=\,
		\Pexp{\mem
			\int dt\,\,
				X_{\tV(t)}
		}
	\,=\,
		\exp\nem\BB{
			X_\chi
		}
	\,.
\end{align}
The function $\chi$ in \eqref{S.sol} is given by
\begin{align}
	\label{eikonal}
	\chi
	\,=\,
		&
		\int dt_1\,\,
			\tV(t_1)
		+
		\frac{1}{2}\mem
		\int_{t_1>t_2}\nem d^2t\,\,
			\pb{\tV(t_1)}{\tV(t_2)}
\\
		&
		+
		\frac{1}{6}\mem
		\int_{t_1>t_2>t_3}\nem d^3t\,\,
			\BB{
				\pb{ \tV(t_1) }{ \pb{ \tV(t_2) }{ \tV(t_3) } }
				+
				\pb{ \tV(t_3) }{ \pb{ \tV(t_2) }{ \tV(t_1) } }
			}
		+ \cdots
	\,.
	\nonumber
\end{align}
The default integral domains are
$(-\infty,+\infty)$.

In sum, we have defined the S-symplectomorphism
$S \in \Diff(\P,\sp)$
in the phase space formulation of classical mechanics
and derived explicit formulae
for its computation as a differential operator
via solving the Liouville equation
in the classical interaction picture.

In a well-defined scattering problem, the limit in \eqref{S} constructively exists
by finiteness of the integral in \eqref{eikonal}.
This stipulates that the interaction Hamiltonian in the interaction picture, $\tV(t)$,
should decay to zero
in the limits $t \to \pm \infty$,
in particular.

The function $\chi$ in \eqref{eikonal}
is an effective Hamiltonian
that reproduces the entire time evolution
from far past to far future
within a unit dimensionless time.
It will be referred to as the \emph{classical eikonal}.

The formula $S^* = \exp(X_\chi)$
manifests the symplecticity of classical scattering.
Probability is conserved by classical scattering.

\subsection{Quantum Interaction Picture}
\label{QIP}

Next, we construct quantum scattering theory in the phase space formulation.

To begin with,
suppose the quantization of
the classical system in \Secs{CIP}{S-SYMP}
in terms of a well-behaved quantization map $\Q : \Cinfty(\P) \to \Ops(\Hilb)$.
Let $(\P,\star)$ be the resulting fuzzy phase space.
The same split in \eqref{Hsplit}
defines time-dependent perturbation theory in the phase space formulation of quantum mechanics,
in which case
the free and interaction Hamiltonians are quantized
with respect to the ordering prescription stipulated by $\Q$:
$\hat{H}^\circ\hnem(t) = \Q(H^\circ\hnem(t))$,
$\hat{V}(t) = \Q(V(t))$

Let $\sUo\hnem(t_1,t_2)$ and $\sU(t_1,t_2)$
be the fuzzy time-evolution diffeomorphisms from time $t_2$ to time $t_1$
due to
the Hamiltonians
$H^\circ\hnem(t)$ and $H(t)$, respectively.
They satisfy the usual axioms:
$\sU(t_1,t_1) = \mathrm{id}$,
$\sU(t_1,t_2) = \sU(t_2,t_1)^{-1}$,
$\sU(t_1,t_2) \circ \sU(t_2,t_3) = \sU(t_1,t_3)$.
Here, we remind ourselves that a fuzzy diffeomorphism
is a $\star$-preserving differential operator.

For any time-dependent function $f(t) \in \Cinfty(\P)$
on the fuzzy phase space,
its \emph{quantum} \emph{interaction picture} image is defined as
\begin{align}
    \label{qip}
    \sUo(t_0,t)\act{
        f(t)
    }
    \,,
\end{align}
where $t_0$ is the fixed chosen time.

This is a natural generalization of \eqref{cip}
in the phase space formulation of quantum mechanics.
Geometrically, \eqref{qip} brings the phase-space function $f(t)$
to the time $t_0$
through
the fuzzy diffeomorphism for
free evolution.
It is not difficult to see that \eqref{qip}
is equivalent 
to the standard definition of the quantum-mechanical interaction picture
familiar from textbooks:
use the intertwining identity in \eqrefs{tw1}{tw2}.

\subsection{Fuzzy S-diffeomorphism}
\label{S-DIFF}

Given the above definition of quantum interaction picture,
the phase space formulation
of quantum scattering theory
unfolds in the following way.

Firstly,
the point of departure
is the
time-dependent perturbation theory
in the phase space formulation of quantum mechanics,
set up in \Sec{QIP}.
Let $\rho(t)$ be a quasiprobability distribution
governed by
the quantum Liouville equation in \eqref{sliouv}.
Let $\trho(t)$
be its quantum interaction picture image.
Consider time evolution between times $t_\pm$,
so 
$
    \rho(t_+)
    =
    \sU(t_+,t_-)\act{
        \rho(t_-)
    }
$.
It follows that
\begin{align}
\begin{split}
    \label{rho-gym.qu}
    \trho(t_+)
    \,&=\,
        \sUo(t_0,t_+)\act{
            \sU(t_+,t_-)\act{
                \sUo(t_-,t_0)\act{
                    \trho(t_-)
                }
            }
        }
    \,.
\end{split}
\end{align}
Provided a well-defined limit
\begin{align}
    \label{S.qu}
    \sS
    \,=\mem
        \lim_{t_\pm \to \pm\infty}
            \sUo(t_0,t_+)
            \circ
            \sU(t_+,t_-)
            \circ
            \sUo(t_-,t_0)
    \,,
\end{align}
\eqref{rho-gym.qu} implies
\begin{align}
    \label{S-act.qu}
    \trho(+\infty)
    \,=\,
        \sS\act{
            \trho(-\infty)
        }
    \,,
\end{align}
meaning that $\sS$ transforms
the quantum state at far past
to the quantum state at far future
in the interaction picture.

Since $\sUo(t_1,t_2)$ and $\sU(t_1,t_2)$
are fuzzy diffeomorphisms for any $t_1,t_2$,
\eqref{S.qu} defines a fuzzy diffeomorphism as well:
\begin{align}
    \label{sS-in-Diff}
    \sS
    \,\,\,\in\,\,\,
    \Diff(\P,\star)
    \,.
\end{align}
We will refer to $\sS$ in \eqref{sS-in-Diff}
as the \emph{fuzzy S-diffeomorphism}.

Secondly,
we desire to the evaluate the fuzzy S-diffeomorphism
explicitly 
as a differential operator.
To this end,
let us take
$\tH(t)$ and $\tV(t)$ as the quantum interaction picture images of $H(t)$ and $V(t)$,
respectively.
By using \eqref{ag.sDiff.b},
it follows that
\begin{align}
\begin{split}
    \label{sliouv.gym}
    \text{\eqref{sliouv}}
    &\qfq
    \sUo(t_0,t)\act{
        \dot{\rho}(t)
    }
    \,=\,
    \spb{
        \sUo(t_0,t)\act{
            H(t)
        }
    }{
        \sUo(t_0,t)\act{
            \rho(t)
        }
    }
    \,,\\
    &\qfq
    \tilde{\dot{\rho}}(t)
    \,=\,
    \spb{
        \tH(t)
    }{
        \trho(t)
    }
    \,,\\
    &\qfq
    \dot{\trho}(t)
    \,=\,
    \spb{
        \tV(t)
    }{
        \trho(t)
    }
    \,,
\end{split}
\end{align}
where the last line uses
a deformed version of \eqref{intdot}
due to quantum interaction picture:
\begin{align}
    \label{intdot.qu}
    \dot{\trho}(t)
    \,=\,
        \tilde{\dot{\rho}}(t)
        + \spb{ \trho(t) }{ \tilde{H}^\circ\hnem(t) }
    \,.
\end{align}

\eqref{sliouv.gym} derives
the partial differential equation satisfied by
$\trho(t)$,
whose solution must reproduce the fuzzy S-diffeomorphism in \eqref{S-act.qu} as
\begin{align}
    \label{liouv.X.int.qu}
    \dot{\trho}(t)
    \,=\,
        \sX_{\tV(t)}\act{
            \trho(t)
        }
    \qiq
    \trho(+\infty)
    \,=\,
        \sS\act{
            \trho(-\infty)
        }
    \,.
\end{align}
This derives
explicit formulae for $\sS$,
via
the Dyson \cite{Dyson:1949ha}
and
Magnus \cite{magnus1954exponential} series:
\begin{align}
	\label{sS.sol}
	\sS
	\,=\,
		\Pexp{\mem
			\int dt\,\,
				\sX_{\tV(t)}
		}
	\,=\,
		\exp\nem\BB{
			\sX_\schi
		}
	\,.
\end{align}
The function $\schi$ in \eqref{sS.sol} is given by
\begin{align}
	\label{eikonal.qu}
	\schi
	\,=\,
		&
		\int dt_1\,\,
			\tV(t_1)
		+
		\frac{1}{2}\mem
		\int_{t_1>t_2}\nem d^2t\,\,
			\spb{\tV(t_1)}{\tV(t_2)}
\\
		&
		+
		\frac{1}{6}\mem
		\int_{t_1>t_2>t_3}\nem d^3t\,\,
			\BB{
				\spb{ \tV(t_1) }{ \spb{ \tV(t_2) }{ \tV(t_3) } }
				+
				\spb{ \tV(t_3) }{ \spb{ \tV(t_2) }{ \tV(t_1) } }
			}
		+ \cdots
	\,.
	\nonumber
\end{align}

In sum, we have defined the fuzzy S-diffeomorphism
$\sS \in \Diff(\P,\star)$
in the phase space formulation of quantum mechanics
and derived explicit formulae
for its computation as a differential operator
via solving the quantum Liouville equation
in the quantum interaction picture.

Again, well-defined scattering problems
stipulate
an appropriate decaying behavior of $\tV(t)$
in the limits $t \to \pm \infty$,
on account of the existence of the limit in \eqref{S.qu}
and the integral expression in \eqref{eikonal.qu}.

The function $\schi$ in \eqref{eikonal.qu}
is an effective Hamiltonian
that reproduces the entire quantum-mechanical time evolution
from far past to far future
within a unit dimensionless time.
It will be referred to as the \emph{quantum eikonal}.

The formula $\sS = \exp(\sX_\schi)$
manifests the fact that $\sS$
is a $\star$-preserving map.
This implies
the conservation of probability by quantum scattering,
based on identities such as the one explicated in \eqref{trace-swap-def}.
Therefore, 
the formula $\sS = \exp(\sX_\schi)$
manifests
unitarity of quantum scattering.
Probability is conserved by quantum scattering.

It is not difficult to show that
the quantum interaction picture smoothly reduces to the classical interaction picture:\footnote{
    See \eqref{intlimit.app}.
    A slight subtlety in the notation $\tilde{V}(t)$ in \eqrefs{sS.sol}{eikonal.qu}
    is clarified in \App{INTCD}.
    This caveat is yet absent
    in physically relevant examples
    featuring quadratic free Hamiltonians.
}
\begin{align}
    \label{intlimit}
    \lim_{\hbar\to0}
        \sUo(t_1,t_2)
    \,=\,
        \Uo(t_1,t_2)^*
    \,.
\end{align}
Therefore, it follows that
\begin{align}
    \lim_{\hbar\to0}\text{
        \eqref{S-act.qu}
    }
    \,&=\,\text{
        \eqref{S-act}
    }\,,\\[-0.15\baselineskip]
    \lim_{\hbar\to0}\text{
        \eqref{liouv.X.int.qu}
    }
    \,&=\,\text{
        \eqref{liouv.X.int}
    }\,,\\[-0.15\baselineskip]
    \lim_{\hbar\to0}\text{
        \eqref{sS.sol}
    }
    \,&=\,\text{
        \eqref{S.sol}
    }\,,\\[-0.15\baselineskip]
    \lim_{\hbar\to0}\text{
        \eqref{eikonal.qu}
    }
    \,&=\,\text{
        \eqref{eikonal}
    }\,.
\end{align}

\subsection{S-matrix}
\label{S-MAT}

Lastly,
for completeness,
we might quickly record the standard, Hilbert-space-based construction of quantum scattering theory.
Let
$\hat{U}^\circ\hnem(t_1,t_2)$
and
$\hat{U}(t_1,t_2)$
be
the time-evolution unitary operators for
the Hamiltonian operators
$\hat{H}^\circ\hnem(t) = \Q(H^\circ\hnem(t))$
and
$\hat{H}(t) = \Q(H(t))$.
The \emph{S-matrix} is defined as the limit
\begin{align}
    \hat{S}
    \,&=\mem
        \lim_{t_\pm \to \pm\infty}
            \hat{U}^\circ\hnem(t_0,t_+)
            \circ
            \hat{U}(t_+,t_-)
            \circ
            \hat{U}^\circ\hnem(t_-,t_0)
    \,.
\end{align}
Explicitly,
the Dyson \cite{Dyson:1949ha}
and
Magnus \cite{magnus1954exponential}
expansions are given by
\begin{align}
    \label{Sop.sol}
    \hat{S}
    \,=\,
		\Pexp{\mem
			\frac{1}{i\hbar}
			\int dt\,\,
				\tilde{\hat{V}}(t)
		\,}
    \,=\,
        \exp\nem\bb{
            \frac{1}{i\hbar}\mem \hat{\chi}
        }
    \,,
\end{align}
where \smash{$\tilde{\hat{V}}(t)$}
is the interaction Hamiltonian operator
in the quantum-mechanical interaction picture
constructed in the standard way,
whose image under $\Q^{-1}$
is precisely $\tV(t)$ in \Sec{S-DIFF}.
In \eqref{Sop.sol},
the \emph{eikonal matrix} $\hat{\chi}$ is
\begin{align}
	\label{eikonal.op}
	\hchi
	\,=\,
		&
		\int dt_1\,\,
			\tilde{\hat{V}}(t_1)
		+
		\frac{1}{2(i\hbar)}\mem
		\int_{t_1>t_2}\nem d^2t\,\,
			\comm{\tilde{\hat{V}}(t_1)}{\tilde{\hat{V}}(t_2)}
\\
		&
		+
		\frac{1}{6(i\hbar)^2}\mem
		\int_{t_1>t_2>t_3}\nem d^3t\,\,
			\BB{
				\comm{ \tilde{\hat{V}}(t_1) }{ \comm{ \tilde{\hat{V}}(t_2) }{ \tilde{\hat{V}}(t_3) } }
				+
				\comm{ \tilde{\hat{V}}(t_3) }{ \comm{ \tilde{\hat{V}}(t_2) }{ \tilde{\hat{V}}(t_1) } }
			}
		+ \cdots
	\,.
	\nonumber
\end{align}
\eqref{Sop.sol}
clearly shows that the eikonal matrix $\hat{\chi}$
is an effective Hamiltonian operator
that generates the entire time evolution 
from past infinity to future infinity
within ``$\mathit{\Delta} t = 1$.''\footnote{
    We have removed a minus sign common in the current literature
    on account of this view.
}

\phantom{.}
\vspace{-0.7\baselineskip}

\section{Classical Limits}
\label{LIMIT}


\Sec{SCA} has defined
various constructs in
classical and quantum scattering theory.
This section establishes exact relations between them,
with a focus on the classical limit.

\subsection{Eikonal Matrix $\to$ Classical Eikonal}
\label{LIM1}

We begin with the relation between
the eikonal matrix $\hat{\chi}$
in \eqref{eikonal.op},
the quantum eikonal $\schi$
in \eqref{eikonal.qu},
and the classical eikonal $\chi$
in \eqref{eikonal}:
\begin{align}
    \label{limitE}
    \hat{\chi}
    \quad{\color{emphcolor}
        \xleftrightarrow[]{
        \qquad\mathclap{\text{
        \footnotesize
            equiv.
        }}\qquad}
    }\quad
    \chi^\star
    \quad{\color{emphcolor}
        \xrightarrow[]{
        \qquad\mathclap{\text{
        \footnotesize
            $\hbar \to 0$
        }}\qquad}
    }\quad
    \chi
    \,.
\end{align}

Firstly,
the quantum eikonal $\chi^\star$
is the equivalent counterpart of
the eikonal matrix $\hat{\chi}$
in the phase space formulation.
To show this,
apply the inverse of a quantization map,
$\Q^{-1} : \Ops(\Hilb) \to \Cinfty(\P)$,
to the formula for $\hat{\chi}$
in \eqref{eikonal.op}.
It follows that
\begin{align}
    \label{limitE.a}
    \Q^{-1}(\hat{\chi})
    \,=\,
        \chi^\star
    \,.
\end{align}
Secondly,
the classical eikonal $\chi$
is the classical limit of the quantum eikonal $\schi$.
The formula in \eqref{eikonal.qu}
clearly establishes that
\begin{align}
    \label{limitE.b}
    \lim_{\hbar\to0}
        \chi^\star
    \,=\,
        \chi
    \,,
\end{align}
as
the deformed Poisson bracket
smoothly reduce to the Poisson bracket
in the $\hbar \to 0$ limit.
\eqrefs{limitE.a}{limitE.b}
together verifies the relation between
$\hat{\chi}$, $\schi$, and $\chi$
stated in \eqref{limitE}.

\subsection{Adjoint Action of S-matrix $\to$ S-symplectomorphism}
\label{LIM2}

Next, we investigate
the relationship between
the S-matrix $\hat{S}$
in \eqref{Sop.sol},
the fuzzy S-diffeomorphism $\sS$
in \eqref{sS.sol},
and the S-symplectomorphism $S$
in \eqref{S.sol}:
\begin{align}
    \label{limitS}
    \Ad_{\hat{S}}
    \quad{\color{emphcolor}
        \xleftrightarrow[]{
        \qquad\mathclap{\text{
        \footnotesize
            equiv.
        }}\qquad}
    }\quad
    S^\star
    \quad{\color{emphcolor}
        \xrightarrow[]{
        \qquad\mathclap{\text{
        \footnotesize
            $\hbar \to 0$
        }}\qquad}
    }\quad
    S^*
    \,.
\end{align}

Firstly,
the relation established in
\eqref{limitE.a}
implies that 
the fuzzy S-diffeomorphism $\sS$
is the equivalent counterpart of
the adjoint action of the S-matrix.
To show this, apply
the intertwining identity in \eqref{tw2}
to the formulae
$\hat{S} = \exp(\hat{\chi}/i\hbar)$
$\sS = \exp(\sX_{\schi})$
and 
$S^* = \exp(X_\chi)$
in \eqrefs{Sop.sol}{S.sol}.
Since \eqref{limitE.a} holds,
it follows that
\begin{align}
    \label{limitS.a}
    \Ad_{\hat{S}}
    \mem\circ\,
    \Q
    \,=\,
        \Q
        \,\circ
        \sS
    \,.
\end{align}
That is, the quantization map $\Q$
intertwines between
the adjoint action of the S-matrix
and
the fuzzy S-diffeomorphism.

Secondly,
the pullback $S^*$ of S-symplectomorphism
is
the classical limit of the fuzzy S-diffeomorphism $\sS$.
Based on the formulae
$\sS = \exp(\sX_{\schi})$
and 
$S^* = \exp(X_\chi)$
in \eqrefs{S.sol}{sS.sol},
the limit established in \eqref{limitE.b}
implies that
\begin{align}
    \label{limitS.b}
    \lim_{\hbar\to0}
        S^\star
    \,=\,
        S^*
    \,.
\end{align}
In particular,
the deformed Hamiltonian vector field
$\sX_{\schi}$,
as a differential operator,
smoothly approaches to
the Hamiltonian vector field
$X_\chi$.
\eqrefs{limitS.a}{limitS.b}
together verifies the relation between
$\hat{S}$, $\sS$, and $S$
stated in \eqref{limitS}.

\subsection{S-matrix $\to$ No Good Limit}
\label{LIM3}

Note the crucial role of the phase space formulation of quantum mechanics
in establishing the relations in \eqrefs{limitE}{limitS}.
For instance,
it does not make sense to
equate a phase-space function $\chi$
with a limit of an operator $\hat{\chi}$.
However, 
it does make sense to
equate a phase-space function $\chi$
with a limit of
a phase space function $\schi$.
In this manner,
the classical limits are established as precise mathematical equalities,
not as mere arguments
appealing to the correspondence principle.

Note also the crucial role of the adjoint action in \eqref{limitS}.
In particular,
suppose we investigate the classical limit of the S-matrix $\hat{S}$ directly.
Its image under $\Q^{-1}$ is given by
\begin{align}
    \label{star-exp}
    \Q^{-1}(\hat{S})
    \,=\,   
        \exp^\star\nem\bb{
            \frac{1}{i\hbar}\mem \schi
        }
    \,=\,
        1
        +
        \sum_{n=1}^\infty\,
            \frac{1}{n!\hem (i\hbar)^n}\,
			\underbrace{
				\schi \star \cdots \star \schi
				{}_{\vphantom{\frac{0}{0}}}
			}_{\text{$n$ times}}
    \,,
\end{align}
whose $\hbar \to 0$ limit is ill-defined.
We also clarify that the $\star$-exponential function 
$\exp^\star$
has appeared for the first time in \eqref{star-exp};
all exponentials so far
have been the ordinary one.
Namely,
compare \eqref{star-exp} with
\begin{align}
    \Q^{-1}\hnem \circ \Ad_{\hat{S}} \mem\circ\, \Q
    \,=\,   
        \exp\nem\BB{
            \sX_\schi
        }
    \,=\,
        1
        +
        \sum_{n=1}^\infty\,
            \frac{1}{n!}\,
                (\sX_\schi)^n
    \,,
\end{align}
which describes the ordinary exponentiation
of a differential operator $\sX_\schi$.










\subsection{Impulse Formulae}
\label{IMPULSE}

We end with a remark on impulse formulae.
The celebrated Kosower, Maybe, O'Connell formalism \cite{KMOC}
provided
a concrete framework
for obtaining the impulse of classical observables
from the S-matrix $\hat{S}$.
The framework of S-symplectomorphism $S$,
however,
provides a purely classical implementation:
the nested bracket formula due to \rcite{eikonaltwistor},
which also traces back to
\rrcite{Gonzo:2024zxo,Damgaard:2023ttc}.
Certainly, 
the analysis of this paper
establishes
how this classical framework
is the faithful $\hbar \to 0$
limit of a quantum framework.

First of all, let us review \rcite{eikonaltwistor}'s nested bracket formula:
\begin{align}
    \label{Del}
    \Del\act{f}
    \,&=\,
        \pb{f}{\chi}
        + \frac{1}{2!}\,
            \pb{\pb{f}{\chi}}{\chi}
        + \frac{1}{3!}\,
            \pb{\pb{\pb{f}{\chi}}{\chi}}{\chi}
        + \cdots
    \,.
\end{align}
This computes the impulse of a classical observable $f \in \Cinfty(\P)$
in classical scattering,
in terms of the classical eikonal $\chi$.
Mathematically, \eqref{Del}
defines a differential operator
$\Del: \Cinfty(\P) \to \Cinfty(\P)$
on phase-space functions
which we may refer to as
\emph{classical} \emph{impulse operator}.
Geometrically, \eqref{Del} can be rewritten as
\begin{align}
\begin{split}
    \label{impulse}
    \Del\act{f}
    \,\,=\,\,
        (S^{-1})^*\act{ f }
        - f
    \,\,&=\,\,
        \sum_{n=1}^\infty\,
            \frac{1}{n!}\,
                (-X_\chi)^n\act{
                    f
                }
    \,,
\end{split}
\end{align}
where
$(S^{-1})^*\act{ f }$ in \eqref{impulse}
computes
the pullback of the function $f$
by the map $S^{-1}$.
Due to the very geometrical meaning 
of the S-symplectomorphism $S$,
this is exactly
the pullback of $f$
from the \textit{final} phase space to the \textit{initial} phase space;
hence $(S^{-1})^*\act{f} - f$ compute the impulse of $f$
in classical scattering.
Therefore,
\eqref{impulse}
provides 
a purely classical derivation
of an impulse formula for classical observables on phase space.

\newpage


Next, let us consider the impulse in quantum scattering.
The
Kosower, Maybe, O'Connell framework \cite{KMOC}
identifies the impulse of an operator $\hat{f} \in \Ops(\Hilb)$ as
the adjoint action
$\hat{S}^{-1}\hem \hat{f}\, \hat{S} = \Ad_{\hat{S}^{-1}}\act{\hat{f}}$.
In the density matrix formulation of quantum mechanics,
this is derived
as follows.
First,
the expectation value of $\hat{f}$
is found by
taking the trace with the density matrix.
Second,
the S-matrix acts on the density matrix 
as the adjoint action $\Ad_{\hat{S}}$.
Third, this is translated to $\Ad_{\hat{S}^{-1}}$
on the operator $\hat{f}$
via the cyclic property of trace.

To carry out this derivation in the phase space formulation of quantum mechanics,
one uses the identity that
the trace of an operator $\hat{f}$ equals
the integral of $\Q^{-1}(\hat{f})$ over the entire phase space
with respect to the Liouville measure $\m$,
as is explored previously 
in \eqrefs{density-norm}{trace-swap-def}.
Consequently,
the phase space formulation computes
the expectation value
of a quantum observable
$f \in \Cinfty(\P)$
by integrating
its star product with the quasiprobability distribution
over $\P$.
In the scattering setup,
the expectation value 
at far future is
\begin{align}
\begin{split}
    \label{smeared.a}
        \int \mu\:\:
            \trho(+\infty)
            \star
            f
    \,\,=
        \int \mu\:\:
            \BB{\sS\act{
                \trho(-\infty)
            }}
            \star
            f
    \,\,=
        \int \mu\:\:
            \trho(-\infty)
            \star
            \BB{\sS^{-1}\act{
                f
            }}
    \,,
\end{split}
\end{align}
where we have used Eqs.\:(\ref{S-act.qu}), (\ref{limitS.a}),
and 
the cyclic property of trace 
(recall \eqref{trace-swap-def}):
\begin{align}
    \label{smeared.b}
    \int \mu\:\:
        \Q^{-1}(\hat{S})
        \star
        \trho(-\infty)
        \star
        \Q^{-1}(\hat{S}^{-1})\mem
        \star
        f
    \,\,=
        \int \mu\:\:
            \trho(-\infty)
            \star
            \Q^{-1}(\hat{S}^{-1})\mem
            \star
            f
            \star
            \Q^{-1}(\hat{S})
    \,.
\end{align}
Therefore,
the expectation value of $f$ at far future
equals the expectation value of $\sS{}^{-1}\act{f}$
evaluated at far past,
for any ensemble state $\trho(t)$
in the quantum interaction picture.

As a result,
the impulse of a quantum observable $f \in \Cinfty(\P)$
can be defined
independently of the smearing ensemble states as
\begin{align}
\begin{split}
    \label{impulse.qu}
    \Del^\star\act{f}
    \,\,=\,\,
        \sS^{-1}\act{ f }
        - f
    \,\,&=\,\,
        \sum_{n=1}^\infty\,
            \frac{1}{n!}\,
                (-\sX_\schi)^n\act{
                    f
                }
    \,,
\end{split}
\end{align}
which defines the \emph{quantum impulse operator} $\Del^\star : \Cinfty(\P) \to \Cinfty(\P)$.
Note that $\trho(t)$ in \eqrefs{smeared.a}{smeared.b}
is merely employed as a dummy object 
in this derivation.
In terms of the quantum eikonal $\schi$,
it is convenient to 
utilize right actions
to write \eqref{impulse.qu} as
\begin{align}
    \label{Del.qu}
    \Del^\star\act{f}
    \,&=\,
        \spb{f}{\schi}
        + \frac{1}{2!}\,
            \spb{\spb{f}{\schi}}{\schi}
        + \frac{1}{3!}\,
            \spb{\spb{\spb{f}{\schi}}{\schi}}{\schi}
        + \cdots
    \,.
\end{align}

Certainly, \eqref{Del.qu} smoothly approaches to \eqref{Del} in the $\hbar\to0$ limit.
In conclusion,
we have shown that \rcite{eikonaltwistor}'s
classical impulse formula
arises as the faithful $\hbar\to0$ limit
of the quantum impulse formula
in the phase space formulation.
In precise terms,
we have established
a yet another set of exact equalities
about classical limit:
\begin{align}
    \label{limitDel}
    \Ad_{\hat{S}^{-1}} - \mathrm{id}
    \quad{\color{emphcolor}
        \xleftrightarrow[]{
        \qquad\mathclap{\text{
        \footnotesize
            equiv.
        }}\qquad}
    }\quad
    \Del^\star
    \quad{\color{emphcolor}
        \xrightarrow[]{
        \qquad\mathclap{\text{
        \footnotesize
            $\hbar \to 0$
        }}\qquad}
    }\quad
    \Del
    \,.
\end{align}
The first arrow in \eqref{limitDel} represents
\begin{align}
    \label{limitDel.a}
    \BB{
        \Ad_{\hat{S}^{-1}}
        - \mathrm{id}
    }
    \mem\circ\,
    \Q
    \,=\,
        \Q
        \,\circ
        \Del^\star
    \,,
\end{align}
which is
implied by \eqref{limitS.a}.
The second arrow in \eqref{limitDel} represents
\begin{align}
    \label{limitDel.b}
    \lim_{\hbar\to0}
        \Del^\star
    \,=\,
    \lim_{\hbar\to0}
    \BB{
        S^\star{}^{-1} - \mathrm{id}
    }
    \,\,=\,\,
        (S^{-1})^* - \mathrm{id}
    \,\,=\,\,
        \Del
    \,,
\end{align}
which is
implied by \eqref{limitS.b}.

\newpage

Note that
both 
$\Del$ and $\Del^\star$
are differential operators
acting on phase-space functions.
Unlike $\Del$,
however, 
$\Del^\star$
generates $\hbar$-dependencies
in general
even when acted on functions with no $\hbar$ dependencies.
Note also that
nothing prevents one from implementing
the Kosower, Maybe, O'Connell formalism \cite{KMOC}
in the phase space formulation:
\begin{align}
    \label{kmoc}
    \hat{S} \,=\, \hat{1} - \frac{1}{i\hbar}\, \hat{T}
    \qiq
    \Del^\star\act{ f }
    \,=\,
        \spb{
            \Q^{-1}(\hat{T})
        }{f}
        + \frac{1}{i\hbar}\, \Q^{-1}(\hat{T}^\dagger)
        \star
        \spb{
            \Q^{-1}(\hat{T})
        }{f}
    \,.
\end{align}
The smoothness of classical limit, however,
is not manifest in this approach.
The so-called superclassical terms arise
from each of
$\spb{
    \Q^{-1}(\hat{T})
}{f}$
and
$
\Q^{-1}(\hat{T}^\dagger)
\star
\spb{
    \Q^{-1}(\hat{T})
}{f}
/i\hbar
$
while $\Q^{-1}(\hat{T}) = - \schi - \schi \star \schi / 2i\hbar - \cdots$
does not admit a nice classical limit
as explored earlier in \Sec{LIM3}.
In contrast,
the adjoint actions in
\eqref{Del.qu} describe a definite classical limit.


\section{Quantum Eikonal from Magnus Expansion}
\label{MAGNUS}

The phase space formulation of classical and quantum scattering theory
now stands well-established,
by virtue of 
the analyses provided by
Secs.\:\ref{PSFOR}, \ref{SCA}, and \ref{LIMIT}.

The eikonals
$\chi$ and $\schi$
deserve a spotlight.
They efficiently encapsulate the complete information about classical and quantum scattering
as scalar functions on phase space,
while also manifesting
the conservation of probability via
$S^* = \exp(X_\chi)$
and
$S^\star = \exp(\sX_\schi)$.
Their explicit formulae
are given
by the Magnus expansions
in \eqrefs{eikonal}{eikonal.qu}.

In this last section,
we urge for a concrete understanding
on the quantum eikonal $\schi$
in terms of its explicit evaluation.
The key observation of this paper
is that
the formula in \eqref{eikonal.qu}
simply $\hbar$-deforms each Poisson bracket in \eqref{eikonal}.
This facilitates
a principled approach
for computing the quantum eikonal
in a diagrammatic language.

In this paper,
we limit our scope up to the third order in the Magnus expansion.
Nevertheless, the loop order can be raised indefinitely
thanks to the systematic formulation
in terms of deformed Poisson brackets.
Also, our approach straightforwardly applies to quantum field theory
by taking an infinite-dimensional phase space,
as shown in \App{QFT}.

For the reader's sake,
the Magnus expansions in \eqrefs{eikonal}{eikonal.qu}
are reproduced below.
The classical eikonal $\chi$ is given at each order as
\begin{subequations}
	\label{CE}
\begin{align}
	\label{CE.1}
    \chia
	\,&=\,
		\int dt_1\,\,
			\tV(t_1)
	\,,\\
	\label{CE.2}
    \chib
    \,&=\,
		\frac{1}{2}\mem
		\int_{t_1>t_2}\nem d^2t\,\,
			\pb{\tV(t_1)}{\tV(t_2)}
    \,,\\
	\label{CE.3}
    \chic
    \,&=\,
		\frac{1}{6}\mem
		\int_{t_1>t_2>t_3}\nem d^3t\,\,
			\BB{
				\pb{ \tV(t_1) }{ \pb{ \tV(t_2) }{ \tV(t_3) } }
				+
				\pb{ \tV(t_3) }{ \pb{ \tV(t_2) }{ \tV(t_1) } }
			}
    \,,
\end{align}
\end{subequations}
while
the quantum eikonal $\schi$ is given at each order as
\begin{subequations}
	\label{QE}
\begin{align}
	\label{QE.1}
	\schia
	\,&=\,
		\int dt_1\,\,
			\tV(t_1)
    \,,\\
	\label{QE.2}
    \schib
    \,&=\,
		\frac{1}{2}\mem
		\int_{t_1>t_2}\nem d^2t\,\,
			\spb{\tV(t_1)}{\tV(t_2)}
    \,,\\
	\label{QE.3}
    \schic
    \,&=\,
		\frac{1}{6}\mem
		\int_{t_1>t_2>t_3}\nem d^3t\,\,
			\BB{
				\spb{ \tV(t_1) }{ \spb{ \tV(t_2) }{ \tV(t_3) } }
				+
				\spb{ \tV(t_3) }{ \spb{ \tV(t_2) }{ \tV(t_1) } }
			}
	\,.
\end{align}
\end{subequations}

\subsection{Classical Eikonal}
\label{CEMAGNUS}

We begin by clarifying the foundations of our diagrammatic framework
for Magnus expansion in phase space.
Readers familiar with the work \cite{eikonalsangmin1}
can directly jump to \Sec{QEMAGNUS.MOYAL}.

Here,
the geometrical premise is 
a Poisson manifold $(\P,\Pi)$
equipped with coordinates $\phi^I$.
The coordinate derivative $\partial_I = \partial/\partial\phi^I$
is a differential operator that
acts on functions $f \in \Cinfty(\P)$
as $\partial_I\act{f}$.
The Poisson bracket is given by
\begin{align}
    \label{pi}
    \pb{f}{g}
    \,=\,
        \Pi^{IJ}
            \,
            \partial_I\act{f}
            \,
            \partial_J\act{g}
    \,,
\end{align}
where $\Pi^{IJ} = -\Pi^{JI}$ are
the components of the \emph{Poisson tensor} $\Pi$.
In symplectic manifolds,
$\Pi^{IJ} = (\omega^{-1})^{IJ}$.
For simplicity, we assume that $\Pi^{IJ}$ are constants.

\subsubsection{Tensor Graphs}
\label{PENROSE}

A graphical representation of \eqref{pi}
is facilitated by utilizing the following two methods of visualization.

The first
is the \emph{differentiation balloon notation}
of Penrose \cite{penrosegraphical1,penrosegraphical2},
incorporated as a part 
in his renowned
graphical notation 
\cite{penrosegraphical1,penrosegraphical2,Penrose:1956tensormethods,penrose1971negdim}
for tensors.
The coordinate derivative $\partial_I$
is represented as a balloon whose tail describes the index $I$:
\begin{align}
    \label{graphical.balloon}
    \quad
    \qwrap{\partial_I\act{\blank}}
    \qpq
    \nem
    \qinc{valign=t,raise=8pt}{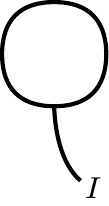}
    \quad.
\end{align}
\vskip-5.5pt\noindent
The Leibniz rule,
$\partial_I\act{f\mem g} = (\partial_I\act{f})\mem g+ f\mem (\partial_I\act{g})$,
is graphically represented as
\begin{align}
    \label{graphical.leibniz}
    \qinc{valign=t,raise=8pt}{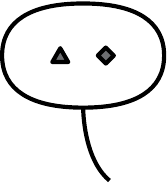}
    \,=\,
    \qinc{valign=t,raise=8pt}{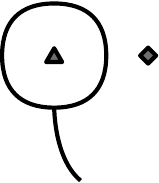}
    \,+\,
    \qinc{valign=t,raise=8pt}{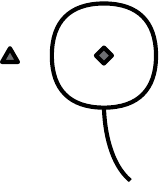}
    \,,
\end{align}
where the triangle and diamond represent
the scalar fields
$f$ and $g$, respectively.
In \eqref{graphical.leibniz},
the explicit index $I$ is omitted
since it essentially serves as a dummy.
The reader is encouraged to adopt—or invent—their preferred intuition for comprehending the visual rule in \eqref{graphical.leibniz}.
For instance, 
one
might imagine the balloon ``digesting'' each component, much like an enzyme acting on a molecular complex.
In \rcite{eikonalsangmin1}, the process is instead envisioned as “popping” the balloon like a bubble, thereby reducing it to smaller pieces.


The second is an \emph{arrow representation for Poisson tensor},
which is an instance of
birdtracks \cite{cvitanovic2008group} notation:
\begin{align}
    \label{graphical.arrow}
    \Pi^{IJ}  \,=\, -\Pi^{JI}
    \qqpqq
    \inc{valign=c}{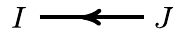}
    \,=\,
    - \inc{valign=c}{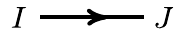}
    \,.
\end{align}
Here, the arrowhead is the symbol for $\Pi$,
on which the two lines representing indices $I,J$ are attached.
Flipping the arrow simply changes the sign,
encoding 
antisymmetry.
In fact,
\eqref{graphical.arrow} 
yields
a simplification of
the Kontsevich \cite{kontsevich} graph notation
for constant $\Pi^{IJ}$.


By incorporating the differentiation balloon
in the birdtracks grammar,
the Poisson bracket in \eqref{pi}
is graphically represented as
\begin{align}
    \label{graphical.pb}
    \qwrap{
        \{\blank\mem,\blank\}
    }\nem
    \qpq
    \qinc{valign=c}{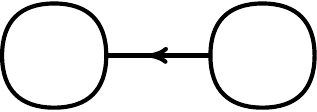}
    \,,
\end{align}
where
the arguments
$f$ and $g$ 
will be inserted in the first and second slots/balloons.
Index contraction
means to
glue the ends of two lines
in the graphical notation
\cite{penrosegraphical1,penrosegraphical2,Penrose:1956tensormethods,penrose1971negdim,cvitanovic2008group}.

Next, a graphical representation of
the Magnus expansion in \eqref{CE}
is facilitated by 
introducing the following two rules.
First,
$\tV(t)$
is represented as a round blob
with label $t$:
\vskip-6pt\noindent
\begin{align}
    \quad
    \tV(t)
    \qqpqq
    \inc{valign=c}{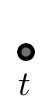}
    \quad.
\end{align}
\vskip-2pt\noindent
Second, the phase-space derivatives of $\tV(t)$ 
are simply denoted as
\begin{subequations}
\label{graphical.abbrV}
\begin{align}
    \qwrap{
        \partial_I\act{
            \tV(t)
        }
    }
    &\qpq
    \qinc{valign=c}{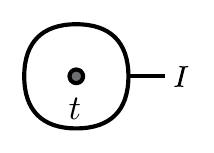}
    \,=\,
    \qinc{valign=c}{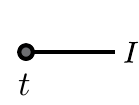}
    \quad,\\
    \qwrap{
        \partial_J\act{
        \partial_I\act{
            \tV(t)
        }}
    }
    &\qpq
    \qinc{valign=c}{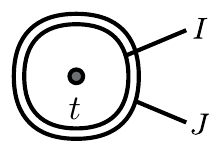}
    \,=\,
    \qinc{valign=c}{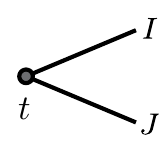}
    \quad,
\end{align}
\end{subequations}
and so on.
This essentially declares each $n$\textsuperscript{th} derivative of $\tV(t)$ as a new tensor in the graphical notation.

As a result, \eqref{CE} is graphically represented as
\begin{subequations}
\label{graphical.chi}
\begin{align}
    \label{graphical.chi1}
    \chia\nem\nem
    &\qqpqq\nem
        \int dt_1\,\,\bbsq{
            \ainc{valign=c}{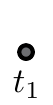}
        }
    \,,\\
    \label{graphical.chi2}
    \chib\nem\nem
    &\qqpqq\nem
        \frac{1}{2}\, 
        \int_{t_1>t_2} d^2t\,\,\bbsq{
            \ainc{valign=c}{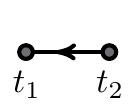}
        }
    \,,\\
    \label{graphical.chi3}
    \chic\nem\nem
    &\qqpqq\nem
        \frac{1}{6}\, 
        \int_{t_1>t_2>t_3} d^3t\,\,\bbsq{
            \ainc{valign=c}{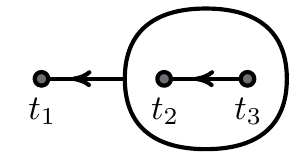}
            +
            \ainc{valign=c}{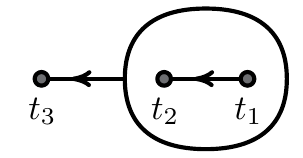}
        }
    \,,
\end{align}
\end{subequations}
where the nested Poisson brackets
have
turned into nested balloons.

\subsubsection{Intermediate Form via Bubble Popping}
\label{BUBBLE}

Regarding explicit evaluation of \eqref{graphical.chi},
the nested brackets are fully expanded out
in the purely diagrammatic language
via the Leibniz rule in \eqref{graphical.leibniz}.
This procedure might be dubbed 
\emph{bubble popping},
following \rcite{eikonalsangmin1}.

Take the first term in the integrand of \eqref{graphical.chi3},
for instance.
The diagrammatic Leibniz rule 
in \eqref{graphical.leibniz}
is applied as
\begin{subequations}
\begin{align}
    \label{graphical.demo-cleib}
    \qinc{valign=c}{br123}
    \,&=\,
    \qinc{valign=c}{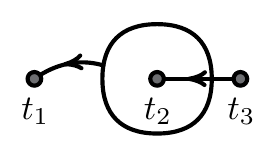}
    \,+\,
    \qinc{valign=c}{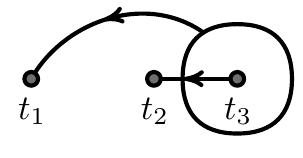}
    \,,
\end{align}
which then boils down to
\begin{align}
    \label{graphical.demo-cleib'}
    \qinc{valign=c}{br123}
    \,&=\,
    \qinc{valign=c}{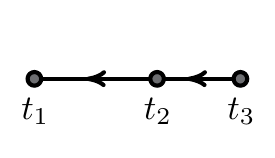}
    \,+\,
    \qinc{valign=c}{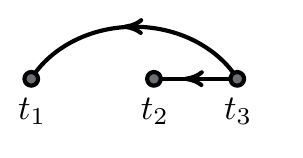}
\end{align}
\end{subequations}
by adopting the simplified notation proposed in \eqref{graphical.abbrV}.
By adding the image under $t_1 \leftrightarrow t_3$ exchange,
it is easy to see that
$\chic$ in \eqref{CE.3} is brought to
\vskip-12pt\noindent
\begin{align}
    \label{graphical.chi3.inter}
    \,\,\,
        \frac{1}{6}\, 
        \int_{t_1>t_2>t_3} d^3t\,\,\bbsq{
            \qwrap{2\: \inc{valign=c}{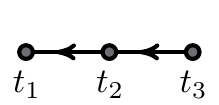}}
            \nem{\kern-0.4em}+{\kern-0.4em} 
            \qinc{valign=c}{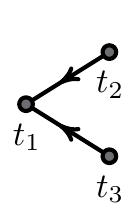}
            \nem{\kern-0.4em}+{\kern-0.4em}
            \qinc{valign=c}{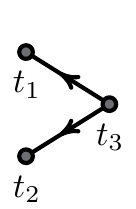}
        }
    \,,
\end{align}
\vskip-2pt\noindent
which we name the \emph{intermediate form}
of \smash{$\chic$}.
Obtaining the intermediate form
of 
$\chin$
means to expand out $n-2$ nontrivially nested Poisson brackets
by using the Leibniz rule in \eqref{graphical.leibniz}:
``pop all balloons.''
Via the abbreviation rule in \eqref{graphical.abbrV},
the outcome is represented as
an integral whose integrand
is a weighted sum of time-labeled graphs.

Generally speaking,
this process
may involve 
flipping some arrows via \eqref{graphical.arrow}.
Also, it should be clarified that
the value of a diagram is left unchanged
by planar isotopy:
\vskip-6pt\noindent
\begin{align}
    \label{graphical.dancing}
    \qinc{valign=c}{br123calc2B}
    \kern-1em
    =
    \qinc{valign=c}{chi3Vr}
    =
    \qinc{valign=c}{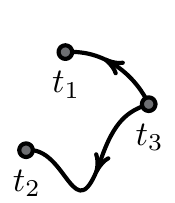}
    =
    \qinc{valign=c}{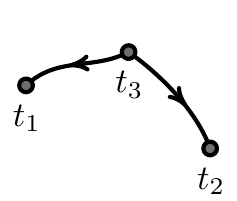}
    .
\end{align}
\vskip-2pt\noindent


\subsubsection{Feynman Graphs}
\label{FEYNMAN}

The graphical notation introduced 
above
is a way of rewriting tensors.
However, we eventually transition to
the \emph{Feynman graph notation}
\cite{feynman1948space}
where each edge is not merely a tensor ({as numerator})
but a propagator ({as numerator with denominator}).

The Heaviside step function 
is defined
such that
\begin{align}
    \label{heaviside}
    \Theta(t_1,t_2)
    \quad=\quad
        \left\{
        \begin{aligned}[c]
        \,\,\,
            1 
            &\quad
            \text{if $t_1 > t_2$}
            \,,\\
        \,\,\,
            0
            &\quad
            \text{if $t_1 < t_2$}
            \,.
        \end{aligned}
        \right.
\end{align}
As explicated at length in \App{QFT}
for both particles and fields,
the \emph{retarded propagator}
of the free theory
arises by
combining the Poisson tensor $\Pi^{I_1I_2}$ 
(as Pauli-Jordan function)
with the step function $\Theta(t_1,t_2)$.
Based on this fact,
an integral of a tensor graph
is converted to a Feynman graph as
\begin{align}
    \label{feynman2}
    \int_{t_1>t_2} d^2t\,\,\bbsq{
        \awrap{
            \inc{valign=c}{V12}
        }
    }
    \,\,\,=\,\,
    \int d^2t\,\,\bbsq{
        \awrap{
            \inc{valign=c}{V12}
            \:\Theta(t_1,t_2)
        }
    }
    \,\,\,=\,\,
        \inc{valign=c}{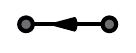}
    \,.
\end{align}
In a Feynman graph
such as the last diagram in \eqref{feynman2},
the vertices are unlabeled
since the times associated to them
are dummy integration variables.
Also, the edges are marked with a different kind of arrowhead
to represent the retarded propagator.

\subsubsection{Final Form via Freeing}
\label{FREEING}

By \eqref{feynman2},
the second-order eikonal $\chib$
in \eqref{graphical.chi2}
is brought to the \emph{final form}
as
\begin{align}
    \chib\nem\nem
    &\qqpqq\nem
    \awrap{
        \frac{1}{2}
        \inc{valign=c}{F2}
    }
    \,.
\end{align}
Obtaining the final form
of the \smash{$n$\textsuperscript{th}-order} eikonal $\chin$
means to 
represent it as a weighted sum of Feynman graphs
where each edge describes the retarded propagator.

To obtain the final form of the third-order eikonal $\chic$,
we recall its intermediate form
obtained in \eqref{graphical.chi3.inter}.
The first term in \eqref{graphical.chi3.inter}
is straightforwardly boiled down to a Feynman graph as
\begin{align}
\begin{split}
    \label{graphical.freeing.3L}
    \int_{t_1>t_2>t_3} d^3t\,\,\bbsq{
        \awrap{
            \inc{valign=c}{chi3L}
        }
    }
    \,\,=\,\,
        \inc{valign=c}{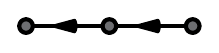}
    \,,
\end{split}
\end{align}
which repeats the exercise in \eqref{feynman2}.


However,
a nontrivial gymnastics is required
for
the second term in \eqref{graphical.chi3.inter}.
The Poisson tensors
connect between times
$(t_1,t_2)$ and $(t_1,t_3)$,
which do not align with the time ordering $t_1 > t_2 > t_3$
prescribed by the integration domain.
Notably, this mismatch can be handled 
by symmetrizing over $t_2 \leftrightarrow t_3$ exchange via change of integration variables:\footnote{
    This gymnastics amounts to the identity
    $
        \Theta(t_1,t_2)\, \Theta(t_2,t_3)
        = 
            \Theta(t_1,t_2)\, \Theta(t_2,t_3)
            +
            \Theta(t_1,t_2)\, \Theta(t_3,t_2)
    $.
}
\vskip-12pt\noindent
\begin{align}
\begin{split}
    \label{graphical.freeing.3V}
    &
    \int_{t_1>t_2>t_3} d^3t\,\,\bbsq{
        \awrap{
            \inc{valign=c}{chi3V}
        }
    }
    \\[-12pt]
    &=\,\,
    \frac{1}{2}\,
    \int_{t_1>t_2>t_3} d^3t\,\,\bbsq{
        \awrap{
            \inc{valign=c}{chi3V}
        }
    }
    \,\,+\,\,\,
    \frac{1}{2}\,
    \int_{t_1>t_3>t_2} d^3t\,\,\bbsq{
        \awrap{
            \inc{valign=c}{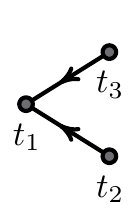}
        }
    }
    \,,\\[-12pt]
    &=\,\,
    \frac{1}{2}\,
    \int_{t_1>t_2,t_1>t_3} d^3t\,\,\bbsq{
        \awrap{
            \inc{valign=c}{chi3V}
        }
    }
    \,\,\,=\,\,\,
        \frac{1}{2}\:
        \inc{valign=c}{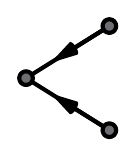}
    \,\,\,
    \,.
\end{split}
\end{align}
In the second equality,
we have used a planar isotopy
\`a la \eqref{graphical.dancing}.

The gymnastics demonstrated in \eqref{graphical.freeing.3V}
will be referred to as
the \emph{freeing} operation,
which induces a factor $1/2$
of a combinatorial origin.
The inverse of this factor,
$2$,
will be referred to as
the \emph{freeing factor}.


As is explained in \rcite{eikonalsangmin1},
the freeing factor is well-formulated in precise mathematical terms.
The V-shaped graph in 
the first line of
\eqref{graphical.freeing.3V}
defines 
a partial ordering between
its time labels:
$\{\mem{ t_1 > t_2 ,\mem t_1 > t_3 }\mem\}$.
This is relevant to its promotion to a Feynman graph:
each retarded propagator encodes a step function,
stipulating $t_1>t_2$ and $t_1>t_3$.
On the other hand,
the integral domain
is given as
a total ordering
between $t_1,t_2,t_3$.
Crucially,
there are two \textit{linear extensions} of
the partial ordering
$\{\mem t_1 > t_2 ,\mem t_1 > t_3 \mem\}$
to a total ordering:
$t_1 > t_2 > t_3$
and
$t_1 > t_3 > t_2$.
The freeing factor $2$ is 
the number of such linear extensions.


\subsubsection{Magnus Coefficient}
\label{CMU}

Eventually, $\chia,\chib,\chic$
are found in the final form as
\begin{subequations}
\label{graphical.CE}
\begin{align}
\label{graphical.CE.a}
    \chia
    &\qpq
        \qinc{valign=c}{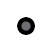}
    \,,\\[12pt]
\label{graphical.CE.b}
    \chib
    &\qpq
            \qwrap{
                \frac{1}{2}\:
                \inc{valign=c}{F2}
            }
    \,,\\
\label{graphical.CE.c}
    \chic
    &\qpq
        \qwrap{
            \frac{1}{3}\:
            \inc{valign=c}{F3L}
        }
        \nem{\kern-0.4em}+{\kern-0.4em}
        \qwrap{
            \frac{1}{12}\:
            \inc{valign=c}{F3V}
        }
        \nem{\kern-0.4em}+{\kern-0.4em}
        \qwrap{
            \frac{1}{12}\:
            \inc{valign=c}{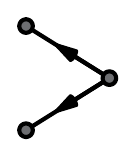}
        }
    \,.
\end{align}
\end{subequations}
\vskip-6pt\noindent
In general,
let us define that
a \emph{classical Magnus graph}
is a connected directed acyclic graph
with $n$ vertices
and zero loops.
The final form of the 
$n$\textsuperscript{th}-order classical eikonal $\chin$
describes
a weighted sum of such graphs:
\begin{align}
    \chin
    \,\qpq\,
        \sum_{
            \G \,\in\, \Mag(n,0)
        }\,
            \m(\G)\,\,
                \G
    \,.
\end{align}
Here, $\Mag(n,0)$ is the set of classical Magnus graphs with $n$ vertices.
In this way, each classical Magnus graph $\G$ 
is assigned with a unique coefficient
$\mu(\G)$,
which may be referred to as the \emph{Magnus coefficient}.
It holds generically true that
$\mu(\G) = \mu(\G^\top)$,
if $\G^\top$ denotes 
the transpose of $\G$:
the graph obtained from $\G$ by reversing 
its orientation structure.
This property is inherited from a $\mathbb{Z}_2$ symmetry in the Magnus expansion
(see \eqref{eikonal}).

With hindsight, it is helpful to multiply
$\m(\G)$
by the symmetry factor $\s(\G)$
of $\G$:
\begin{align}
    \label{omega}
    \omega(\G)
    \,=\,
        \mu(\G)\, \s(\G)
    \,,
\end{align}
which could be called \emph{reduced Magnus coefficient}.
For instance, \eqref{graphical.CE} describes that
\begin{align}
\begin{split}
    \omega\mem\BB{
        \inc{valign=c}{F1}
    }
    \,=\,
        1
    &\,,\quad
    \omega\mem\BB{
        \inc{valign=c}{F3L}
    }
    \,=\,
        \frac{1}{3}
    \,,\\
    \omega\mem\BB{
        \inc{valign=c}{F2}
    }
    \,=\,
        \frac{1}{2}
    &\,,\quad
    \omega\mem\BB{
        \inc{valign=c}{F3V}
    }
    \,=\,
    \omega\mem\BB{
        \inc{valign=c}{F3Vr}
    }
    \,=\,
        \frac{1}{6}
    \,.
\end{split}
\end{align}
\vskip-4pt\noindent
Note that some graphs do not contribute to the classical eikonal
by having zero coefficient,
as is substantiated at order $n=4$:
\vskip-8pt\noindent
\begin{align}
\begin{split}
    \omega\mem\BB{
        \inc{valign=c}{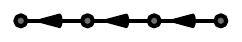}
    }
    \,=\,
        \frac{1}{4}
    &\,,\quad
    \omega\mem\BB{
        \inc{valign=c}{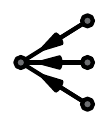}
    }
    \,=\,
        0
    \,,\\[-6pt]
    \omega\mem\BB{
        \inc{valign=c}{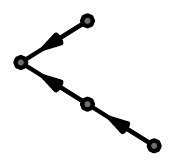}
    }
    \,=\,
    \omega\mem\BB{
        \inc{valign=c}{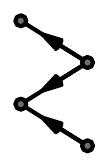}
    }
    \,=\,
        \frac{1}{12}
    &\,,\quad
    \omega\mem\BB{
        \inc{valign=c}{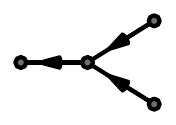}
    }
    \,=\,
        \frac{1}{6}
    \,.
\end{split}
\end{align}
\vskip-4pt\noindent
Also, some graphs start to exhibit negative coefficients from order $n=5$.

The coefficients of classical Magnus graphs
have been well-studied.
\rcite{eikonalsangmin1} shows that
the reduced Magnus coefficient $\omega(\G)$
generalizes a graph function
previously defined by Murua \cite{Murua_2006}
for rooted tree graphs
(up to conventional sign factor $(-1)^{|\G|-1}$).
In the Hopf-algebraic framework \cite{Calaque_2011},
$\omega(\G)$ 
serves as an inverse to the graph function $e(\G)$
in the antipode sense,
where $e(\G)$ is the freeing factor for $\G$
divided by $|\G|!$.

\subsection{Quantum Eikonal (Symmetric Ordering)}
\label{QEMAGNUS.MOYAL}

Provided the preliminary discussion in \Sec{CEMAGNUS},
this subsection computes
the quantum eikonal
for systems quantized with the Moyal star product.

Here, the geometrical premise is 
a linear Poisson manifold $(\P,\Pi)$.
In terms of linear coordinates $\phi^I$,
the Moyal star product is defined as
\begin{align}
	\label{starM}
	f \star g
	\,=\,
			f
			\,
			\exp\nem
            \bb{\,
                \overleftarrow{\partial_I}\,
                \b\mem\Pi^{IJ}\mem
                \overrightarrow{\partial_J}
            \mem}
			\,
			g
	\,.
\end{align}
Then $(\P,\star)$ is the quantization of the classical system $(\P,\Pi)$
in a symmetric ordering prescription:
\eqref{starM} treats all coordinates $\phi^I$
on the same footing
(is ignorant of polarization data).
For brevity, we have desired to take the deformation parameter as
\begin{align}
    \label{beta}
    \beta
    \,=\,
        i\hbar/2
    \,.
\end{align}
\eqref{starM} reproduces \eqref{moyal-def}
for $\P = \R^2$ with $\Pi = \partial_x \wedge \partial_p$.

\subsubsection{Graphical Notation}
\label{PENROSE-MOYAL}

The deformed Poisson bracket
due to the Moyal star product in \eqref{starM} is
\begin{align}
	\label{spbM}
    \sum_{k=0}^\infty\,
        \frac{\beta^{2k}}{(2k{\,+\mem}1)!}\,\,
            \BB{
            \partial_{I_1}\act{
            \cdots
                \partial_{I_{2k+1}}\act{
                    f
                }
            }
            }
			\,
            \Pi^{I_1J_1}
            \cdots
            \Pi^{I_{2k+1}J_{2k+1}}
			\mem
            \BB{
            \partial_{J_1}\act{
            \cdots
                \partial_{J_{2k+1}}\act{
                    g
                }
            }
            }
    \,,
\end{align}
which reduces to \eqref{spb.moyal}
for $\P = \R^2$.
For an intuitive diagrammatic representation,
we introduce a fuzzy line as a new graphic element:
\begin{align}
	\label{graphical.spbM}
    \qwrap{\spb{\blank}{\blank}}
    \qpq
    \qinc{valign=c}{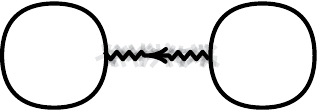}
    \,.
\end{align}
That is,
the deformed Poisson bracket in \eqref{graphical.spbM}
is 
literally
the Poisson bracket 
in \eqref{graphical.pb}
\textit{made fuzzy}.
With this notation, \eqref{spbM}
translates to
\begin{align}
    \label{graphical.Moyal}
    \,\,
    &
    \inc{valign=c}{spb}
    \\[-5pt]
    &
    =\,
    \ainc{valign=c}{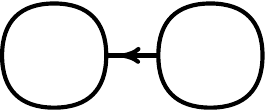}
    + \awrap{
        \frac{\b^2}{3!}\,\,
            \inc{valign=c}{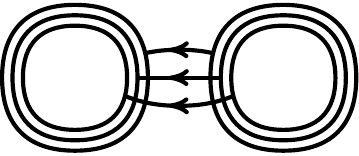}
    }
    + \awrap{
        \frac{\b^4}{5!}\,\,
            \inc{valign=c}{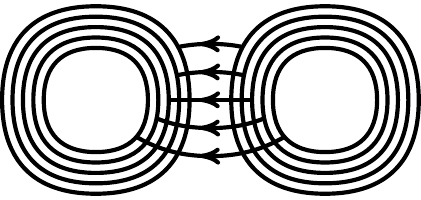}
    }
    + \awrap{\cdots}
    \,,
    \nonumber
\end{align}
whose right-hand side 
is a direct application of
the graphical notation set up in \Sec{PENROSE}.



For simplicity of our notation,
we also regard that
\vskip-12pt\noindent
\begin{align}
    \qinc{valign=c}{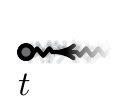}
    \,\,\,\text{means}\quad
    \qinc{valign=c}{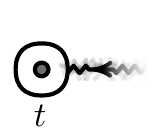}
    \,,
\end{align}
\vskip-10pt\noindent
meaning that the balloon enclosing $\tV(t)$
can be omitted for fuzzy lines as well
(cf.\:\eqref{graphical.abbrV}).


With this understanding,
\eqref{QE} is graphically represented as
\begin{subequations}
\label{graphical.schi}
\begin{align}
    \label{graphical.schi1}
    \schia\nem\nem
    &\qqpqq\nem
        \int dt_1\,\,\bbsq{
            \ainc{valign=c}{V1}
        }
    \,,\\
    \label{graphical.schi2}
    \schib\nem\nem
    &\qqpqq\nem
        \frac{1}{2}\, 
        \int_{t_1>t_2} d^2t\,\,\bbsq{
            \ainc{valign=c}{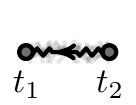}
        }
    \,,\\
    \label{graphical.schi3}
    \schic\nem\nem
    &\qqpqq\nem
        \frac{1}{6}\, 
        \int_{t_1>t_2>t_3} d^3t\,\,\bbsq{
            \ainc{valign=c}{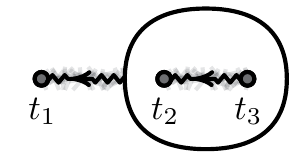}
            +
            \ainc{valign=c}{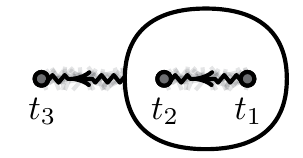}
        }
    \,.
\end{align}
\end{subequations}
Compare this with \eqref{graphical.chi}.

\subsubsection{Evaluation}
\label{QEMAGNUS.MOYAL.EVAL}

We now perform the diagrammatic evaluation of the quantum eikonal in \eqref{graphical.schi}.
Once all fuzzy lines are unpacked 
into ordinary lines,
the procedure takes two steps as before:
bubble popping (Leibniz rule) and freeing (counting linear extensions).

The evaluation of \eqrefs{graphical.schi1}{graphical.schi2}
is straightforward;
hence we consider the third-order quantum eikonal in \eqref{graphical.schi3}.

To begin with,
all the fuzzy lines must be unpacked via \eqref{graphical.Moyal}.
The first term in the integrand of \eqref{graphical.schi3},
for instance,
is unpacked as
\vskip-4pt\noindent
\begin{align}
    \label{graphical.demo-a}
    \kern-1.1em
    \inc{valign=c}{br123}
    \,+\,\,
    \frac{\b^2}{3!}\:\Bigg[{\,
        \inc{valign=c}{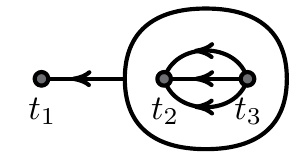}
        \,\,\nem\nem+\nem\nem\nem
        \inc{valign=c}{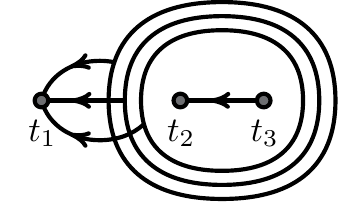}
    \,\,\,\,}\Bigg]
    \,+\, \O(\b^4)
    \,.
    \kern-0.6em
\end{align}
The $\O(\b^0)$ part is of course
the classical term evaluated in \eqref{graphical.demo-cleib}.
The $\O(\b^2)$ part,
on the other hand,
describes the leading quantum contribution.

\begin{subequations}
Next, we implement the bubble popping.
For example,
take the $\O(\b^2)$ part of \eqref{graphical.demo-a}.
There are two diagrams.
By applying the Leibniz rule in \eqref{graphical.leibniz},
the first diagram is evaluated as
\vskip-12pt\noindent
\begin{align}
    \label{graphical.demo-b1}
    \inc{valign=c}{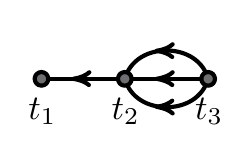}
    \,+\,
    \inc{valign=c}{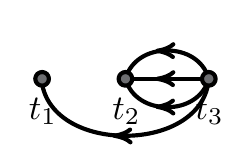}
    \,,
\end{align}
whereas the second diagram is evaluated as
\vskip-10pt\noindent
\begin{align}
    \label{graphical.demo-b2}
    \inc{valign=c}{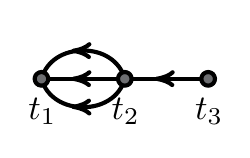}
    \,+\,
    \kern6pt
    {3\: \inc{valign=c}{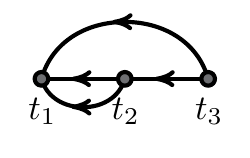}}
    \,+\,
    \kern6pt
    {3\: \inc{valign=c}{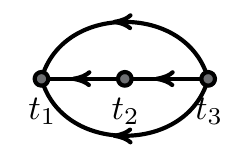}}
    \,+\,
    \inc{valign=c}{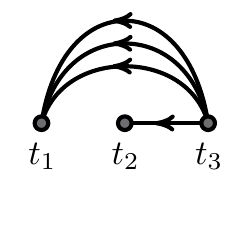}
    \,.
\end{align}
\end{subequations}
\vskip-14pt\noindent
Then the intermediate form of
the third-order quantum eikonal 
at two loops
is found via
the sum of
\eqrefs{graphical.demo-b1}{graphical.demo-b2}
plus its image under
$t_1 \leftrightarrow t_3$ exchange.

Finally, we implement the freeing.
Take the sum of the last diagrams in
\eqrefs{graphical.demo-b1}{graphical.demo-b2},
for instance.
Upon integration over the domain
$t_1>t_2>t_3$, it is
\vskip-4pt\noindent
\begin{align}
    \label{freeing-demo.13b}
    \int_{t_1>t_2>t_3} d^3t\,\,\bbsq{
        \nem\nem
        \bwrap{
            \inc{valign=c}{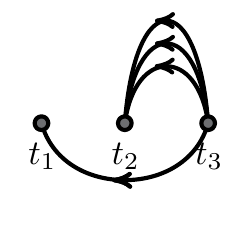}
        }\nem\nem
    }
    \,\,\,+\,\,\,
    \int_{t_2>t_1>t_3} d^3t\,\,\bbsq{
        \nem\nem
        \bwrap{
            \inc{valign=c}{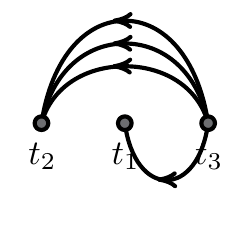}
        }\nem\nem
    }
    \,,
\end{align}
\vskip-12pt\noindent
where a change of integration variables is used.
By using planar isotopy,
\eqref{freeing-demo.13b} becomes
\vskip-12pt\noindent
\begin{align}
    \label{freeing-demo.13c}
    \int_{t_1>t_3, t_2>t_3} d^3t\,\,\bbsq{
        \nem\nem
        \bwrap{
            \inc{valign=c}{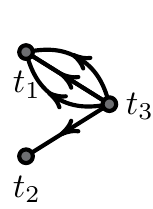}
        }\nem\nem
    }
    \,\,\,
    \,\,\,=\,\,\,
        \binc{valign=c}{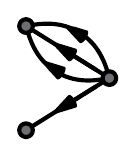}
    \,.
\end{align}
\vskip-4pt\noindent

From the calculations detailed above,
it is not difficult to see that
the final form of the third-order quantum eikonal $\schic$ at two loops
describes the sum
\vspace{-3pt}
\begin{align}
\begin{split}
    \label{graphical.ME2}
    \kern-1.2em
    {\kern6pt \frac{\beta^2}{6} \inc{valign=c}{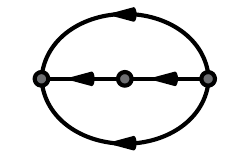}}
    &\,+\,
    {\kern6pt \frac{\beta^2}{12} \inc{valign=c}{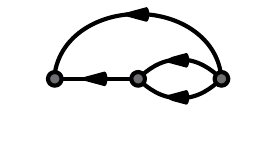}}
    \,+\,
    {\kern6pt \frac{\beta^2}{18} \inc{valign=c}{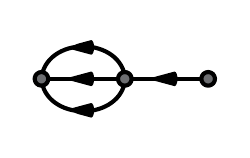}}
    \,+\,
    {\kern6pt \frac{\beta^2}{36} \inc{valign=c}{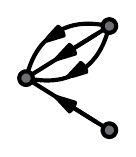}}
    \\
    &\,+\,
    {\kern6pt \frac{\beta^2}{12} \inc{valign=c}{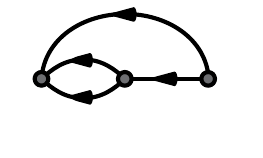}}
    \,+\,
    {\kern6pt \frac{\beta^2}{18} \inc{valign=c}{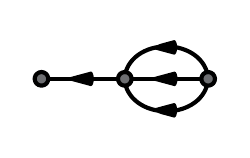}}
    \,+\,
    {\kern6pt \frac{\beta^2}{36} \inc{valign=c}{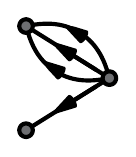}}
    {\kern6pt}
    \,.
    \kern-0.9em
\end{split}
\end{align}
\vskip-6pt\noindent
This concludes a demonstration for
the diagrammatic evaluation of the quantum eikonal in Moyal quantization (symmetric ordering).

Crucially, \eqref{graphical.ME2}
describes a collection of directed \textit{acyclic} graphs.
Cyclic graphs are forbidden
since any chain of retarded propagators (step functions)
evaluates to zero:
they are incompatbile with time ordering.

By working in the same fashion,
it can be found that
the third-order quantum eikonal $\schic$
describes
the sum of the following two groups of terms:
\begin{subequations}
\label{M3}
\begin{align}
    \label{M3.desc}
    &
    \lrp{
    \begin{aligned}[c]
        \frac{1}{3}\,
        \sum_{ 2\nmid a,\hem 2\nmid b }
        \frac{\b^{a+b-2}}{a!\mem b!}\,
            \inc{valign=c}{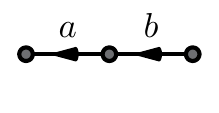}
    &
    {}
        \,+\,
        \frac{1}{12}\,
        \sum_{ 2\nmid a }
        \frac{\b^{2a-2}}{a!^2}\,
            \inc{valign=c}{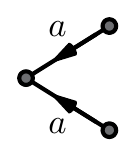}
        \,+\,
        \frac{1}{12}\,
        \sum_{\substack{
             2\nmid a,\hem 2\nmid b,\\
             a\neq b 
        }}
        \frac{2\b^{a+b-2}}{a!\mem b!}\,
            \inc{valign=c}{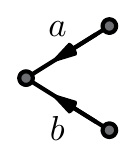}
        \\
    &
    {}
        \,+\,
        \frac{1}{12}\,
        \sum_{ 2\nmid a }
        \frac{\b^{2a-2}}{a!^2}\,
            \inc{valign=c}{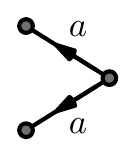}
        \,+\,
        \frac{1}{12}\,
        \sum_{\substack{
             2\nmid a,\hem 2\nmid b,\\
             a\neq b 
        }}
        \frac{2\b^{a+b-2}}{a!\mem b!}\,
            \inc{valign=c}{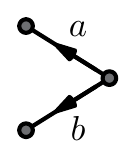}
    \end{aligned}
    }
    \,,\\
    \label{M3.prim}
    &
    \lrp{\,
        \frac{\b^{a+b+c-2}}{6}\,
        \sum_{
            2\nmid (a+b)
            ,\hem
            2\nmid c
        }
        \frac{1}{a!\mem b!\mem c!}\,
            \inc{valign=c}{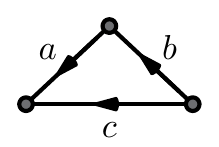}
        \,+\,
        \frac{\b^{a+b+c-2}}{3}\,
        \sum_{
            2\nmid a ,\hem 2\nmid b,\hem
            2\mid c
        }
        \frac{1}{a!\mem b!\mem c!}\,
            \inc{valign=c}{MTabc}
    }
    \,.
\end{align}
\end{subequations}
Here, $a,b,c \geq 1$ are positive integers indicating the multiplicity of each edge.
We have explicitly verified \eqref{M3}
up to $\ell = 101$ loops
by numerical means.


The astute reader will point out that
the first group of terms,
\eqref{M3.desc},
simply equals
the third-order classical eikonal $\chic$
in \eqref{graphical.CE.c}
if all lines are promoted to fuzzy lines:
\vspace{-8pt}
\begin{align}
    \Fuz(\chic)
    \,\,\,\,\qpq
    \label{graphical.fuz3}
        \qwrap{
            \frac{1}{3}\:
            \inc{valign=c}{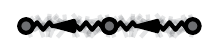}
        }
        \nem{\kern-0.4em}+{\kern-0.4em}
        \qwrap{
            \frac{1}{12}\:
            \inc{valign=c}{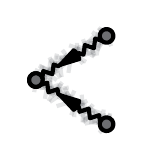}
        }
        \nem{\kern-0.4em}+{\kern-0.4em}
        \qwrap{
            \frac{1}{12}\:
            \inc{valign=c}{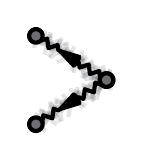}
        }
    \,.
\end{align}
\vskip-8pt\noindent
Let us recall to \eqref{graphical.fuz3}
as the \emph{fuzzification} of $\chic$.
In contrast,
the second group of terms, \eqref{M3.prim},
does not arise in this way.
Hence a qualitative difference is implied between
\eqrefs{M3.desc}{M3.prim}.
At $\O(\b^2)$,
\eqref{graphical.fuz3}
reproduces 
the third and fourth columns
in \eqref{graphical.ME2}
but not the first and second columns.

In fact,
we can consider 
the following
noncommutative diagram.
\begin{align}
	\label{eq:cd-fuzzification}
	\adjustbox{valign=c}{\begin{tikzpicture}
	    \node[empty] (O) at (0,0) {};
	    \node[empty] (Y) at (0, 2.8) {};
	    \node[empty] (y) at (0, 1.0) {};
	    \node[empty] (X) at (6.6, 0) {};
	    \node[b] (a) at ($(O)$) {
            \eqref{graphical.chi}
        };
	    \node[b] (b) at ($(O)+(Y)$) {
            \eqref{graphical.schi}
        };
	    \node[b] (A) at ($(O)+(X)$) {
            $\chin$
        };
	    \node[b] (B) at ($(O)+(X)+(Y)-(y)$) {
            $\Fuz(\chin)$
        };
	    \node[b] (B') at ($(O)+(X)+(Y)$) {
            $\schin$
        };
	    \draw[->] (a)--(b) node[midway,left] {\itshape%
            fuzzify%
            \:%
        };
	    \draw[->] (A)--(B) node[midway,right] {\itshape%
            \,%
            fuzzify%
        };
	    \draw[->] (a)--(A) node[midway,below] 
        {\itshape
            pop bubbles
        };
	    \draw[->] (b)--(B') node[midway,above] {\itshape
            unpack then
            pop bubbles
        };
	\end{tikzpicture}}
\end{align}
While $\schin$ is the desired answer
for the quantum eikonal,
$\Fuz(\chin)$
is what one obtains by
replacing every edge in the final form of the classical eikonal $\chin$
with the fuzzy line.
Crucially,
we find $\schin \neq \Fuz(\chin)$
from $n = 3$.


The mismatch
$\schin - \Fuz(\chin)$
precisely originates from the fact that
the Lebniz rule in \eqref{graphical.leibniz}
does not apply for balloons attached with fuzzy lines,
since \eqref{graphical.Moyal} describe
a collection of higher-derivative operators.
Thus one must unpack
all fuzzy lines to ordinary lines
via \eqref{graphical.Moyal}
before initiating the bubble popping process,
as is clarified at the beginning of \Sec{QEMAGNUS.MOYAL.EVAL}.
In other words,
$\Fuz(\chin)$
is what one would get
if the Leibniz rule
is mistakenly applied for balloons attached with fuzzy lines.

\subsubsection{Magnus-Moyal Coefficient}
\label{QMU.MOYAL}

Let us define that
a \emph{Magnus graph of the first kind}
is a connected directed acyclic graph.
The final form of the 
$n$\textsuperscript{th}-order quantum eikonal $\schin$
in Moyal quantization
is represented as 
a weighted sum
of such graphs:
\begin{align}
    \schin
    \,\qpq\,    
        \sum_{\ell=0}^\infty\,
        \sum_{
            \G \mem\in\mem \Mag(n,\ell)
        }\,
            \b^\ell\mem
            \m_1(\G)\,\,
                \G
    \,,
\end{align}
where $\Mag(n,\ell)$ is the set of Magnus graphs of the first kind
with $n$ vertices and $\ell$ loops.
In this way, each $\G \in \Mag(n,\ell)$ 
is assigned with a unique coefficient
$\m_1(\G)$.
This defines
the \emph{Magnus-Moyal coefficient} of $\G$.

From the property of the deformed Poisson bracket,
it is immediate that 
this Magnus-Moyal coefficient coincides
with the Magnus coefficient defined in \Sec{CMU}
at zero loops:
$\m_1(\G) = \m(\G)$ if $\G \in \Mag(n,0)$.
Another immediate fact is that $\m_1(\G) = 0$
at odd loop orders: $2\nmid\ell$.
The $\mathbb{Z}_2$ symmetry
$\m(\G^\top) = \m(\G)$
continues to hold.

\newpage

Again, it is desirable to work with
\emph{reduced Magnus-Moyal coefficient}:
\begin{align}
    \label{omega1}
    \omega_1(\G)
    \,=\,
        \mu_1(\G)\, \s(\G)
    \,.
\end{align}
For example,
\eqref{M3} describes that
\begin{subequations}
\label{M3!}
\begin{align}
    \label{M3L!}
    \kern-0.5em
    \omega_1\hem\BB{
        \inc{valign=c}{MLab}
    }
    \,\,&=\,\,
    \omega\mem\BB{
        \inc{valign=c}{F3L}
    }\,
    \awrap{\cdot}
        \left\{
        \begin{aligned}[c]
        \,\,\,
            1
            &\quad
            \text{if $a,b$ are both odd
            }
            \\
        \,\,\,
            0
            &\quad
            \text{otherwise}
        \end{aligned}
        \right\}
    \,,\\
    \label{M3V!}
    \omega_1\hem\BB{
        \inc{valign=c}{MVab}
    }
    \,\,&=\,\,
    \omega\mem\BB{
        \inc{valign=c}{F3V}
    }\,
    \awrap{\cdot}
        \left\{
        \begin{aligned}[c]
        \,\,\,
            1
            &\quad
            \text{if $a,b$ are both odd
            }
            \\
        \,\,\,
            0
            &\quad
            \text{otherwise}
        \end{aligned}
        \right\}
    \,,\\
    \label{M3T!}
    \kern-0.5em
    \omega_1\hem\BB{
        \inc{valign=c}{MTabc}
    }
    \,\,&=\,\,
        \left\{
        \begin{aligned}[c]
        \,\,\,
            1/6
            &\quad
            \text{if $(a,b,c)$ is
                (odd,even,odd) or
                (even,odd,odd)
            }
            \\
        \,\,\,
            1/3
            &\quad
            \text{if $(a,b,c)$ is
                (odd,odd,even)
            }
            \\
        \,\,\,
            0
            &\quad
            \text{otherwise}
        \end{aligned}
        \right\}
    \,,
    \kern-0.5em
\end{align}
\end{subequations}
compactly summarizing all Magnus-Moyal coefficients
at three vertices and all loops.

The equalities in \eqrefs{M3L!}{M3V!}
encode 
our earlier observation that
a part of $\schic$
simply recycles the classical eikonal $\chic$
via fuzzification.
Hence
\eqref{M3T!}
describes the ``genuinely quantum'' part of $\schic$
that cannot be straightforwardly 
derived from $\chic$.




For a precise mathematical formulation,
let $\Prim(n,\ell)$
be the subset of $\Mag(n,\ell)$
whose elements are simple.
A graph is simple if it has 
at most one edge between any two vertices.
Elements of $\Prim(n,\ell)$ are called \emph{primaries of the first kind}.
Accordingly,
$\G \in \Mag(n,\ell)$
is a \emph{descendant} of
$\G_0 \in \Prim(n,\ell_0)$
if $\G$ arises by giving an edge multiplicity to $\G_0$.
The set of descendants of $\G_0 \in \Mag(n,\ell_0)$
is denoted as $\Desc(\G_0)$.
For example,
\begin{align}
    \label{descin}
    \ainc{valign=c}{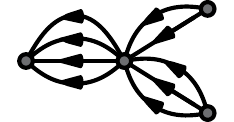}
    \quad\in\quad
    \Desc\bb{
        \ainc{valign=c}{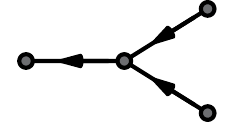}
    }
    \,
    \,.
\end{align}
%
Meanwhile, 
$\G \in \Mag(n,\ell)$ is a \emph{Moyal-descendant} of 
$\G_0 \in \Mag(n,0)$
if every edge multiplicity given to $\G_0$
for obtaining $\G$
is odd.
For example, \eqref{descin}
does not describe a Moyal descendant.
%
Finally,
$\G \in \Mag(n,\ell)$
is \emph{trivial}
if it 
is a descendant of
a classical Magnus graph:
$\G \in \Desc(\G_0)$
for $\G_0 \in \Mag(n,0)$.
$\G \in \Mag(n,\ell)$ is \emph{nontrivial} if it is not trivial.
Fuzzification $\Fuz$
is a linear map
that outputs a weighted sum of 
trivial Moyal-descendants.

With these definitions,
it can be shown for any $n$ that
\begin{align}
    \label{ME.primary}
    \schin
    \,-\,
    \Fuz(\chin)
    \,\qpq\,
        \sum_{k=1}^\infty\,
            \sum_{
            \substack{
                \G \mem\in\mem \Mag(n,2k)
                \\
                \text{nontrivial}
            }
            }   
                \b^{2k} \cdot
                \frac{\omega_1(\G)}{\s(\G)}\,\,
                    \G
    \,,
\end{align}
from which \eqrefs{M3L!}{M3V!} follow as
\begin{subequations}
\begin{align}
    \text{$\G$ trivial Moyal-descendant of $\G_0$}
    &\qiq
    \omega_1(\G) \,=\, \omega(\G_0)
    \,,\\
    \text{$\G$ trivial but not Moyal-descendant}
    &\qiq
    \omega_1(\G) \,=\, 0
    \,.
\end{align}
\end{subequations}
Thus,
it suffices to 
determine the coefficients of 
nontrivial graphs.
At $n=3$,
the triangle in \eqref{M3T!}
is the only topology of a nontrivial primary.
At $n=4$,
there are four topologies
with one ($
    \inc{valign=c,scale=0.6,raise=1.35pt}{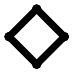}
$, $
    \inc{valign=c,scale=0.6,raise=1.35pt}{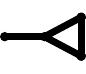}
$), two ($
    \inc{valign=c,scale=0.6,raise=1.35pt}{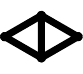}
$), and three ($
    \inc{valign=c,scale=0.6,raise=1.35pt}{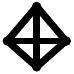}
$) loops
when orientation is ignored.
At $n=5$, there are $18$ such topologies.

\subsection{Quantum Eikonal (Normal Ordering)}
\label{QEMAGNUS.WICK}

Finally,
this subsection computes
the quantum eikonal
for systems quantized with the Wick star product,
such as quantum field theories.

Here, the geometrical premise is 
a K\"ahler vector space
or a suitable generalization.
This is a linear Poisson manifold $(\P,\Pi)$
endowed with a rank-two tensor $W = W^{IJ}\mem \partial_I \otimes \partial_J$,
which we call the \emph{Wightman tensor}.
Importantly, it is related to the Poisson tensor as
\begin{align}
    \label{W-to-Pi}
    \frac{1}{i}\,\BB{
        W^{IJ} - W^{JI}
    }
    \,=\,
        \Pi^{IJ}
    \,.
\end{align}
We suppose linear coordinates $\phi^I$ such that the components $W^{IJ}$ are constants.
The Wick star product is defined as
\begin{align}
	\label{starW}
	f \star g
	\,=\,
			f
			\,
			\exp\nem
            \bb{\,
                \overleftarrow{\partial_I}\,
                \hbar\mem W^{IJ}\mem
                \overrightarrow{\partial_J}
            \mem}
			\,
			g
	\,.
\end{align}
\eqref{starW} reproduces \eqref{wick-def}
for $\P \cong \C^1$ 
by taking $W = \partial_{a} \otimes \partial_{\ba}$,
which shows that the physical content of the Wightman tensor
is the very Wick contraction between $a$ and $\ba$.
See \App{WIGHTMAN}
for more details on the mathematical setup
of this subsection.

\subsubsection{Graphical Notation}
\label{PENROSE-WICK}

We shall begin by constructing the graphical notation for Wick star product.
As in \Sec{CEMAGNUS},
we start with tensor graphs
and then transition to Feynman graphs.

The Wightman tensor will be represented
as a red line:
\begin{align}
    \label{graphical.W}
    W^{IJ}\nem\hnem
    \qqpqq
    \inc{valign=c}{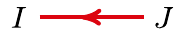}
    \,.
\end{align}
The Wightman tensor does not exhibit any index symmetry.
\eqref{W-to-Pi} translates to
\begin{align}
    \label{graphical.W-to-Pi}
    \awrap{\frac{1}{i}\,\BB{
        \ainc{valign=c}{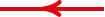}
        -
        \ainc{valign=c}{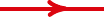}
    }}
    \,=\,
        \ainc{valign=c}{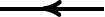}
    \,.
\end{align}
The Wick star product in \eqref{starW}
then describes
\begin{align}
    f \star g 
    \qpq
    \ainc{valign=c}{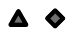}
    +
    \awrap{\frac{\hbar^1}{1!}\,
        \inc{valign=c}{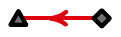}}
    +
    \awrap{\frac{\hbar^2}{2!}\,
        \inc{valign=c}{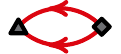}}
    +
    \awrap{\frac{\hbar^3}{3!}\,
        \inc{valign=c}{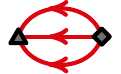}}
    + \O(\hbar^4)
    \,.
\end{align}
Here,
the weights $1/1!, 1/2!, 1/3!, \cdots$
serve as symmetry factors
for edge multiplicity.

The deformed Poisson bracket
$\pb{f}{g}^\star$
due to the Wick star product in \eqref{starW} is
\begin{align}
	\label{spbW}
    \frac{1}{i}\mem
    \sum_{\ell=0}^\infty\,
        \frac{\hbar^\ell}{(\ell{\,+\mem}1)!}\,\,
            \BB{
            \partial_{I_1}\act{
            {\cdots\mem}
                \partial_{I_{\ell+1}}\act{
                    f
                }
            }
            }
			\,
            \bb{
            \begin{aligned}[c]
                W^{I_1J_1}
                {\cdots}
                W^{I_{\ell+1}J_{\ell+1}}
                \\[-2.5pt]
                -
                W^{J_1I_1}
                {\cdots}
                W^{J_{\ell+1}I_{\ell+1}}
            \end{aligned}
            }
			\mem
            \BB{
            \partial_{J_1}\act{
            {\cdots\mem}
                \partial_{J_{\ell+1}}\act{
                    g
                }
            }
            }
    \,,
\end{align}
which reduces to \eqref{spb.wick}
for $\P = \C^1$.
In the graphical notation,
\eqref{spbW} translates to
\begin{align}
    \label{graphical.spbW.fg}
    \frac{1}{i}\,
        \BB{
            \inc{valign=c}{WB1}
            -
            \inc{valign=c}{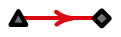}
        }
    + \frac{\hbar}{2!\hem i}\,
        \BB{
            \inc{valign=c}{WB2}
            -
            \inc{valign=c}{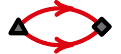}
        }
    + \frac{\hbar^2}{3!\hem i}\,
        \BB{
            \inc{valign=c}{WB3}
            -
            \inc{valign=c}{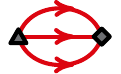}
        }
    + \O(\hbar^4)
    \,.
\end{align}
Physically, \eqref{graphical.spbW.fg}
visualizes
the pings and pongs of quantum excitations
exchanged between $f$ and $g$.

\newpage

As a bi-differential operator,
the deformed Poisson bracket
$\spb{\blank}{\blank}$
is represented as
\begin{align}
    \label{graphical.Wick}
    \ainc{valign=c}{spb}
    \,=\,
    \:\:
    &
    \frac{1}{i}\,
    \bbsq{\,
        \ainc{valign=c}{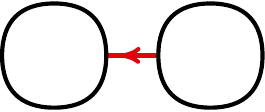}
        -
        \ainc{valign=c}{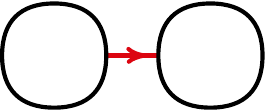}
    \,}
    \\
    &
    + \frac{\hbar}{2!\hem i}\,
    \bbsq{\,
        \ainc{valign=c}{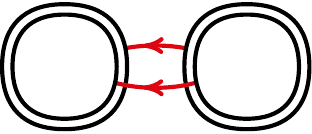}
        -
        \ainc{valign=c}{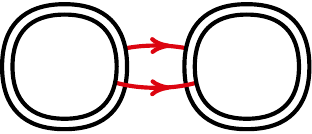}
    \,}
    \nonumber
    \\
    &
    + \frac{\hbar^2}{3!\hem i}\,
    \bbsq{\,
        \ainc{valign=c}{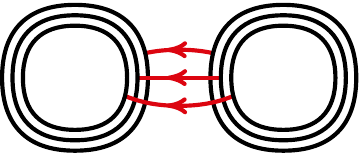}
        -
        \ainc{valign=c}{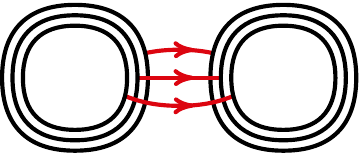}
    \,}
    \nonumber
    \\
    &
    + \O(\hbar^3)
    \,,
    \nonumber
\end{align}
where we have employed the notation
in \eqref{graphical.spbM}
again.

Surely,
the diagrammatic Leibniz rule in \eqref{graphical.leibniz}
applies to balloons attached to red lines.
The Magnus expansion for the quantum eikonal
is again given by \eqref{graphical.schi},
yet with the definition in \eqref{graphical.Wick}.

Lastly, we describe the transition to Feynman graphs.
As is explicated at length
for both particles and fields in \App{QFT},
the 
\emph{positive-frequency retarded propagator}
arises by
combining the Wightman tensor $W^{I_1I_2}$ 
(as Wightman function)
with the step function $\Theta(t_1,t_2)$.
The positive-frequency retarded propagator
is represented
as a red arrow:
\begin{subequations}
\label{F2}
\begin{align}
    \label{Fp2}
    \ainc{valign=c}{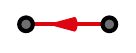}
    \,&=\,\,\,
        \int d^2t\,\,\,
            \partial_{I_1}\act{\tV(t_1)}\,
            \BB{\,
                W^{I_1I_2}\,
                \Theta(t_1,t_2)
            \,}
            \,\partial_{I_2}\act{\tV(t_2)}
    \,.
\end{align}
Similarly,
the 
\emph{negative-frequency retarded propagator}
arises by
combining the transposed Wightman tensor $W^{I_2I_1}$
with the step function $\Theta(t_1,t_2)$.
The negative-frequency retarded propagator
is represented
as a blue arrow:
\begin{align}
    \label{Fm2}
    \ainc{valign=c}{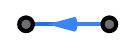}
    \,&=\,\,\,
        \int d^2t\,\,\,
            \partial_{I_1}\act{\tV(t_1)}\,
            \BB{\,
                W^{I_2I_1}\,
                \Theta(t_1,t_2)
            \,}
            \,\partial_{I_2}\act{\tV(t_2)}
    \,.
\end{align}
\end{subequations}
To clarify, these are exactly the positive- and negative-frequency parts of the retarded propagator 
defined in \Sec{FEYNMAN},
up to customary $\pm i$ factors:
\begin{align}
    \label{decomp}
    \inc{valign=c}{F2}
    \,=\,
    -i\,
        \inc{valign=c}{Fp2}
    +i\,
        \inc{valign=c}{Fm2}
    \,.
\end{align}
To be explicit, 
the retarded propagator is
\begin{align}
    \label{Pi2}
    \ainc{valign=c}{F2}
    \,&=\,\,\,
        \int d^2t\,\,\,
            \partial_{I_1}\act{\tV(t_1)}\,
            \BB{\,
                \Pi^{I_1I_2}\,
                \Theta(t_1,t_2)
            \,}
            \,\partial_{I_2}\act{\tV(t_2)}
    \,.
\end{align}
A crucial feature of the quantum eikonal
is that it cannot be represented solely in terms of the retarded propagator in \eqref{Pi2},
unlike as in the classical case.
Namely, the retarded propagator must resolve into more fundamental building blocks,
\eqrefs{Fp2}{Fm2}.

Note that the colors red and blue
indicate whether
the index flow in the Wightman tensor
aligns with the time flow in the step function.
In \eqref{Fp2}, they are aligned as
$I_1 \hhnem\leftarrow I_2$ and
$t_1 \nem\leftarrow t_2$.
In \eqref{Fp2}, they are anti-aligned as
$I_1 \nem\rightarrow I_2$ and
$t_1 \nem\leftarrow t_2$.
In fact,
this analysis reveals that
there lie two independent notions of ordering
in our graphical notation:
\textit{operator ordering} and \textit{time ordering}.
This point will be revisited several times.

\newpage


Finally, the counterpart of 
\eqref{feynman2} is
\begin{subequations}
\label{feynmanW2}
\begin{align}
    \label{feynmanW2p}
    \int_{t_1>t_2} d^2t\,\,\bbsq{
        \awrap{
            \inc{valign=c}{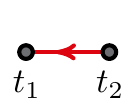}
        }
    }
    &\,\,\,=\,\,
    \int d^2t\,\,\bbsq{
        \awrap{
            \inc{valign=c}{W12}
            \:\Theta(t_1,t_2)
        }
    }
    \,\,\,=\,\,
        \inc{valign=c}{Fp2}
    \,,\\
    \label{feynmanW2m}
    \int_{t_1>t_2} d^2t\,\,\bbsq{
        \awrap{
            \inc{valign=c}{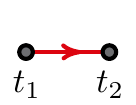}
        }
    }
    &\,\,\,=\,\,
    \int d^2t\,\,\bbsq{
        \awrap{
            \inc{valign=c}{W12r}
            \:\Theta(t_1,t_2)
        }
    }
    \,\,\,=\,\,
        \inc{valign=c}{Fm2}
    \,.
\end{align}
\end{subequations}
Here,
the arrows put on the tensor graphs
represent the propagating direction of quantum excitations:
the index flow of Wightman tensors.
On the other hand,
the arrows put on the Feynman graphs
represent the time direction:
the time flow stipulated by step functions.

\subsubsection{Evaluation}
\label{QEMAGNUS.WICK.EVAL}

We now perform the diagrammatic evaluation of the quantum eikonal in \eqref{graphical.schi}
by using the deformed Poisson bracket in \eqref{graphical.Wick}.

The computation is trivial
at orders $n=1$ and $n=2$.
In particular,
$\schib$ is found as
\vspace{-4pt}
\begin{align}
    \label{graphical.schi2W}
    \frac{1}{2i}\,
    \BB{
        \inc{valign=c}{Fp2}
        -
        \inc{valign=c}{Fm2}
    }
    \,+\, 
    \frac{\hbar}{4i}\,
    \BB{
        \inc{valign=c}{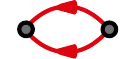}
        -
        \inc{valign=c}{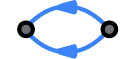}
    }
    + \O(\hbar^3)
    \,\,\,=\,
        \frac{1}{2} 
        \inc{valign=c}{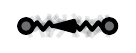}
    \,.
\end{align}
Again, fuzzification reproduces the entire answer at this order:
$\schib = \Fuz(\chib)$.
Hence it suffices to focus on the third-order quantum eikonal,
$\schic$.
The procedure is very much the same
as in \Sec{QEMAGNUS.MOYAL.EVAL}:
unpacking, bubble popping, and freeing.
Yet, the process can be streamlined
a bit more
by recalling the lessons learned in 
\Secs{QEMAGNUS.MOYAL.EVAL}{QMU.MOYAL}.

We begin with unpacking.
The counterpart of
\eqref{graphical.demo-a}
is given by
\begin{align}
\begin{split}
    \label{graphical.demo-w}
    \ainc{valign=c}{sbr123}
    =\,
    \,
    &
    \inc{valign=c}{br123}
    \\
    &
    \,+\,\,
    \frac{\hbar}{2\hem i^2}\left[{\,
    \begin{aligned}[c]
    &
        \inc{valign=c}{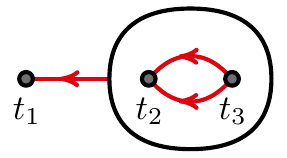}
        +
        \inc{valign=c}{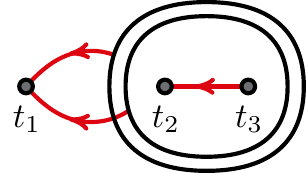}
        \\
    &
        -
        \inc{valign=c}{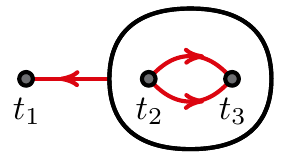}
        -
        \inc{valign=c}{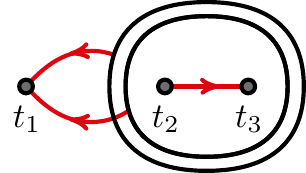}
        \\
    &
        + \cdots
    \end{aligned}
    \,\,\,\,}\right]
    \\
    &
    \,+\,\,
    \O(\hbar^2)
    \,,
\end{split}
\end{align}
where the one-loop part
describes a sum of
eight terms.

Next, we demonstrate the bubble popping
at one loop.
Due to the lessons in
\Secs{QEMAGNUS.MOYAL.EVAL}{QMU.MOYAL},
we expect that the result will be classified into three categories:
linear (L),
V-shaped (V),
and triangle (T).
Regarding this point,
the first and second columns
in the bracketed group of terms in \eqref{graphical.demo-w}
play different roles:
\begin{subequations}
\begin{align}
    \label{graphical.br12}
    \,\inc{valign=c}{br1p2p}
    \,-\, \inc{valign=c}{br1p2m}
    \,+\, \cdots
    &\quad\xrightarrow[\quad]{}\quad
    \text{
        L, V categories
    }
    \,,\\
    \label{graphical.br21}
    \inc{valign=c}{br2p1p} 
    \,
    \,-\, \inc{valign=c}{br2p1m}
    \mem
    \,+\, \cdots
    &\quad\xrightarrow[\quad]{}\quad
    \text{
        L, T categories
    }
    \,.
\end{align}
\end{subequations}
With the expectation that the T category of diagrams
will be only of a genuine interest,
we extract the triangle diagrams 
from \eqref{graphical.br21}.
The result is the sum
\begin{align}
    \label{graphical.Tdemo}
    + \inc{valign=c}{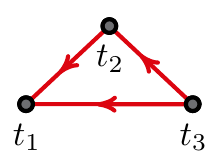}
    - \inc{valign=c}{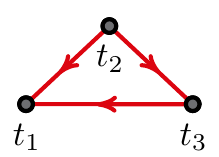}
    - \inc{valign=c}{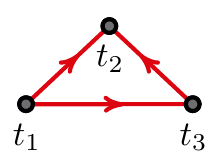}
    + \inc{valign=c}{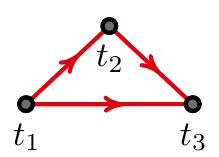}
    \,,
\end{align}
multiplied by an overall coefficient $2$.

Crucially, the graphs in
\eqref{graphical.Tdemo}
are all \textit{acyclic},
even though a loop is formed.
This is because
the very structure of 
the Wick star product (normal ordering)
strictly forbids 
exchanging quantum excitations in a cycle.
We shall remind ourselves that
our graphical calculus
have been computing operator products
in essence,
which describe one-dimensional arrays (strings) of objects.
For instance, the third triangle diagram 
in \eqref{graphical.Tdemo}
arises as
\begin{align}
    \label{vevdemo}
    \inc{valign=c}{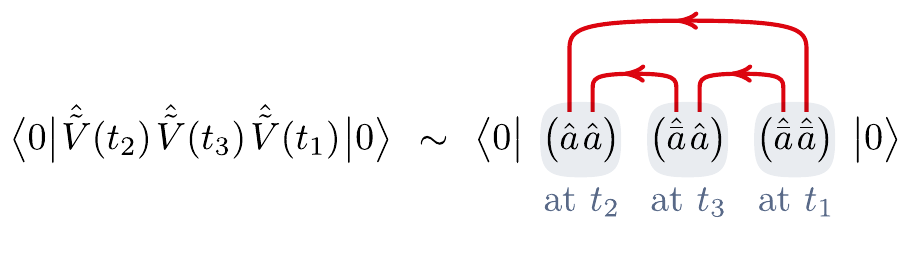}
    \,,
\end{align}
\vskip-16pt\noindent
which clearly shows that
a cycle cannot occur.
The red lines are
the Wick contractions;
the directions of arrows
are pulled back from the ordering for 
a linear array of operators.


With this remark made,
we find the image of \eqref{graphical.Tdemo}
under $t_1 \leftrightarrow t_3$ exchange.
Using planar isotopy, it boils down to
\begin{align}
    \label{graphical.Tdemo-mirror}
    + \inc{valign=c}{T11}
    - \inc{valign=c}{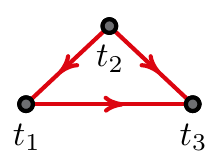}
    - \inc{valign=c}{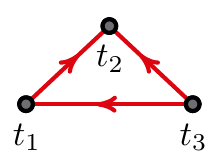}
    + \inc{valign=c}{T00}
    \,.
\end{align}
Eventually,
the sum of \eqrefs{graphical.Tdemo}{graphical.Tdemo-mirror}
derives that
the triangle contribution to
the intermediate form of $\schic$
at one loop
is given by 
the integration of the following
over the domain $t_1>t_2>t_3$.
Incorporating
the overall coefficient $1/6$ in \eqref{graphical.schi3},
we find
\begin{align}
\begin{split}
    \label{graphical.Tdemo-res}
    &
    \frac{\hbar}{6\hem i^2}\,\lrp{
    \begin{aligned}[c]
    &
        + 2 \inc{valign=c}{T00}
        + 2 \inc{valign=c}{T11}
    \\[-4pt]
    &
        - \inc{valign=c}{T010}
        - \inc{valign=c}{T100}
        - \inc{valign=c}{T011}
        - \inc{valign=c}{T101}
    \end{aligned}
    }
    \,.
\end{split}
\end{align}

Finally, we transition to Feynman graphs.
For triangles, there is no room for freeing
since the retarded propagators
will completely resolve the ordering between vertices.
Hence the triangle contribution to the final form of $\schic$ is found as
\vspace{-6pt}
\begin{align}
\begin{split}
    \label{graphical.Tdemo-feyn}
    &
    \frac{\hbar}{3\hem i^2}\,\bb{
        \inc{valign=c}{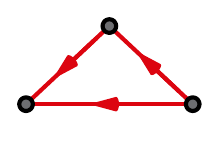}
        + \inc{valign=c}{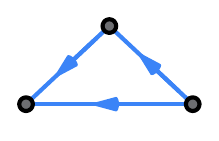}
    }
    \\[-4pt]
    &
    - \frac{\hbar}{6\hem i^2}\,\bb{
        \inc{valign=c}{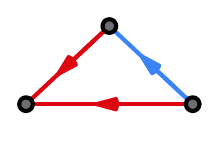}
        + \inc{valign=c}{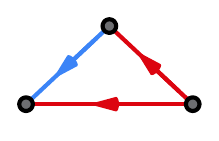}
        + \inc{valign=c}{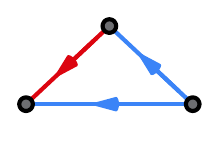}
        + \inc{valign=c}{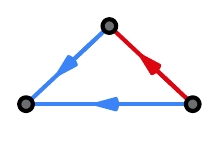}
    }
    \,,
\end{split}
\end{align}
\vskip-2pt\noindent
while
computation shows that
every L or V term
arises from the fuzzification
$\Fuz(\chic)$.

This concludes a demonstration for
the diagrammatic evaluation of the quantum eikonal in Wick quantization (normal ordering).
By working in the same fashion,
it can be found that
the final form of $\schic$ is given by
\vspace{-4pt}
\begin{align}
    \label{W3}
    &
    \lrp{
        \,\,
        \wrap{
            \frac{1}{3}
            \inc{valign=c}{fuzF3L}
        }
        +
        \wrap{
            \frac{1}{12}
            \inc{valign=c}{fuzF3V}
        }
        +
        \wrap{
            \frac{1}{12}
            \inc{valign=c}{fuzF3Vr}
        }
        \vphantom{\inc{valign=c}{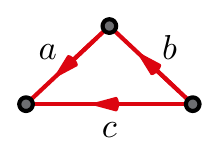}}
        \nem\nem\nem
    }
    \nonumber
    \\
    &
    \,\,\,{}
    + \,\sum_{a,b,c}\,\,
    \frac{\hbar^{a+b+c-2}}{3\hem i^2\mem a!\mem b!\mem c!}\,\lrp{
        \inc{valign=c}{Tabcp}
        + \inc{valign=c}{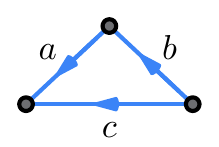}
    }
    \\[-2pt]
    &
    \,\,\,{}
    - \,\sum_{a,b,c}\,\,
    \frac{\hbar^{a+b+c-2}}{6\hem i^2\mem a!\mem b!\mem c!}\,\lrp{
        \inc{valign=c}{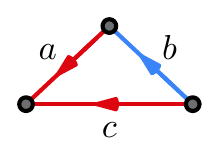}
        + \inc{valign=c}{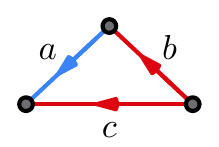}
        + \inc{valign=c}{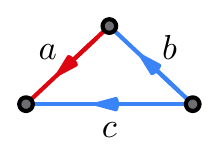}
        + \inc{valign=c}{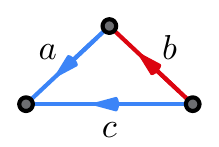}
    }
    \,,
    \nonumber
\end{align}
where $a,b,c \geq 1$ are positive integers
indicating the multiplicity of each edge.
Note that the trivial and nontrivial parts
due to fuzzification
are cleanly split.
We have explicitly verified \eqref{W3}
up to $\ell = 22$ loops
by numerical means.

\subsubsection{Magnus-Wick Coefficient}
\label{QMU.WICK}

In \Sec{CMU},
we defined
classical Magnus graphs
as connected directed acyclic graphs with zero loops.
In fact,
a classical Magnus graph
$\G \in \Mag(n,0)$
can be regarded as a pair $\G = (\X, \toT)$.
$\X$ is a connected unoriented graph with no loops,
while $\toT$ is an acyclic orientation structure on $\X$.
The orientation structure $\toT$
defines
a partial ordering 
between the vertices of $\G$,
which encodes time ordering.

In \Sec{QMU.MOYAL},
we defined
Magnus graphs of the first kind
as connected directed acyclic graphs.
In fact,
a Magnus graph of the first kind $\G \in \Mag(n,\ell)$
can be regarded as a pair $\G = (\X, \toT)$.
$\X$ is a connected unoriented graph,
while $\toT$ is an acyclic orientation structure on $\X$.
The orientation structure $\toT$
defines
a partial ordering 
between the vertices of $\G$,
which encodes time ordering.

Here, let us define that
a \emph{Magnus graph of the second kind}
is a pair $\G = {(\X,\toT,\toO)}$.
$\X$ is a connected unoriented graph,
while $\toT$ and $\toO$
are two independent acyclic orientation structures on $\X$.
Physically,
$\toT$
encodes the \emph{time ordering}
whereas 
$\toO$
encodes the \emph{operator ordering}.
The set of Magnus graphs of the second kind
with $n$ vertices and $\ell$ loops
will be denoted as $\Mag^2(n,\ell)$.

The presence of an additional orientation structure $\toO$
in Magnus graphs of the second kind
traces back to the fact that
Wick quantization (normal ordering)
has demanded polarization
as an additional geometric structure
on the phase space.
Moyal quantization (symmetric ordering),
in contrast,
stipulates no such structure.
Physically,
Moyal quantization employs
a symmetric propagator between quantum fluctuations,
sensitive only to time ordering.
For Wick quantization, however,
one additionally needs to specify 
which end is $a$ (annihilation)
and which end is $\ba$ (creation)
when contracting quantum fluctuations.

To understand why and how
our notation for Feynman graphs
in \Secs{PENROSE-WICK}{QEMAGNUS.WICK.EVAL}
has described 
Magnus graphs of the second kind,
recall the remark in \Sec{PENROSE-WICK} that
the index flow due to the Wightman tensor $W^{IJ}$
and the time ordering due to the step functions in the retarded propagators
define separate
acyclic orientations.
Recall also \eqref{vevdemo}
to understand how the index flow of the Wightman tensor
describes
the propagating direction of quantum fluctuations.

\begin{figure}[t]
    \centering
    \begin{align*}
        \ainc{valign=c}{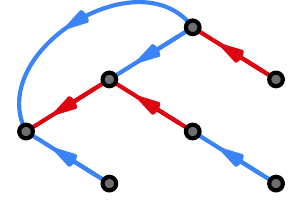}
        \nem
        &\quad\xleftrightarrow[\quad]{}\quad
        \bb{
            \ainc{valign=c}{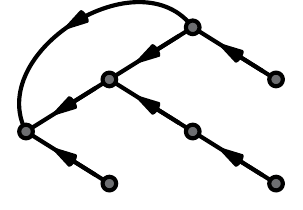}
        ,\,
            \ainc{valign=c}{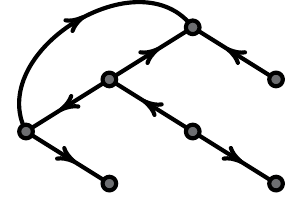}
        }
        \,,\\
        \ainc{valign=c}{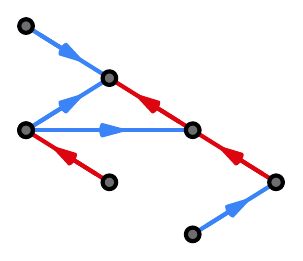}
        \nem
        &\quad\xleftrightarrow[\quad]{}\quad
        \bb{
            \ainc{valign=c}{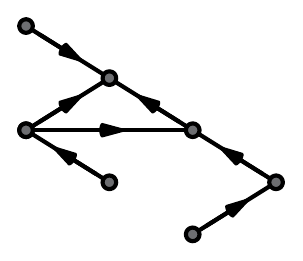}
        ,\,
            \ainc{valign=c}{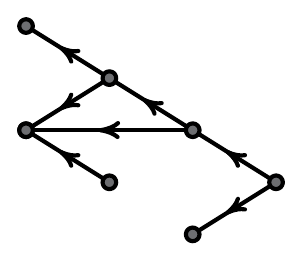}
        }
        \,.
    \end{align*}
    \caption{
        The bichrome notation
        for doubly directed graphs
        stands as
        a technique of visualization
        that aims to
        efficiently display two orientation structures 
        within a single diagram.
        In this figure,
        the same $\G \in \Mag^2(7,1)$
        is drawn,
        while emphasizing
        the time ordering $\toT$ 
        (first row)
        or the operator ordering $\toO$
        (second row).
    }
    \label{breakdown}
\end{figure}

\begin{figure}[t]
    \centering
    \begin{align*}
        \ainc{valign=c}{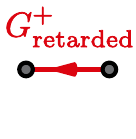}
        \,=\,\,\,
        \int d^2t\,\,\,
            \partial_{I_1}\act{\tV(t_1)}\,
            \,
            \bb{
                \ainc{valign=c}{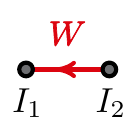}
            \times
                \ainc{valign=c}{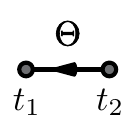}
            }\,
            \,\partial_{I_2}\act{\tV(t_2)}
        &\,\,\,\,=\,\,\mem
        \text{``\,\inc{valign=c}{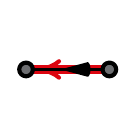}\mem''}
        \\
        \ainc{valign=c}{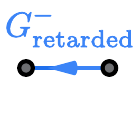}
        \,=\,\,\,
        \int d^2t\,\,\,
            \partial_{I_1}\act{\tV(t_1)}\,
            \,
            \bb{
                \ainc{valign=c}{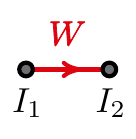}
            \times
                \ainc{valign=c}{biTh}
            }\,
            \,\partial_{I_2}\act{\tV(t_2)}
        &\,\,\,\,=\,\,\mem
        \text{``\,\inc{valign=c}{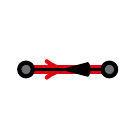}\mem''}
    \end{align*}
    \caption{
        A graphical representation of \eqrefs{Fp2}{Fm2}.
        Ideally, we would have drawn 
        two arrows per each edge
        to visualize 
        a doubly directed graph.
    }
    \label{breakdownG}
\end{figure}

To be explicit about these points,
we have provided an example of the breakdown
$\G = (\X, \toT, \toO)$
in \fref{breakdown}.
Also,
in \fref{breakdownG},
we have shown the explicit equivalence between
the bichrome notation we have been employing
and the implementation of Magnus graphs of the second kind
as doubly directed acyclic graphs.

The final form of the 
$n$\textsuperscript{th}-order quantum eikonal $\schin$
in Wick quantization
describes
a weighted sum
of Magnus graphs of the second kind:
\begin{align}
    \schin
    \,\qpq\,    
        \sum_{\ell=0}^\infty\,
        \sum_{
            \G \mem\in\mem \Mag^2(n,\ell)
        }\,
            \frac{\hbar^\ell}{i^{n-1}}\mem
            \m_2(\G)\,\,
                \G
    \,.
\end{align}
In this way, each $\G \in \Mag^2(n,\ell)$ 
is assigned with a unique coefficient
$\m_2(\G)$,
which defines the \emph{Magnus-Wick coefficient} of $\G$.

Again, it is desirable to define
\emph{reduced Magnus-Wick coefficient}:
\begin{align}
    \label{omega2}
    \omega_2(\G)
    \,=\,
        \mu_2(\G)\, \s(\G)
    \,.
\end{align}
Of course, the symmetry factor $\s(\G)$ concerns both $\toT$ and $\toO$.
\eqref{W3} boils down to
\begin{subequations}
\label{W3!}
\begin{align}
    \label{W3L!}
    &
    \omega_2\mem\BB{
        \inc{valign=c}{MLab}
    }
    \mem\,=\,\mem
    \omega\mem\BB{
        \inc{valign=c}{F3L}
    }\,
    \cdot\,
        \left\{
        \begin{aligned}[c]
        \,\,\,
            +1
            &\quad
            \text{if monochrome
            }
            \\
        \,\,\,
            -1
            &\quad
            \text{otherwise}
        \end{aligned}
        \right\}
    \,,\\
    &
    \label{W3V!}
    \omega_2\mem\BB{
        \inc{valign=c}{MVabr}
    }
    \mem\,=\,\mem
    \omega_2\mem\BB{
        \inc{valign=c}{MVab}
    }
    \mem\,=\,\mem
    \omega\mem\BB{
        \inc{valign=c}{F3V}
    }\,
    \cdot\,
        \left\{
        \begin{aligned}[c]
        \,\,\,
            +1
            &\quad
            \text{if monochrome
            }
            \\
        \,\,\,
            -1
            &\quad
            \text{otherwise}
        \end{aligned}
        \right\}
    \,,\\
    \label{W3T!}
    &
    \omega_2\mem\BB{
        \inc{valign=c}{MTabc}
    }
    \mem\,=\,\mem
        \left\{
        \begin{aligned}[c]
        \,\,\,
            1/3
            &\quad
            \text{if monochrome
            }
            \\
        \,\,\,
            -1/6
            &\quad
            \text{otherwise
            }
        \end{aligned}
        \right\}
    \,,
\end{align}
\end{subequations}
where $a,b,c$ are edge multiplicities.
To clarify, 
the arguments of $\omega_2$ in \eqref{W3!}
are meant to be 
colored with red and blue.
That is, we have stipulated only the time ordering $\toT$
on the left-hand side.



It should be clear that
$\G = (\X,\toT,\toO) \in \Mag^2(n,\ell)$
is \emph{monochrome} iff $\toT = \pm\, \toO$.
Specifically,
$\G$
is \emph{monochrome red} iff $\toT = +\, \toO$
and 
\emph{monochrome blue} iff $\toT = -\, \toO$.

It should be also clear that
a multiplied edge
in any Magnus graph
must have the same orientation structure
due to the acyclicity condition.
In particular, the same color must be shared
to qualify as a Magnus graph of the second kind:
\begin{align}
    \inc{valign=c}{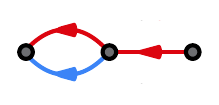}
    \,\,\notin\,\,
        \Mag^2(3,1)
    \,,\quad
    \inc{valign=c}{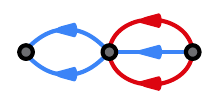}
    \,\,\notin\,\,
        \Mag^2(3,3)
    \,,\quad
    \cdots
    \,.
\end{align}
On a related note,
only six coloring schemes exist for
the triangle diagrams in \eqref{W3T!}.
The forbidden cases are
\begin{align}
    \ainc{valign=c}{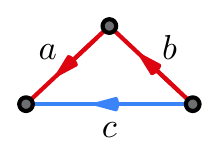}
    \,,\quad
    \ainc{valign=c}{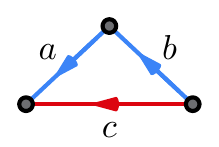}
    \,.
\end{align}
The allowed cases are shown in \eqref{W3}.

\newpage

\eqref{W3!}
provides a compact summary of the Magnus-Wick coefficients
at three vertices and all loops.
Again, the equalities in \eqrefs{W3L!}{W3V!}
reflect the fact that
a part of $\schic$
simply recycles the classical eikonal $\chic$
via fuzzification
as stated in \eqref{W3}.

For a precise mathematical formulation
of this observation,
let $\Prim^2(n,\ell)$
be the subset of $\Mag^2(n,\ell)$
whose elements are simple.
Elements of $\Prim^2(n,\ell)$ are called \emph{primaries of} \emph{the second kind}.
Accordingly,
$\G \in \Mag^2(n,\ell)$
is a \emph{descendant} of
$\tilde{\G}_0 \in \Prim^2(n,\ell_0)$
if $\G$ arises by giving an edge multiplicity to $\tilde{\G}_0$.
The set of descendants of $\tilde{\G}_0 \in \Mag^2(n,\ell_0)$
is denoted as $\Desc(\tilde{\G}_0)$.
For example,
\begin{align}
    \label{cescin}
    \ainc{valign=c}{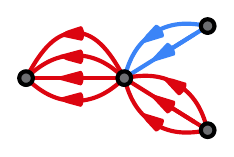}
    \quad\in\quad
    \Desc\bb{
        \ainc{valign=c}{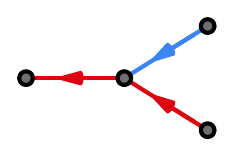}
    }
    \,
    \,.
\end{align}
An element of $\Mag(n,\ell)$
is either a descendant or a primary.

Let $\G_0 = (\X,\toT) \in \Mag(n,0)$
be a classical Magnus graph.
Then $\tilde{\G}_0 = {(\X,\toT,\toO)} \in \Mag^2(n,0)$
is said to be
a \emph{bicoloring} of $\G_0$
for an acyclic orientation structure $\toO$ on $\X$.
Specifically,
let $\gamma : \Mag(n,0) \to \Mag^2(n,0)$ be the bicoloring map
such that $\tilde{\G}_0 = \gamma(\G_0)$.
The \emph{red number} of $\gamma$
is 
the number of edges 
on which $\toT = +\, \toO$
is denoted as $n_+(\gamma)$.
The \emph{blue number} of $\gamma$
is
the number of edges 
on which $\toT = -\, \toO$
is denoted as $n_-(\gamma)$.
These are the number of reds and blues which the bicoloring $\gamma$ chooses.

Any primary of the second kind
$\tilde{\G}_0 \in \Prim^2(n,\ell)$
is a bicoloring of 
a primary of the first kind
$\G_0 \in \Prim(n,\ell)$.
A unique bicoloring map $\gamma$ exists such that
$\tilde{\G}_0 = \gamma(\G_0)$.

Finally,
$\G \in \Mag^2(n,\ell)$
is \emph{trivial}
if it is a descendant of
a bicoloring of a classical Magnus graph:
$\G \in \Desc(\gamma(\G_0))$
for $\G_0 \in \Mag(n,0)$.
$\G \in \Mag^2(n,\ell)$ is \emph{nontrivial} if it is not trivial.

In the context of Wick quantization,
fuzzification $\Fuz$
is a linear map
that outputs a weighted sum of 
trivial Magnus graphs of the second kind.
That is, a classical Magnus graph
$\G_0 \in \Mag(n,0)$
is mapped to
a weighted sum of 
elements from
$\Desc(\gamma(\G_0))$.

With these definitions,
it holds at any $n$ that
\begin{align}
    \label{WE.primary}
    \schin
    \,-\,
    \Fuz(\chin)
    \,\qpq\,
        \sum_{\ell=1}^\infty\,
            \sum_{
            \substack{
                \G \mem\in\mem \Mag(n,\ell)
                \\
                \text{nontrivial}
            }
            }
                \frac{\hbar^\ell}{i^{n-1}} \cdot
                \frac{\omega_2(\G)}{\s(\G)}\,\,
                    \G
    \,,
\end{align}
%
which implies
\begin{align}
    \text{$\tilde{\G}_0 = \gamma(\G_0)$ for $\G_0 \in \Mag(n,0)$}
    &\qiq
    \omega_2(\tilde{\G}_0) \,=\, (-1)^{n_-(\gamma)}\mem \omega(\G_0)
    \,.
\end{align}
Furthermore, 
our explorations up to the order $n=3$
has confirmed that
\begin{align}
    \text{$\G \in \Desc(\tilde{\G}_0)$}
    &\qiq
    \omega_2(\G) \,=\, \omega_2(\tilde{\G}_0)
    \,,
\end{align}
where
$\G \in \Mag^2(n,\ell)$ 
and $\tilde{\G}_0 \in \Prim^2(n,\ell_0)$.

It might suffice to 
determine the coefficients of 
nontrivial primaries,
the insight of which may provide some guidance
when exploring
higher orders.

\newpage

\section{Summary and Outlook}

In this paper,
we established
classical and quantum scattering theory
in the phase space formulation.
The classical and quantum scattering processes
are geometrically interpreted as
symplectomorphisms
and fuzzy diffeomorphisms
mapping the initial phase space to the final phase space.
The logarithm of these maps via Magnus expansion
defines
the classical and quantum eikonals
as phase-space functions,
serving as generators of scattering.
Consequently, exact relations are established between
\begin{align}
\begin{array}{llll}
    \hat{S}
    &\,\,
    \text{
        S-matrix 
    }\,,
    &\quad
    \hat{\chi}
    &\,\,
    \text{
        Eikonal Matrix 
    }
    \,,\\
    S^\star
    &\,\,
    \text{
        Fuzzy S-diffeomorphism 
    }\,,
    &\quad
    \chi^\star
    &\,\,
    \text{
        Quantum Eikonal 
    }
    \,,\\
    S
    &\,\,
    \text{
        S-symplectomorphism
    }\,,
    &\quad
    \chi
    &\,\,
    \text{
        Classical Eikonal
    }
    \,.
\end{array}
\end{align}
The first row yields the second row
through adjoint actions
while using
quantization map as an intertwiner.
The second row yields the third row
through the literal $\hbar \to 0$ limit.
The first and second columns are in
exponential/log relationships
up to necessary refinements.
The details are reviewed in \fref{S-summary}.

\begin{figure}
    \centering
    \vphantom{.}\kern0.71em
	\adjustbox{valign=c}{\begin{tikzpicture}
	    \node[empty] (O) at (0,0) {};
	    \node[empty] (X) at (5.0, 0) {};
	    \node[empty] (Y) at (0, -1.58) {};
	    \node[b] (z1) at ($(O)$) {\smash{$
        \mathllap{
            \hat{S}
        }
        \mathrlap{\color{domaincolor}{}
            \,\in\, \Aut(\Hilb)
        }
        $}};
	    \node[b] (z2) at ($(O)+(Y)$) {\smash{$
        \mathllap{
            \Q^{-1}(\hat{S})
        }
        \mathrlap{\color{domaincolor}{}
            \,\in\, \A_\star\formal{\hbar}
        }
        $}\vphantom{\Big|}};
	    \node[b] (z3) at ($(O)+2*(Y)$) {\smash{
            \color{lgray}{}
            (dead end)
        }\vphantom{\Big|}};
	    \node[b] (a1) at ($(O)+(X)$) {\smash{$
        \mathllap{
            \Ad_{\hat{S}}
        }
        \mathrlap{\color{domaincolor}{}
            \,\in\, \Aut(\Ops(\Hilb))
        }
        $}};
	    \node[b] (a2) at ($(O)+(X)+(Y)$) {\smash{$
        \mathllap{
            \sS
        }
        \mathrlap{\color{domaincolor}{}
            \,\in\, \Aut(\A_\star\formal{\hbar})
        }
        $}\vphantom{\Big|}};
	    \node[b] (a3) at ($(O)+(X)+2*(Y)$) {\smash{$
        \mathllap{
            S^*
        }
        \mathrlap{\color{domaincolor}{}
            \,\in\, \Aut(\A_{\pb{\,\,}{\,\,}})
        }
        $}\vphantom{\Big|}};
	    \node[b] (b1) at ($(O)+2*(X)$) {\smash{$
        \mathllap{
            \ad_{\hat{\chi}/i\hbar}
        }
        \mathrlap{\color{domaincolor}{}
            \,\in\, \Der(\Ops(\Hilb))
        }
        $}};
	    \node[b] (b2) at ($(O)+2*(X)+(Y)$) {\smash{$
        \mathllap{
            \sX_\schi
        }
        \mathrlap{\color{domaincolor}{}
            \,\in\, \Der(\A_\star\formal{\hbar})
        }
        $}\vphantom{\Big|}};
	    \node[b] (b3) at ($(O)+2*(X)+2*(Y)$) {\smash{$
        \mathllap{
            X_\chi
        }
        \mathrlap{\color{domaincolor}{}{}
            \,\in\, \Der(\A_{\pb{\,\,}{\,\,}})
        }
        $}\vphantom{\Big|}};
	    \draw[->,columnactioncolor] (z1)--(z2) node[midway,right] {\small\color{columnactioncolor}
            $\Q^{-1}$
        };
	    \draw[->,columnactioncolor] (a1)--(a2) node[midway,right] {\small\color{columnactioncolor}
            Intertwine by
            $\Q^{-1}$
        };
	    \draw[->,columnactioncolor] (b1)--(b2) node[midway,right] {\small\color{columnactioncolor}
            Intertwine by
            $\Q^{-1}$
        };
	    \draw[->,columnactioncolor] (z2)--(z3) node[midway,right] {\small\color{columnactioncolor}
            $\hbar{\hem\mem\to\mem\hhem}0$
        \vphantom{\Big|}};
	    \draw[->,columnactioncolor] (a2)--(a3) node[midway,right] {\small\color{columnactioncolor}
            $\hbar{\hem\mem\to\mem\hhem}0$
        \vphantom{\Big|}};
	    \draw[->,columnactioncolor] (b2)--(b3) node[midway,right] {\small\color{columnactioncolor}
            $\hbar{\hem\mem\to\mem\hhem}0$
        \vphantom{\Big|}};
	    \node[b] (x1) at ($(O)+0.615*(X)-0.035*(Y)$) {\smash{\color{rowactioncolor}$
            \xrightarrow[\:\:\:\qquad\mathclap{\text{\small
                action
            }}\qquad\,]{\:\:\:\qquad\mathclap{\text{\small
                adjoint
            }}\qquad\,}
        $}};
	    \node[b] (x2) at ($(O)+0.615*(X)-0.035*(Y)+1.08*(Y)$) {\smash{\color{rowactioncolor}$
            \xrightarrow[\:\:\:\qquad\mathclap{\text{\small
                via $\star$
            }}\qquad\,]{\:\:\:\qquad\mathclap{\text{\small
                conjugation
            }}\qquad\,}
        $}};
	    \node[b] (y1) at ($(O)+1.68*(X)-0.08*(Y)$) {\smash{\color{rowactioncolor}$
            \accentset{\,\,\text{\small $\log$}\,}{-{\kern-0.65em}-{\kern-0.65em}\rightharpoonup}
        $}};
	    \node[b] (y1') at ($(O)+1.68*(X)$) {\smash{\color{rowactioncolor}$
            \underaccent{\,\,\,\,\,\text{\small $\exp$}}{\leftharpoondown{\kern-0.65em}-{\kern-0.65em}-}
        $}};
	    \node[b] (y1) at ($(O)+1.68*(X)-0.08*(Y)+1.08*(Y)$) {\smash{\color{rowactioncolor}$
            \accentset{\,\,\text{\small $\log$}\,}{-{\kern-0.65em}-{\kern-0.65em}\rightharpoonup}
        $}};
	    \node[b] (y1') at ($(O)+1.68*(X)+1.08*(Y)$) {\smash{\color{rowactioncolor}$
            \underaccent{\,\,\,\,\,\text{\small $\exp$}}{\leftharpoondown{\kern-0.65em}-{\kern-0.65em}-}
        $}};
	    \node[b] (y1) at ($(O)+1.68*(X)-0.08*(Y)+2.08*(Y)$) {\smash{\color{rowactioncolor}$
            \accentset{\,\,\text{\small $\log$}\,}{-{\kern-0.65em}-{\kern-0.65em}\rightharpoonup}
        $}};
	    \node[b] (y1') at ($(O)+1.68*(X)+2.08*(Y)$) {\smash{\color{rowactioncolor}$
            \underaccent{\,\,\,\,\,\text{\small $\exp$}}{\leftharpoondown{\kern-0.65em}-{\kern-0.65em}-}
        $}};
	\end{tikzpicture}}
    \kern0.25em
    \vphantom{.}
    \caption{
        Summary of exact relations
        in classical and quantum scattering theory.
        Here, we pursue 
        mathematical precision.
        Classical observables in the phase space formulation
        are identified as 
        elements of the Poisson algebra $\A_{\pb{\,\,}{\,\,}}$.
        Quantum observables in the phase space formulation
        are identified as
        formal power series in $\hbar$
        whose coefficients are elements of the noncommutative ring $\A_\star$.
        Also, $\Aut$ and $\Der$
        are notations for
        the sets of automorphisms and derivations,
        respectively.
    }
    \label{S-summary}
\end{figure}

The power of the phase space formulation
shines at both conceptual and practical realms.
Conceptually,
the phase space formulation
facilitates 
the very rigorous comparison between 
classical and quantum scattering theories
by formulating quantum-mechanical operators 
in the same ``data type''
as classical observables:
phase-space functions.
Moreover,
the phase space formulation
also manifests the well-definedness of the classical limit
in terms of deformed Poisson bracket and Hamiltonian vector fields.

Practically,
the phase space formulation
straightforwardly establishes the fact that
the quantum eikonal
arises by simply deforming each Poisson bracket
in the Magnus formula for the classical eikonal.
This leads to a systematic and principled approach to computing the quantum eikonal
to arbitrary number of loops
at any order,
as is concretely demonstrated in detail in \Sec{MAGNUS}.
The generality of this formalism
ensures that
both
symmetric and normal orderings
are handled within the same framework,
providing a complete operational definition of
Magnus coefficients
for both Moyal and Wick quantizations.

Specifically,
it is shown that 
the Magnus expansion
in deformation quantization
defines
two new graph functions
$\omega_1(\G)$ and $\omega_2(\G)$
on singly and doubly directed acyclic graphs.
For concreteness,
a detailed analysis is given for graphs with three vertices.
The relevant graphs split into
linear, V-shaped, and triangle categories
based on the idea of descendants and primaries,
the last of which being only nontrivial.
The triangle category describes
numbers $1/3$ or $\pm 1/6$,
depending on
the parity of edge multiplicities
(for Moyal quantization)
or 
the sign flip associated with the transposed Wightman tensor
(for Wick quantization).

In a mathematician's perspective,
this work connects between various branches of mathematics
from
symplectic geometry and
noncommutative algebra
to
combinatorics and graph theory.
In a physicist's perspective,
this work promotes and formulates
intuitive analogies into precise equalities,
clarifies good and bad pathways for the classical limit,
and derives explicit results 
about
the systematic evaluation of the quantum eikonal.

We end with a few remarks
on generalizations and future directions.

First of all,
it should be clear that
the results of this paper
applies to a general class of systems.
In the main parts of this paper,
we have presumed the following conditions
for the sake of concreteness in our exposition:
\begin{enumerate}[label=(\alph*)]
    \item 
        The classical phase space is a symplectic manifold.
    \item 
        The quantum system is realized concretely
        on a Hilbert space
        so that the phase space formulation
        arises by a well-behaved quantization map.
    \item
        The phase space is a real plane or a complex plane.
    \item   
        The star product is associative.
\end{enumerate}
However, these premises can all be relaxed
except (d).
The minimal requirement is
a fuzzy phase space
with an associative star product
whose classical limit yields any Poisson manifold.
The definitions of $S,\sS,\chi,\schi$
all survive,
as well as the Dyson/Magnus formulae.

\begin{enumerate}[label=(\alph*)]
    \item 
In Poisson manifolds,
$S \in \Diff(\P,{\pb{\,\,\,}{\,\,}})$ is the Poisson diffeomorphism
from the initial phase space to the final phase space.
The classical eikonal $\chi \in \A_{\pb{\,\,}{\,\,}}$
is defined by $S^* = \exp(X_\chi)$.
The Liouville measure is lost,
but the Hamiltonian equations of motion
or the Liouville equation
can still be posited and studied.
Physically speaking,
one can define and compute
$S$, $\sS$, $\chi$, $\schi$
without requiring an action principle
\cite{Kim:2025sey}.

\item
In formal (not necessarily strict) deformation quantization,
$\sS \in \Aut(\A_\star\formal{\hbar})$ is the formal fuzzy diffeomorphism
from the initial fuzzy phase space
to the final fuzzy phase space.
The quantum eikonal $\schi \in \A_\star\formal{\hbar}$
is 
a formal power series in $\hbar$,
defined by $\sS = \exp(\sX_\schi)$.
Here, one loses 
the convergence of power series in $\hbar$
or the Hilbert space $\Hilb$.
Physically speaking,
one can define and compute
$S$, $\sS$, $\chi$, $\schi$
for
systems whose
perturbation theory
yields only asymptotic series
in the loop expansion.

\item
The phase can be nonlinear, compact, or curved.
In particular,
it is well-established that
the phase space formulation for spin
takes the two-sphere
$\P = \mathrm{S}^2$
as the phase space
via an explicit, well-behaved quantizer $\Q$
\cite{varilly1989moyal}.
The compactness of the phase space 
corresponds to the finite dimension $2j+1$ of the spin-$2j$ Hilbert space.

\end{enumerate}

Regarding (d),
our framework has stipulated
Jacobi identity and the associativity of star product
as crucial axioms.
Relaxing them will lead to rather radical generalizations
but still could be of a physical interest,
regarding
magnetic monopoles
\cite{Szabo:2017yxd,Bakas:2013jwa}.

Eventually,
the above generalizations at the geometry level
will boil down to
generalizations at the combinatorics and graph theory level.
For example,
suppose one quantizes 
a generic Poisson manifold $\P$
by the Kontsevich quantization formula \cite{kontsevich}:
non-constant $\Pi^{IJ}$.
Then the Magnus expansions
describe weighted sum of Kontsevich graphs
multiplied by factors of step functions,
which will define \emph{Magnus-Kontsevich} coefficients.
What are the Magnus-Kontsevich coefficients at, say,
three vertices and one loop?
What is the all-orders formula?
Do the ideas about descendants and primaries generalize?
The linear Poisson manifold $\P = T^*\mathfrak{g}$
for a Lie algebra $\mathfrak{g}$
would provide a nice cubic specialization.
Physically, these explorations might connect with
the perturbation theory of 
open strings \cite{cattaneo-felder}.

The recursion relations and Hopf-algebraic formulations
for the quantum eikonal
remain to be an open problem,
although those are the natural future avenues
as per \rcite{eikonalsangmin1}.


Another open avenue
is Magnus coefficients for 
symplectic perturbations
\cite{wlf-ps,eikonalsangmin2},
relevant to magnetic couplings.
This means to implement interactions by modifying the symplectic structure
instead of the Hamiltonian
\cite{souriau1970structure,dyson1990feynman},
which induces several subtleties.
The believed formula for the classical eikonal
may need to be proven,
while partial progress has been made by \rcite{eikonalsangmin2}.


It may be interesting to investigate
global properties of S-symplectomorphisms.
Strictly speaking,
our formulae for the eikonals
have worked within local patches.

The treatment of field-theoretical S-matrices in this paper
is based on second quantization,
as is elaborated in \App{QFT}.
It will be interesting to investigate if a first-quantized framework is viable,
in which case the map $S$ in general
may describe a morphism in the symplectic category:
``S-morphism.''

\bigskip\noindent
\textbf{Acknowledgements}.|
The author is grateful to
Li Guo, Jung-Wook Kim, Sungsoo Kim, Jianrong Li, and Toby Saunders-A'Court
for discussions,
from which he learned many insights
about Magnus expansion.
J.-H.K. is supported by the Department of Energy (Grant No.~DE-SC0011632) and by the Walter Burke Institute for Theoretical Physics.

\newpage

\newpage
\appendix


\section{Appendices}

\subsection{More on Interaction Pictures}
\label{INTCD}

In \Secs{CIP}{QIP},
we defined the classical and quantum interaction picture images as
\begin{align}
    \label{CIP,QIP}
    \tilde{f}(t)
    \,=\,
        \Uo(t_0,t)^*\act{
            f(t)
        }
    \,,\quad
    \tilde{f}^\star\hnem(t)
    \,=\,
        \sUo(t_0,t)\act{
            f(t)
        }
    \,.
\end{align}
A subtle point worth clarifying
is that they differ in general,
as explicit examples show:
\begin{align}
    \label{intx}
    \sUo(t_0,t)
    \,\neq\,
    \Uo(t_0,t)^*
    \,\,\qiq\,\,
    \exists\mem f(t)
    \quad\text{s.t.}\quad
    \tilde{f}^\star\hnem(t)
    \,\neq\,
    \tilde{f}(t)
    \,.
\end{align}
However, it still holds that
the latter
retrieves the former
in the classical limit:
\begin{align}
    \label{intlimit.app}
    \lim_{\hbar\to0}
        \sUo(t_1,t_2)
    \,=\,
        \Uo(t_1,t_2)^*
    \qiq
    \lim_{\hbar\to0}
        \tilde{f}^\star\hnem(t)
    \,=\,
        \tilde{f}(t)
    \,\,\,
    \forall\hem f(t)
    \,.
\end{align}
This can be shown by using the Magnus \cite{magnus1954exponential} series
for $\sUo(t_1,t_2)$
and $\Uo(t_1,t_2)^*$.
Note that \eqref{intlimit.app}
is a concretely formulated equation about differential operators.

A summary can be given in terms of the following (non)commutative diagram.
\begin{align}
	\label{eq:cd-intx}
	\adjustbox{valign=c}{\begin{tikzpicture}
	    \node[empty] (O) at (0,0) {};
	    \node[empty] (X) at (6.0, 0) {};
	    \node[empty] (x) at (2.1, 0) {};
	    \node[empty] (Y) at (0, -2.4) {};
	    \node[b] (a) at ($(O)$) {$\hat{f}(t)$};
	    \node[b] (b) at ($(O)+(X)$) {$f(t)$};
	    \node[b] (A) at ($(O)+(Y)$) {$\smash{\tilde{\hat{f}}(t)}\vphantom{\tilde{f}}$};
	    \node[b] (B) at ($(O)+(X)+(Y)-(x)$) {$\tilde{f}^\star(t)$};
	    \node[b] (B') at ($(O)+(X)+(Y)$) {$\tilde{f}(t)$};
	    \draw[<-] (a)--(b) node[midway,above] {$\Q$};
	    \draw[<-] (A)--(B) node[midway,above] {$\Q$};
	    \draw[->] (a)--(A) node[midway,left] 
        {$
            \Ad_{\hat{U}^\circ\hnem(t_0,t)}\,
        $};
	    \draw[->] (b)--(B') node[midway,right] {$
            \,\Uo(t_0,t)^*
        $};
        \draw[->,color=emphcolor] (B)--(B') node[midway,above] {\color{emphcolor}
            $\hbar {\to} 0$
        };
	    \draw[->] (b)--(B) node[midway,left] {$
            \sUo(t_0,t)\:\mem
        $};
	\end{tikzpicture}}
\end{align}
Namely, 
quantization does not commute with interaction picture,
although classical limit closes the square:
\begin{align}
    \Q 
    \,\circ\,
    \nem
    \BB{\text{
        Quantum Interaction Picture
    }}
    \:\neq\:
        \BB{\text{
            Classical Interaction Picture
        }}
        \nem
        \,\circ\,
        \Q
    \,.
\end{align}

Crucially, however,
physically relevant examples
tend to achieve 
the equality
$\tilde{f}^\star(t) = \tilde{f}(t)$
by prescribing a simple free Hamiltonian.
First,
suppose quantization in symmetric ordering:
\Sec{QMAP}.
A quadratic free Hamiltonian,
such as $p^2\nem/2m$,
keeps the free evolution linear
as $x \mapsto x + pt/m$, $p \mapsto p$,
in which case
the interaction picture preserves the symmetric ordering.
Second,
suppose quantization in normal ordering:
\Sec{NORMAL}.
The harmonic free Hamiltonian in \eqref{1d.H0}
keeps the free evolution holomorphic in the sense that no mixing between $a$ and $\ba$ develops:
$a \mapsto a\mem \mathe^{-i\omega t}$,
$\ba \mapsto \ba\mem \mathe^{+i\omega t}$
(that is, no Bogoliubov $\beta$ coefficient).
As a result, the interaction picture preserves the normal ordering.
%
More explicitly,
one can verify
$\sUo(t_0,t) = \Uo(t_0,t)^*$
directly by showing that
$\sX_{\ba a} = X_{\ba a}$.

The precise criterion for 
the equality
$\tilde{f}^\star(t) = \tilde{f}(t)$
is
the preservation of operator ordering
by free time evolution.
We may impose that
\textit{the free theory respects the quantization}
in this precise sense,
as an optional consistency condition
in the quantum theory of scattering
with a nicely behaving classical limit.




Despite this caveat,
the formulae in \eqrefs{sS.sol}{eikonal.qu}
is exact
since $\tilde{V}(t)$ in \Sec{S-DIFF}
denotes
$\sUo(t_0,t)\act{ V(t) }
$:
the symbol of
$\tilde{\hat{V}}(t)$
($\neq \hat{\tilde{V}}(t)$).
Generally speaking,
this develops $\hbar$ corrections from 
$\tilde{V}(t)$ in \Sec{S-SYMP},
which denotes
$\Uo(t_0,t)^*\act{ V(t) }$.
However,
one should feel free to
regard that we have presumed the consistency
between free theory and quantization.


\subsection{In Background Perturbation Theory}
\label{BFT}

This appendix provides
an optional analysis
on background field theory computations.
This is in consideration of 
the exploration given in \rcite{Haddad:2025cmw},
where the vacuum expectation value of the eikonal matrix $\hat{\chi}$
is compared with the classical eikonal.
In our framework,
such vacuum expectation values essentially compute the symbol of operators
via $\Q^{-1}$.
Thus,
provided a correct interpretation of notations,
our conclusions will be
\begin{align}
    \label{expval-chi}
    \schi
    \,&=\,
        \expval{\,\hat{\chi}\,}_\text{bkgd}
    \qiq
    \chi
    \,=\,
        \lim_{\hbar\to0}
        \expval{\,\hat{\chi}\,}_\text{bkgd}
    \,,
\end{align}
as well as
\begin{align}
    \label{expval-S}
    \exp^\star\nem\bb{
        \frac{1}{i\hbar}\mem \schi
    }
    \,=\,
        \expval{\,\hat{S}\,}_\text{bkgd}
    \,.
\end{align}

First of all, we review the fact that 
star products admit path integral derivations:
recall the brief comments given around \eqrefs{spb.wick}{spb.moyal}.
For the Moyal star product,
a worldline path integral derivation
is nicely detailed
in \rcite{deformation-quantification}
(see also \rcite{cattaneo-felder}).

For the Wick star product,
we may take the formula in \eqref{dequantizer.normal}
as the starting point.
Its right-hand side can be reinterpreted as a path integral
\begin{align}
    \label{PI.ab.1}
    \int_a^{\ba} \D{\l}\hem \D{\bl}\,\,
        \exp\nem\bb{\,
            \frac{1}{\hbar}\mem
            \bb{
                \ba\mem \bigbig{
                    \lambda(t_+) - a
                }
                - \int^{t_+}_{t_-} dt\,\,
                    \bl(t)\mem \dot{\l}(t)
            }
        }
        \,\,
        f(\l(t_0),\bl(t_0))
    \,,
\end{align}
the boundary conditions for which are
$\l(t_-) = a$
and
$\bl(t_+) = \ba$.
Here, $t_+$, $t_-$, and $t_0$
fixed arbitrary times
such that $t_+ > t_0 > t_-$.
This freedom is due to the fact that
the action in \eqref{PI.ab.1} defines a topological (reparametrization-invariant) theory
\cite{deformation-quantification}:
phase space action with zero Hamiltonian.

By expanding around a static background as
$\l(t) = a + \delta a(t)$
and
$\bl(t) = \ba + \delta \ba(t)$,
one then finds that \eqref{dequantizer.normal}
can be recast to
\begin{align}
    \label{expval.ab}
    \Q^{-1}(\hat{f})(a,\ba)
    \,=\,
    \Bexpval{
        f\bigbig{
            a + \delta a(t_0)
            ,
            \ba + \delta \ba(t_0)
        }
    }
    \,,
\end{align}
which employs the path integration
\begin{align}
    \label{Bexpval.ab}
    \Bexpval{\color{emphcolor}
        (\,\cdots)
    }
    \,=\,
        \int \D{\d a}\hem \D{\d\ba}\,\,
        \exp\nem\bb{
            -\frac{1}{\hbar}\mem
                \int^{t_+}_{t_-} dt\,\,
                    \delta\ba(t)\mem \delta\dot{a}(t)
        }\,
        {\color{emphcolor}
            (\,\cdots)
        }
\end{align}
with boundary conditions
$\delta a(t_-) = 0$,
$\delta \ba(t_+) = 0$.
As a result, 
the propagator is given by
\begin{align}
    \label{prop.ab}
    \Bexpval{
        \delta a(t_1)\, \delta a(t_2)
    }
    \hem=\mem
        0
    \,,\,\,\,\,
    \Bexpval{
        \delta a(t_1)\, \delta \ba(t_2)
    }
    \hem=\mem
        \hbar\,
            \Theta(t_1,t_2)
    \,,\,\,\,\,
    \Bexpval{
        \delta \ba(t_1)\, \delta \ba(t_2)
    }
    \hem=\mem
        0
    \,,
\end{align}
where $\Theta(t_1,t_2)$ 
equals 
$1$ if $t_1>t_2$ and $0$ if $t_1<t_2$.
Finally,
the operator ordering is implemented 
as the time ordering
in the path integral as
\begin{align}
    \label{cattaneo.ab}
    \Q^{-1}(\hat{f}\hem\hat{g})(a,\ba)
    \,=\,
        \Bexpval{
            f\bigbig{
                a + \delta a(t_1)
                ,
                \ba + \delta \ba(t_1)
            }
            \,
            g\bigbig{
                a + \delta a(t_2)
                ,
                \ba + \delta \ba(t_2)
            }
        }
    \,,
\end{align}
where $t_1$ and $t_2$ are arbitrary times such that
$t_+ > t_1 > t_2 > t_-$.
Again,
such a freedom arises because
the path integral is sensitive to
only the ordering between $t_1$ and $t_2$
as the topological information
\cite{deformation-quantification}.
Evaluating \eqref{cattaneo.ab}
with the propagator in \eqref{prop.ab}
readily reproduces the Wick star product in \eqref{wick-def},
which sums over vacuum bubble diagrams
arising from joining $f$ and $g$ as two vertices.

\newpage

In the operator language,
\eqref{expval.ab}
could be described as
\begin{align}
\begin{split}
    \label{opl.ab}
    \Q^{-1}(\hat{f})(a,\ba)
    \,&=\,
        \Bra{0}
        \normal{
            f\bigbig{
                a + \widehat{\d a}
                ,
                \ba + \widehat{\d \ba}
            }
        }
        \Ket{0}
    \,,
\end{split}
\end{align}
where $a,\ba$ are merely c-number variables;
it is the fluctuation fields
that are quantized.
The vacuum ket (bra) in \eqref{opl.ab}
is annihilated by $\widehat{\d a}$ ($\widehat{\d\ba}$).

With this understanding,
consider abbreviating
\eqrefsor{expval.ab}{opl.ab} to
\begin{align}
    \label{abbrv.ab}
    \Q^{-1}(\hat{f})
    \,=\,
        \expval{\,\hat{f}\,}_\text{bkgd}
    \,.
\end{align}
In the path integral perspective,
the definition of the right-hand side in \eqref{abbrv.ab}
is to sum over vacuum bubble diagrams
in a \textit{background field theory} computation,
which is indeed what \rcite{Haddad:2025cmw} performs.
We clarify that this path integral is with respect to the topological (zero-Hamiltonian) action.
In the operator perspective,
the definition of the right-hand side in \eqref{abbrv.ab}
is to compute the vacuum expectation value 
with the crucial stipulation that
both the quantization and vacuum are
implemented
with respect to the \textit{fluctuation} fields.

With this clarification,
we find \eqref{expval-chi}
from $\chi^\star = \Q^{-1}(\hat{\chi})$
in \eqref{limitE.a}.
For \eqref{expval-S},
consider the relation derived in \eqref{star-exp}.

A remark
regarding \eqref{expval-S}
is that
its path integral representation
can take either of the following
as the phase space action:
\begin{align}
    \label{PI.Ieff}
    I_\text{eff}[x,p]
    &\,=\,
    \int_0^1 dt\,\,
    \bb{
        \frac{1}{2}\mem\BB{
            p(t)\, \dot{x}(t) - \dot{p}(t)\, x(t)
        }
        - \chi^\star(x(t),p(t),t)
    }
    \,,\\
    \label{PI.I}
    I[x,p]
    &\,=\,
    \int_{-\infty}^{+\infty} dt\,\,
    \bb{
        \frac{1}{2}\mem\BB{
            p(t)\, \dot{x}(t) - \dot{p}(t)\, x(t)
        }
        - \tV^\star(x(t),p(t),t)
    }
    \,.
\end{align}
This is indeed consistent with the fact that
$\schi$ is the effective Hamiltonian
reproducing the time evolution within unit time.
Specifically, it can be seen that
\begin{align}
    \int \D{x}\hem\D{p}\,\,
        \mathe^{
            -I_\text{eff}[x,p]
            /i\hbar
        }
    \,=\,
    \expval{\,
        \exp(\hat{\chi}/i\hbar)
    \,}_\text{bkgd}
    \,=\,
    \expval{\,
        \hat{S}
    \,}_\text{bkgd}
    \,=\,
        \int \D{x}\hem\D{p}\,\,
            \mathe^{
                -I[x,p]
                /i\hbar
            }
    \,,
\end{align}
as the exponential $\exp(\hat{\chi}/i\hbar)$
merges with the exponentiated topological part of the action
in the path integral.
%
%
For the free theory
$H^\circ = p^2\nem/2m$,
for instance,
\eqref{PI.I} will
precisely describe
the perturbation theory of \rcite{Haddad:2025cmw}
that expands around a straight-line trajectory,
provided a proper treatment of
the interaction picture \cite{eikonalsangmin1}.

Finally, note the subtle difference between
\begin{subequations}
\begin{align}
    \label{explog.a}
    i\hbar\,
    \expval{\,\log \hat{S}\,}_\text{bkgd}
    \,&=\,
        i\hbar\mem \log^\star\nem \Q^{-1}(\hat{S})
    \,=\,
        \chi^\star
    \,,\\
    \label{explog.b}
    i\hbar\mem \log
    \expval{\,\hat{S}\,}_\text{bkgd}
    \,&=\,
        i\hbar\mem \log \Q^{-1}(\hat{S})
    \,,
\end{align}
\end{subequations}
where $\log^\star$ is an inverse of $\exp^\star$
via $\log^\star \circ\, \Q^{-1} = \Q^{-1} \,\circ\, \log$.
As is shown in \eqref{explog.a},
$\hbar$-deformed functions such as $\log^\star$ or $\exp^\star$
translate to
ordinary functions
acting inside the vacuum expectation value.
As is shown in \eqref{explog.b},
however, ordinary functions such as $\log$ or $\exp$
simply act outside the expectation value.
As should be clear from \eqref{abbrv.ab},
$\expval{\,\,\,}_\text{bkgd}$
is essentially synonymous to $\Q^{-1}$:
extracting the symbol.

\newpage

\subsection{More on Wightman Tensor}
\label{WIGHTMAN}

In \Sec{QEMAGNUS.WICK},
an oscillator is viewed as a Hamiltonian system that takes a K\"ahler vector space
as the phase space,
on which
the Wightman tensor is defined
to yield the Wick star product.
This appendix elaborates on the details of this construction.

Concretely, suppose $N$-dimensional harmonic oscillator.
The phase space is 
the K\"ahler vector space
$\P = \C^N$
with complex coordinates $a_\a$.
The \emph{K\"ahler triple} is
formed by
metric $A$,
symplectic form $\sp$,
and complex structure $J$
such that
$A(X,Y) = \sp(J(X),Y)$:
\begin{align}
    A
    \,&=\,
        \delta^{\bar{\b}\a}\mem
        \BB{
            d\ba_\wrap{\bar{\b}} \otimes d a_\a
            +
            da_\a \otimes d\ba_\wrap{\bar{\b}}
        }
    \,,\\
\label{1sp}
    \sp
    \,&=\,
        i\mem \BB{
            d\ba^\a \otimes da_\a
            -
            da_\a \otimes d\ba^\a
        }
    \,,
\end{align}
where $\ba_\wrap{\bar{\b}} = [a_\b]^*$.
Note that one takes
$\ba^\a = \ba_\wrap{\bar{\b}}\mem \delta^{\bar{\b}\a}$
via the metric.

The inverse metric $A^{-1}$ is the pointwise inverse of $A$:
\begin{align}
    A^{-1}
    \,=\,
            \frac{\partial}{\partial a_\a} \otimes \frac{\partial}{\partial \ba^\a}
        +
            \frac{\partial}{\partial \ba^\a} \otimes \frac{\partial}{\partial a_\a}
    \,.
\end{align}
\begin{subequations}
The Poisson tensor $\Pi$ is the pointwise inverse of $\omega$:
\begin{align}
    \Pi
    \,=\,
        -i\, 
            \frac{\partial}{\partial a_\a} \otimes \frac{\partial}{\partial \ba^\a}
        +i\, 
            \frac{\partial}{\partial \ba^\a} \otimes \frac{\partial}{\partial a_\a}
    \,.
\end{align}
Crucially, the \emph{Wightman tensor} $W$ is defined as
\begin{align}
    W
    \,=\,
        \frac{1}{2}\,\BB{
            A^{-1} + i\mem \Pi
        }
    \,=\,
        \frac{\partial}{\partial a_\a} \otimes \frac{\partial}{\partial \ba^\a}
    \,.
\end{align}
This is the pointwise inverse of the hermitian form 
$\langle\blank,\nem\hnem\blank\rangle$ on $\C^N$,
in fact.
Its transpose is
\begin{align}
    W^\top
    \,=\,
        \frac{1}{2}\,\BB{
            A^{-1} - i\mem \Pi
        }
    \,=\,
        \frac{\partial}{\partial \ba^\a} \otimes \frac{\partial}{\partial a_\a}
    \,.
\end{align}
Note the relation
\begin{align}
    \Pi
    \,=\,
        \frac{1}{i}\,\BB{
            W - W^\top
        }
    \,.
\end{align}
\end{subequations}

\begin{subequations}
\label{x}
The Poisson bracket is defined as
(regard $I_1,I_2$ as abstract indices)
\begin{align}
    \label{xPB}
    \pb{f}{g}
    \,=\,
    f\mem
        \,\BB{
        \overleftarrow{\partial_{I_1}}\,
            \Pi^{I_1I_2}\,
        \,\overrightarrow{\partial_{I_2}}
        }\,
    g
    \,=\,
        -i\,
        \frac{\partial f}{\partial a_\a}
        \frac{\partial g}{\partial \ba^\a}
        +i\,
        \frac{\partial f}{\partial \ba^\a}
        \frac{\partial g}{\partial a_\a}
    \,,
\end{align}
which arises from the Poisson tensor.
The \emph{Wightman bracket} is defined as
\begin{align}
    \label{xWB}
    \pb{f}{g}^+
    \,=\,
    f\mem
        \,\BB{
        \overleftarrow{\partial_{I_1}}\,
            W^{I_1I_2}\,
        \,\overrightarrow{\partial_{I_2}}
        }\,
    g
    \,=\,
        \frac{\partial f}{\partial a_\a}
        \frac{\partial g}{\partial \ba^\a}
    \,,
\end{align}
which arises from the Wightman tensor.
The \emph{transposed Wightman bracket} 
is defined as
\begin{align}
    \label{xWBr}
    \pb{f}{g}^-
    \,=\,
    f\mem
        \,\BB{
        \overleftarrow{\partial_{I_1}}\,
            W^{I_2I_1}\,
        \,\overrightarrow{\partial_{I_2}}
        }\,
    g
    \,=\,
        \frac{\partial f}{\partial \ba^\a}
        \frac{\partial g}{\partial a_\a}
    \,=\,
        \pb{g}{f}^+
    \,,
\end{align}
which arises from the transpose $(W^\top)^{I_1I_2} = W^{I_2I_1}$ of the Wightman tensor.
Then we have
\begin{align}
    \label{xdecomp}
    \pb{f}{g}
    \,=\,
        \frac{1}{i}\,\BB{\mem
            \pb{f}{g}^+
            -
            \pb{f}{g}^-
        }
    \,.
\end{align}
\end{subequations}

The Wick star product in \eqref{starW}
is an exponentiation of the Wightman bracket in \eqref{xWB}.
That said,
the Wightman bracket $\pb{f}{g}^+$
is the $\O(\hbar^1)$ part of $f \star g$.

\subsection{Field Theory}
\label{QFT}

This appendix shows that 
the results in
Sections \ref{PSFOR},\,\ref{SCA},\,\ref{LIMIT},\,and\,\ref{MAGNUS}
universally apply to both particles and fields.
Especially, we show that 
the final form of eikonals in \Sec{MAGNUS}
has precisely described
Feynman diagrams in the retarded causality prescription,
in both quantum mechanics and quantum field theory.

\begin{subequations}
To recapitulate,
the definitions of the retarded propagator
and its positive- and negative-frequency parts
were proposed in \eqrefss{feynman2}{F2}{Pi2},
with their relation stated in \eqref{decomp}.
For the reader's sake, we reproduce them below:
\label{recap}
\begin{align}
    \label{recap0}
    \ainc{valign=c}{F2}
    \,&=\,\,\,
        \int d^2t\,\,\,\,
            \tV(t_1)\,
            \BB{\,
            \overleftarrow{\partial_{I_1}}\,
                \Pi^{I_1I_2}\,
                \Theta(t_1,t_2)
            \,\overrightarrow{\partial_{I_2}}
            \,}
            \,\tV(t_2)
    \,,\\
    \label{recapp}
    \ainc{valign=c}{Fp2}
    \,&=\,\,\,
        \int d^2t\,\,\,\,
            \tV(t_1)\,
            \BB{\,
            \overleftarrow{\partial_{I_1}}\,
                W^{I_1I_2}\,
                \Theta(t_1,t_2)
            \,\overrightarrow{\partial_{I_2}}
            \,}
            \,\tV(t_2)
    \,,\\
    \label{recapm}
    \ainc{valign=c}{Fm2}
    \,&=\,\,\,
        \int d^2t\,\,\,\,
            \tV(t_1)\,
            \BB{\,
            \overleftarrow{\partial_{I_1}}\,
                W^{I_2I_1}\,
                \Theta(t_1,t_2)
            \,\overrightarrow{\partial_{I_2}}
            \,}
            \,\tV(t_2)
    \,,
\end{align}
where
\begin{align}
    \label{recap.decomp}
    \inc{valign=c}{F2}
    \,=\,
    \frac{1}{i}\mem \BB{
        \inc{valign=c}{Fp2}
    -
        \inc{valign=c}{Fm2}
    }
    \,.
\end{align}
\end{subequations}
Since we treat interactions perturbatively,
it suffices to examine free theories.
We also use the notations
$\dbar \xi = d\xi/2\pi$,
$\deltabar(\xi) = 2\pi\hem \delta(\xi)$.

Recalling a famous quote by Sidney Coleman,
we first revisit
the one-dimensional harmonic oscillator 
defined by \eqrefss{omega.1d}{1d.H0}{wick-def}.
The phase space is $\P = \R^2 \cong \C^1$.
The classical and quantum interaction pictures 
coincide as shown in \App{INTCD}.
The free time evolution $\Uo(t,0)$
acts on the phase space simply as a phase rotation,
$a \mapsto a\mem \mathe^{-i\omega t}$,
$\ba \mapsto \ba\mem \mathe^{i\omega t}$,
where we take the fixed time slice as $t_0 = 0$ for simplicity.
We then take the position variable $x$
in \eqref{ab-xp}
as a function on the phase space.
Its interaction picture image
$\tx(t)$
splits into the
\emph{positive- and negative-} \emph{frequency parts} as
\begin{align}
    \tx(t)
    \,=\,
		\frac{1}{\sqrt{2m\omega}}\,
            a\mem \mathe^{-i\omega t}
        \,+\,
		\frac{1}{\sqrt{2m\omega}}\,
            \ba\mem \mathe^{+i\omega t}
    \,.
\end{align}
The Wick star product in \eqref{wick-def} implies
\begin{align}
    \label{app.1W}
    \tx(t_1)
    \mem\star\mem
    \tx(t_2)
    \,=\,
        \tx(t_1)\mem \tx(t_2)
    +
        \hbar\mem
        W(t_1,t_2)
    \,,
\end{align}
which derives the \emph{Wightman function} $W(t_1,t_2)$.
The Poisson bracket in \eqref{omega.1d} implies
\begin{align}
    \label{app.1PJ}
        \pb{\tx(t_1)}{\tx(t_2)}
    \,=\,
        \frac{1}{i}\,\BB{
            W(t_1,t_2)
            -
            W(t_2,t_1)
        }
    \,=\,
        \Delta(t_1,t_2)
    \,,
\end{align}
which derives
the \emph{Pauli-Jordan function} $\Delta(t_1,t_2) = -\Delta(t_2,t_1)$.
Explicitly,
\begin{align}
    \label{app.1explicit}
    W(t_1,t_2)
    \,=\,
        \frac{1}{2m\omega}\,
            \mathe^{-i\omega(t_1-t_2)}
    \,,\quad
    \Delta(t_1,t_2)
    \,=\,
        - \frac{
            \sin(\hem
                \omega(t_1{\mem-\,}t_2)
            )
        }{m\omega}
    \,.
\end{align}

From the clarifications made in \App{WIGHTMAN},
it should be clear that 
\eqrefs{app.1W}{app.1PJ}
have computed
the Wightman and Poisson brackets,
\begin{align}
\begin{split}
    \label{app.1brs}
    \tx(t_1)
    \,
    \BB{\,
    \overleftarrow{\partial_{I_1}}\,
        W^{I_1I_2}
    \,\overrightarrow{\partial_{I_2}}
    \,}
    \,
    \tx(t_2)
    \,&=\,
        W(t_1,t_2)
    \,,\\
    \tx(t_1)
    \,
    \BB{\,
    \overleftarrow{\partial_{I_1}}\,
        \Pi^{I_1I_2}
    \,\overrightarrow{\partial_{I_2}}
    \,}
    \,
    \tx(t_2)
    \,&=\,
        \Delta(t_1,t_2)
    \,.
\end{split}
\end{align}
Here, $I_1,I_2$ are
indices taking values in $\{1,2\}$,
which understands
the phase space concretely
as a real manifold $\R^2$.

With this understanding,
we consider the case in which the interaction Hamiltonian
arises from a potential $v(x)$ in position space,
so $\tV(x,p;t) = v(\tx(x,p;t))$.
By using chain rule,
\eqref{recap} then
boils down to
\begin{subequations}
\label{recap.1}
\begin{align}
    \label{recap0.1}
    \ainc{valign=c}{F2}
    \,&=\,\,\,
        \int d^2t\,\,\,\,
            v'\hnem(\tx(t_1))\,
            \BB{
                \Delta(t_1,t_2)\,
                \Theta(t_1,t_2)
            }\,
            v'\hnem(\tx(t_2))
    \,,\\
    \label{recapp.1}
    \ainc{valign=c}{Fp2}
    \,&=\,\,\,
        \int d^2t\,\,\,\,
            v'\hnem(\tx(t_1))\,
            \BB{
                W(t_1,t_2)\,
                \Theta(t_1,t_2)
            }\,
            v'\hnem(\tx(t_2))
    \,,\\
    \label{recapm.1}
    \ainc{valign=c}{Fm2}
    \,&=\,\,\,
        \int d^2t\,\,\,\,
            v'\hnem(\tx(t_1))\,
            \BB{
                W(t_2,t_1)\,
                \Theta(t_1,t_2)
            }\,
            v'\hnem(\tx(t_2))
    \,,
\end{align}
\end{subequations}
which should be clear from \eqref{app.1brs}.

Crucially, the bracketed terms in 
\eqrefss{recap0.1}{recapp.1}{recapm.1}
are respectively
the \emph{retarded propagator},
the \emph{positive-frequency retarded} \emph{propagator},
and
the \emph{negative-frequency} \emph{retarded propagator}:
\begin{subequations}
\label{Gret}
\begin{align}
    \label{Gret0}
    G_\text{ret}(t_1,t_2)
    \,&=\,
        \Delta(t_1,t_2)\, \Theta(t_1,t_2)
    \,,\\
    \label{Gretp}
    G_\text{ret}^+(t_1,t_2)
    \,&=\,
        W(t_1,t_2)\, \Theta(t_1,t_2)
    \,,\\
    \label{Gretm}
    G_\text{ret}^-(t_1,t_2)
    \,&=\,
        W(t_2,t_1)\, \Theta(t_1,t_2)
    \,.
\end{align}
To clarify, $G_\text{ret}^\pm(t_1,t_2)$
are the positive- and negative-frequency parts of
$G_\text{ret}(t_1,t_2)$
up to the customary $\pm i$ factors:
\begin{align}
    \label{Gret.decomp}
    G_\text{ret}(t_1,t_2)
    \,=\,
        \frac{1}{i}\,\BB{
            G_\text{ret}^+(t_1,t_2)
            -
            G_\text{ret}^-(t_1,t_2)
        }
    \,.
\end{align}
\end{subequations}

To see this,
note first
that the Wightman and Pauli-Jordan functions are
\textit{homogeneous} solutions to the harmonic oscillator equations of motion,
which one can directly verify
from their explicit form given in
\eqref{app.1explicit}:
\begin{align}
    m\mem
    \bb{
        - \frac{\partial^2}{{\partial t_1}^2}
        - \omega^2
    }\,
        W(t_1,t_2)
    \,=\,
        0
    \,,\quad
    m\mem
    \bb{
        - \frac{\partial^2}{{\partial t_1}^2}
        - \omega^2
    }\,
        \Delta(t_1,t_2)
    \,=\,
        0
    \,.
\end{align}
Similarly, it follows that
$G_\text{ret}(t_1,t_2)$ in \eqref{Gret0}
is the retarded Green's function
solving the \textit{inhomogeneous} equation
\begin{align}    
    m\mem
    \bb{
        - \frac{\partial^2}{{\partial t_1}^2}
        - \omega^2
    }\,
        G_\text{ret}(t_1,t_2)
    \,=\,
        \delta(t_1{\mem-\,}t_2)
    \,.
\end{align}

Consequently,
\eqref{recap.1}
precisely describes
the integrals
that one obtains in a covariant perturbation theory
in the retarded causality prescription.
This establishes that
the left-hand sides in \eqref{recap.1}
are Feynman diagrams
in the precise sense,
constructed with retarded propagators.

\newpage

It is easy to generalize the above result
to the case of an $N$-dimensional oscillator,
in which case the indices $I_1,I_2$
are valued in a set of $2N$ integers.
By promoting them to continuous labels,
one obtains field theories
without much difficulty.

Concretely,
suppose the Klein-Gordon field in $d$ spacetime dimensions.
As a Hamiltonian system,
it describes the infinite-dimensional phase space
$\P = T^*(\Cinfty(\R^{d-1}))$
whose coordinates are
$\phi : \vex \mapsto \phi^\vex$
and 
$\pi : \vex \mapsto \pi^\vex$,
where $\vex \in \R^{d-1}$.
In particular, the derivation of
the Poisson bracket and Hamiltonian
from the relativistic Lagrangian
is a textbook matter.
To establish the interpretation as an oscillator system,
one considers the following change of coordinates
as a generalization of \eqref{ab-xp}:
\begin{align}
\begin{split}
    \label{a2.ab-xp}
    \phi^\vx
    \,&=\,
    \int
        \frac{\dbar^{d-1}\nem k}{2\omega(\vk)}\,\,
        \BB{
            a_\vk\, \mathe^{i\vk\cdot\vx}
            +
            \ba^\vk\, \mathe^{-i\vk\cdot\vx}
        \mem}
    \,,\\
    \pi^\vx
    \,&=\,
    \int
        {\dbar^{d-1}\nem k}
        \,\,
        \frac{1}{2i}\mem
        \BB{
            a_\vk\, \mathe^{i\vk\cdot\vx}
            -
            \ba^\vk\, \mathe^{-i\vk\cdot\vx}
        \mem}
    \,.
\end{split}
\end{align}
Here, a dispersion relation
\smash{$\omega(\vk) = (\vk{}^2 + \m^2)^{1/2}$} is assumed
for a rest-mass parameter $\m$.
As is well-known, the Poisson bracket is then given by
\begin{align}
    \label{a2.pb}
    \BB{\,
    \overleftarrow{\partial_{I_1}}\,
        \Pi^{I_1I_2}
    \,\overrightarrow{\partial_{I_2}}
    \,}
    \,&=\,
    -i\mem
        \int \dbar^{d-1}\nem k\,\,\mem
    2\omega(\vk)\,
    \bb{
        \overleftarrow{\frac{\delta}{\delta a_\vk}}
        \mem
        \overrightarrow{\frac{\delta}{\delta \ba^\vk}}
        -
        \overleftarrow{\frac{\delta}{\delta \ba^\vk}}
        \mem
        \overrightarrow{\frac{\delta}{\delta a_\vk}}
    }
    \,.
\end{align}
Here,
$I_1,I_2$ 
may be identified with
the continuous indices $\vx_1,\vx_2$
taking values in $\R^{d-1}$.
Yet more elegantly,
one can regard them as
abstract indices given for the infinite-dimensional manifold $\P$.

Next, one quantizes the system in the normal ordering prescription,
in which case the Wick contraction rule is
\begin{align}
    \label{a2.star}
    a_{\vk_1} \nem\star\mem \ba^{\vk_2}
    \,&=\,
        a_{\vk_1} \ba^{\vk_2}
        +
        2\hbar\mem \omega(\vk_1)\,
        \deltabar^{(d-1)}(
            \vk_1 {\mem-\,} \vk_2
        )
    \,.
\end{align}
From the discussion around \eqref{wick-made-easy},
it should be clear that
\eqref{a2.star} is a mere rewriting of the textbook equation,
\begin{align}
        \hat{a}_{\vk_1} 
        \hem
        \hat{\ba}^{\vk_2}
    \,&=\,
    \normal{
        \hat{a}_{\vk_1}\hem \hat{\ba}^{\vk_2}
    }
        +
        2\hbar\mem \omega(\vk_1)\,
        \deltabar^{(d-1)}(
            \vk_1 {\mem-\,} \vk_2
        )
    \,.
\end{align}
Generalizing \eqref{wick-def},
the Wick star product is given by
\begin{align}
    \label{a2.star}
    \star
    \,\,\,\,=\,\,\,
    \exp\nem
    \BB{\,
    \overleftarrow{\partial_{I_1}}\,
        \hbar\mem W^{I_1I_2}
    \,\overrightarrow{\partial_{I_2}}
    \,}
    \mem\,&=\,\,\mem
    \exp\nem
    \bbsq{\,\mem
        \hbar
            \int \dbar^{d-1}\nem k\,\,\mem
        2\omega(\vk)\,
        \bb{
            \overleftarrow{\frac{\delta}{\delta a_\vk}}
            \mem
            \overrightarrow{\frac{\delta}{\delta \ba^\vk}}
        }
    \mem
    }
    \,.
\end{align}

The free Hamitonian is
\begin{align}
    \label{a2.H0}
    H^\circ
    \,&=\,
        \int \frac{\dbar^{d-1}\nem k}{2\omega(\vk)}\,\,
        \BB{
            \omega(\vk)\,\hem \ba^\vk a_\vk
        }
    \,.
\end{align}
In turn,
the interaction picture image of
the ``position variable'' $\phi$ in \eqref{a2.ab-xp}
is
\begin{align}
    \label{a2.intphi}
    \tphi^\vex(t)
    \,=
        \int
            \frac{\dbar^{d-1}\nem k}{2\omega(\vk)}\,\,
            \BB{
                a_\vk\, \mathe^{-i\omega(\vk)\hem t + i\vk\cdot\vx}
                +
                \ba^\vk\, \mathe^{i\omega(\vk)\hem t - i\vk\cdot\vx}
            \mem}
    \,,
\end{align}
yielding
the split of the time-evolved free field into
\emph{positive- and negative-frequency parts}.

\newpage

The Wick star product in \eqref{a2.star} implies
\begin{align}
    \label{phi.W}
    \tphi^{\vx_1}(t_1)
    \mem\star\mem
    \tphi^{\vx_2}(t_2)
    \,=\,
        \tphi^{\vx_1}(t_1)\mem \tphi^{\vx_2}(t_2)
    +
        \hbar\mem
        W((t_1,\vx_1),(t_2,\vx_2))
    \,,
\end{align}
where the \emph{Wightman function} 
is defined as a function $W(x_1,x_2)$ of two spacetime points.
The Poisson bracket in \eqref{a2.pb} then implies
\begin{align}
\begin{split}
    \label{phi.PJ}
    \pb{
        \tphi^{\vx_1}(t_1)
    }{
        \tphi^{\vx_2}(t_2)
    }
    \,&=\,
        \frac{1}{i}\,\BB{
            W((t_1,\vx_1),(t_2,\vx_2))
            -
            W((t_2,\vx_2),(t_1,\vx_1))
        }
    \,,\\
    \,&=\,
        \Delta((t_1,\vx_1),(t_2,\vx_2))
    \,,
\end{split}
\end{align}
where the \emph{Pauli-Jordan function}
is defined as a function $\Delta(x_1,x_2) = -\Delta(x_2,x_1)$ of two spacetime points.
Explicitly,
\begin{align}
\begin{split}
\label{app.2explicit}
    W(x_1,x_2)
    \,&=\,
    \int
        \dbar^{d}\nem k\,\,
            \deltabar(k^2 {\mem+\,} \mu^2)\,
            \Theta(k^0)\,
            \mathe^{ik(x_1-x_2)}
    \,,\\
    \Delta(x_1,x_2)
    \,&=\,
    \frac{1}{i}\mem
    \int
        \dbar^{d}\nem k\,\,
            \deltabar(k^2 {\mem+\,} \mu^2)\,
            \sgn(k^0)\,
            \mathe^{ik(x_1-x_2)}
    \,.
\end{split}
\end{align}
Again,
\eqrefs{phi.W}{phi.PJ}
have computed
the Wightman and Poisson brackets,
\begin{align}
\begin{split}
    \label{app.2brs}
    \tphi^{\vx_1}(t_1)
    \,
    \BB{\,
    \overleftarrow{\partial_{I_1}}\,
        W^{I_1I_2}
    \,\overrightarrow{\partial_{I_2}}
    \,}
    \,
    \tphi^{\vx_2}(t_2)
    \,&=\,
        W((t_1,\vx_1),(t_2,\vx_2))
    \,,\\
    \tphi^{\vx_1}(t_1)
    \,
    \BB{\,
    \overleftarrow{\partial_{I_1}}\,
        \Pi^{I_1I_2}
    \,\overrightarrow{\partial_{I_2}}
    \,}
    \,
    \tphi^{\vx_2}(t_2)
    \,&=\,
        \Delta((t_1,\vx_1),(t_2,\vx_2))
    \,.
\end{split}
\end{align}
It should be clear that \eqref{app.2brs}
has merely rewritten the textbook equations
about Wick contraction and commutator
between fields at different points.

With this understanding,
we consider the typical situation
in which the interaction Hamiltonian
arises from a potential $v : \R \to \R$:
\begin{align}
    V
    \,=\,
        \int d^{d-1}\nem x\,\,\,
            v\bigbig{
                \phi^\vx
            }
    \qiq
    \tV(t)
    \,=\,
        \int d^{d-1}\nem x\,\,\,
            v\bigbig{\hem
                \tphi^\vx(t)
            }
    \,.
\end{align}
By using chain rule,
\eqref{recap} then
boils down to
\begin{subequations}
\label{recap.2}
\begin{align}
    \label{recap0.2}
    \inc{valign=c}{F2}
    \,\,&=\,\,\mem
        \int_{t_1,t_2,\vx_1,\vx_2} 
            v'\hnem\bigbig{
                \tphi^{\vx_1}(t_1)
            }\,
            \BB{
                \Delta((t_1,\vx_1),(t_2,\vx_2))
                \,
                \Theta(t_1,t_2)
            }\,
            v'\hnem\bigbig{
                \tphi^{\vx_2}(t_2)
            }
    \,,\\
    \label{recapp.2}
    \inc{valign=c}{Fp2}
    \,\,&=\,\,\mem
        \int_{t_1,t_2,\vx_1,\vx_2}
            v'\hnem\bigbig{
                \tphi^{\vx_1}(t_1)
            }\,
            \BB{
                W((t_1,\vx_1),(t_2,\vx_2))
                \,
                \Theta(t_1,t_2)
            }\,
            v'\hnem\bigbig{
                \tphi^{\vx_2}(t_2)
            }
    \,,\\
    \label{recapm.2}
    \inc{valign=c}{Fm2}
    \,\,&=\,\,\mem
        \int_{t_1,t_2,\vx_1,\vx_2}
            v'\hnem\bigbig{
                \tphi^{\vx_1}(t_1)
            }\,
            \BB{
                W((t_2,\vx_2),(t_1,\vx_1))
                \,
                \Theta(t_1,t_2)
            }\,
            v'\hnem\bigbig{
                \tphi^{\vx_2}(t_2)
            }
    \,,
\end{align}
\end{subequations}
which follows from \eqref{app.2brs}.
Here, we have abbreviated
$\int dt_1\mem dt_2\mem d^{d-1}\nem x_1\mem d^{d-1}\nem x_2$
as
$\int_{t_1,t_2,\vx_1,\vx_2}$.

Crucially,
the bracketed terms in 
\eqrefss{recap0.2}{recapp.2}{recapm.2}
describe respectively
the \emph{retarded propagator},
the \emph{positive-frequency retarded} \emph{propagator},
and
the \emph{negative-frequency retarded propagator}:
\begin{subequations}
\label{Gret}
\begin{align}
    \label{Gret0}
    G_\text{ret}(x_1,x_2)
    \,&=\,
        \Delta(x_1,x_2)\, \Theta(x_1^0,x_2^0)
    \,,\\
    \label{Gretp}
    G_\text{ret}^+(x_1,x_2)
    \,&=\,
        W(x_1,x_2)\, \Theta(x_1^0,x_2^0)
    \,,\\
    \label{Gretm}
    G_\text{ret}^-(x_1,x_2)
    \,&=\,
        W(x_2,x_1)\, \Theta(x_1^0,x_2^0)
    \,.
\end{align}
To clarify, $G_\text{ret}^\pm(x_1,x_2)$
are the positive- and negative-frequency parts of
$G_\text{ret}(x_1,x_2)$
up to the customary $\pm i$ factors:
\begin{align}
    \label{Gret.decomp}
    G_\text{ret}(x_1,x_2)
    \,=\,
        \frac{1}{i}\,\BB{
            G_\text{ret}^+(x_1,x_2)
            -
            G_\text{ret}^-(x_1,x_2)
        }
    \,.
\end{align}
\end{subequations}

To see this,
note first
that the Wightman and Pauli-Jordan functions are
\textit{homogeneous} solutions to the Klein-Gordon equation,
which one can directly 
verify from the on-shell support $\deltabar(k^2{\mem+\,} \mu^2)$ in
\eqref{app.2explicit}:
\begin{align}
    \BB{
        \partial_1^2 - \mu^2
    }\,
        W(t_1,t_2)
    \,=\,
        0
    \,,\quad
    \BB{
        \partial_1^2 - \mu^2
    }\,
        \Delta(t_1,t_2)
    \,=\,
        0
    \,.
\end{align}
Similarly, it follows that
$G_\text{ret}(t_1,t_2)$ in \eqref{Gret0}
is the retarded Green's function
solving the following \textit{inhomogeneous} Klein-Gordon equation,
which is a well-known fact:
\begin{align}    
    \BB{
        \partial_1^2 - \mu^2
    }\,
        G_\text{ret}(x_1,x_2)
    \,=\,
        \delta^{(d)}\hnem(x_1{\mem-\,}x_2)
    \,.
\end{align}

Eventually,
we switch to the conventional notation in quantum field theory via
\begin{align}
    \tphi^\vx(t)
    \,=\,
    \phi_\mathrm{I}(x)
    \transition{where}
    x \,=\, (t,\vx)
    \,\in\, \R^{1,d-1}
    \,,
\end{align}
where the subscript $\mathrm{I}$ 
signifies ``interaction picture.''
Then \eqref{recap.2} is brought to
\begin{subequations}
\label{recap.QFT}
\begin{align}
    \label{recap0.QFT}
    \ainc{valign=c}{F2}
    \,&=\,\,\,
        \int d^dx_1\mem d^dx_2\,\,\,
            v'\hnem(\phi_\mathrm{I}(x_1))\,
                G_\text{ret}(x_1,x_2)\,
            v'\hnem(\phi_\mathrm{I}(x_2))
    \,,\\
    \label{recapp.QFT}
    \ainc{valign=c}{Fp2}
    \,&=\,\,\,
        \int d^dx_1\mem d^dx_2\,\,\,
            v'\hnem(\phi_\mathrm{I}(x_1))\,
                G_\text{ret}^+(x_1,x_2)\,
            v'\hnem(\phi_\mathrm{I}(x_2))
    \,,\\
    \label{recapm.QFT}
    \ainc{valign=c}{Fm2}
    \,&=\,\,\,
        \int d^dx_1\mem d^dx_2\,\,\,
            v'\hnem(\phi_\mathrm{I}(x_1))\,
                G_\text{ret}^-(x_1,x_2)\,
            v'\hnem(\phi_\mathrm{I}(x_2))
    \,,
\end{align}
\end{subequations}
Evidently,
\eqref{recap.QFT}
describes nothing but
Feynman diagrams in the retarded causality prescription.
Specifically, 
these are precisely the integrals that
one encounters when
computing the S-matrix 
in the operator formalism
via the interaction picture.
This establishes that
the left-hand sides in \eqref{recap.1}
have described
Feynman diagrams
in the precise sense,
constructed with 
the quantum field theoretical
retarded propagators.

In conclusion,
we have established that
the final forms of eikonals
in \Sec{MAGNUS}
are Feynman diagrams in the retarded causality prescription,
for both quantum mechanics and quantum field theory.
We have simply examined single-multiplicity lines 
since it is easy to derive the same conclusion for multiplied lines as well.

To clarify, however, these diagrams compute operators.
That is, the output is operator-valued.
It is not a matrix element.
For instance, take the cubic potential $v(\phi) = g\hem \phi^3\hnem/3!$.
Then
the second-order classical eikonal, $\chib$,
is found from
\eqref{recap0.QFT} as
\begin{align}
    \label{chi2.QFT}
    \frac{1}{2}
    \ainc{valign=c}{F2}
    =\,\,\,\,
        \frac{g^2}{8}\,
        \int d^dx_1\mem d^dx_2\,\,\,
            (\phi_\mathrm{I}(x_1))^2\,
                G_\text{ret}(x_1,x_2)\,
            (\phi_\mathrm{I}(x_2))^2
    \,.
\end{align}
This is simply a normal-ordered operator
described in the phase space formulation:
\eqref{chi2.QFT} is equivalent to
Eq.\,(4.11) of \rcite{Pichini:2025igz}.

Finally, 
the demonstration of this appendix
should also clarify that 
our results for the final form of eikonals
were manifestly Lorentz covariant:
just interpret the diagrams as in \eqref{recap.QFT}.

\newpage

In fact,
an elaborate framework 
known as Peierls bracket
\cite{Peierls:1952cb,Marolf:1993zk,Duetsch:2002yp}
or
covariant phase space \cite{Witten:1986qs}
can readily manifest covariance in all intermediate steps in our formalism:
perturbative algebraic quantum field theory \cite{Fredenhagen:2012pAQFT}.
Namely, we take $\P$ as the space of solutions to free field equations
and work with its Poisson and Wightman structures
which arise through the Pauli-Jordan and Wightman functions,
which avoids the $1 + (d{\,-\mem}1)$ decomposition.
There is not enough space to delve into this direction in this paper,
but we hope to do so in a future work.

\bibliography{references.bib}

\end{document}